\documentclass[prx,twocolumn,superscriptaddress]{revtex4}

\usepackage{times,amsmath,amsfonts,amssymb}
\usepackage{epstopdf}
\usepackage{graphicx}
\usepackage{float}

\usepackage[usenames]{color}
\usepackage{epsf}
\usepackage{ulem} 

\definecolor{grey}{rgb}{.6,.6,.6}

\newcommand{\cpm}[1]{#1}
\newcommand{\corr}[1]{#1}
\newcommand{\corrnew}[1]{#1}

\definecolor{darkgreen}{rgb}{0, 0.8, 	0.5}
\definecolor{darkgreen}{rgb}{0.2, 0.7, 0.3}
\definecolor{purple}{rgb}{0.7, 0.1, 0.3}

\newcommand{\g}{| g \rangle}
\newcommand{\e}{| e \rangle}
\newcommand{\beq}{\begin{equation}}
\newcommand{\eeq}{\end{equation}}
\newcommand{\bea}{\begin{eqnarray}}
\newcommand{\eea}{\end{eqnarray}}
\newcommand{\upa}{\uparrow}
\newcommand{\downa}{\downarrow}
\newcommand{\Upa}{\Uparrow}
\newcommand{\Downa}{\Downarrow}
\newcommand{\ket}[1]{\left\vert#1\right\rangle}
\newcommand{\bra}[1]{\left\langle#1\right\vert}

\renewcommand{\phi}{\varphi}

\renewcommand\appendix{\par
  \setcounter{section}{0}
  \setcounter{subsection}{0}
  \setcounter{figure}{0}
  \setcounter{table}{0}
  \renewcommand\thesection{Appendix \Alph{section}}
  \renewcommand\thefigure{A\arabic{figure}}
  \renewcommand\thetable{A\arabic{table}}
}

\DeclareFontFamily{U}{euc}{}
\DeclareFontShape{U}{euc}{m}{n}{<-6>eurm5<6-8>eurm7<8->eurm10}{}
\DeclareSymbolFont{AMSc}{U}{euc}{m}{n}
\DeclareMathSymbol{\umu}{\mathord}{AMSc}{"16}

\begin{document}
\title{Exploring the Kondo model in and out of equilibrium with alkaline-earth atoms}

\author{M\'arton Kan\'asz-Nagy} 
\affiliation{Department of Physics, Harvard University, Cambridge MA 02138, United States}
\author{Yuto Ashida}
\affiliation{Department of Physics, University of Tokyo, 7-3-1 Hongo, Bunkyo-ku, Tokyo
113-0033, Japan}
\author{Tao Shi} 
\affiliation{Max-Planck-Institut f\"ur Quantenoptik, Hans-Kopfermann-Strasse. 1, 85748 Garching, Germany}
\affiliation{CAS Key Laboratory of Theoretical Physics, Institute of Theoretical Physics, Chinese Academy of Sciences, P.O. Box 2735, Beijing 100190, China}
\author{C\u at\u alin Pa\c scu  Moca} 
\affiliation{MTA-BME Exotic Quantum Phases "Momentum" Research Group and Department of Theoretical Physics, Budapest University of Technology and Economics, 1111 Budapest, Hungary}
\affiliation{Department of Physics, University of Oradea, 410087, Oradea, Romania}
\author{Tatsuhiko N. Ikeda} 
\affiliation{Institute for Solid State Physics, University of Tokyo, Kashiwa, Chiba 277-8581, Japan}
\author{Simon F\"olling}
\affiliation{Fakult\"at f\"ur Physik, Ludwig-Maximilians-Universit\"at, Schellingstraße 4, 80799 M\"unchen, Germany}
\affiliation{Max-Planck-Institut f\"ur Quantenoptik, Hans-Kopfermann-Straße 1, 85748 Garching, Germany}
\author{J. Ignacio Cirac}
\affiliation{Max-Planck-Institut f\"ur Quantenoptik, Hans-Kopfermann-Strasse. 1, 85748 Garching, Germany}
\author{Gergely Zar\'and} 
\affiliation{MTA-BME Exotic Quantum Phases "Momentum" Research Group and Department of Theoretical Physics, Budapest University of Technology and Economics, 1111 Budapest, Hungary}
\author{Eugene A. Demler} 
\affiliation{Department of Physics, Harvard University, Cambridge MA 02138, United States}

\begin{abstract}
\corr{We propose a scheme to realize the Kondo model with tunable anisotropy using alkaline-earth atoms in an optical lattice. The new feature of our setup is Floquet engineering of interactions using time-dependent Zeeman shifts, that can be realized either using state-dependent optical Stark shifts or magnetic fields. The properties of the resulting Kondo model strongly depend on the anisotropy of the ferromagnetic interactions. In particular,  easy-plane couplings give rise to Kondo singlet formation even though microscopic interactions are all ferromagnetic. We discuss both equilibrium and dynamical properties of the system that can be measured with ultracold atoms, including the impurity spin susceptibility, the impurity spin relaxation rate, as well as the equilibrium and dynamical spin correlations between the impurity and the ferromagnetic bath atoms. We analyze the non-equilibrium time evolution of the system using a variational non-Gaussian approach, which allows us to explore coherent dynamics over both short and long timescales, as set by the bandwidth and the Kondo singlet formation, respectively. In the quench-type experiments, when the Kondo interaction is suddenly switched on, we find that real-time dynamics shows crossovers reminiscent of poor man's renormalization group flow used to describe equilibrium systems. For bare easy-plane ferromagnetic couplings, this allows us to follow the formation of the Kondo screening cloud as the dynamics crosses over from ferromagnetic to antiferromagnetic behavior. On the other side of the phase diagram, our scheme makes it possible to measure quantum corrections to the well-known Korringa law describing the temperature dependence of the impurity spin relaxation rate. Theoretical results discussed in our paper can be measured using currently available experimental techniques.}
\end{abstract}

\date{\normalsize{\today}}

\maketitle

\section{Introduction} \label{sec:introduction}

The Kondo effect is a ubiquitous phenomenon in electron systems. It was originally studied in the context of the anomalous temperature dependence of resistivity of metals, which arises from electron scattering on magnetic impurities~\cite{anderson1961localized, Kondo1964, Abrikosov1965PerturbativeKondo}. Subsequent experimental and theoretical work showed that in systems with \corr{a periodic lattice of} localized spins and itinerant electrons the Kondo effect gives rise to a whole new family of strongly correlated electron systems, the so-called heavy fermion materials~\cite{Graebner1975OriginalHeavyFermionPaper, doniach1977kondo, si1999quantum, coleman2001fermi, si2001locally}.
Strong enhancement of the quasiparticle mass in these materials has its origin in the formation of Kondo singlets~\cite{Nozieres1980KondoEffectInRealMetals, tsvelik1993HeavyFermionKondo, Coleman2001HeavyFermions, HewsonBook}. 
Some of the most intriguing examples of the non-Fermi liquid behavior of electrons have been observed in the vicinity of the quantum critical point between the heavy fermion phase and the magnetically ordered state~\cite{si1999quantum, Lohneysen2007FermiLiquidInstabilities}. In mesoscopic systems Kondo effect also takes on a central role; in particular, transport through small quantum dots in the Coulomb blockade regime for odd occupation numbers is \corrnew{strongly affected by} the formation of Kondo resonances~\cite{glazman1988resonant, ng1988site, GoldhaberGordon1998KondoExperiment, Cronenwett1998TunableKondoEffectInQuantumDots, kouwenhoven2001revival,  jeong2001kondo, mebrahtu2012quantum}. 
\corrnew{Through its equivalence to the spin-boson problem, the Kondo model also describes the process of decoherence and dissipation in many-body quantum systems~\cite{deBruyn1982influence, zwerger1983dynamics, Leggett1987DissipativeTwoStateSystem,  Chakravarty1995DissipativeDynamics, Strong1997TransitionBetweenCoherenceAndIncoherence, costi1998scaling}, and macroscopic quantum tunneling~\cite{Voss1981MacrosopicQuantumTunneling, Chakravarty1982QuantumTunnelingInSuperconductors, han1991observation, Friedman2000MacroscopicQuantumSuperposition, VanDerWal2000MacroscopicQuantumSuperposition}.}
 
From the conceptual point of view, the Kondo effect provides a striking example of the non-perturbative effect of interactions in many-body systems. \corrnew{The antiferromagnetic} Kondo interaction is a relevant perturbation to the Fermi liquid phase, so even for small Kondo scattering the system "flows" to the strongly coupled fixed point at low temperatures~\cite{Anderson1970PoorMansScaling}. The character of the low-temperature fixed point cannot be captured within a simple mean-field approximation and low energy properties of the system are very different from those of the original free electrons. Formation of the Kondo resonance at the Fermi energy intrinsically has a many-body character as manifested, for example, by the anomalous Wilson ratio~\cite{nozieres1974fermi}. 
Accurate theoretical analysis from the spin susceptibility to the specific heat of the Kondo system is possible either in the high temperature/energy limit, where interactions can be treated perturbatively or at very low temperatures where one can start from the low energy fixed point. 
Many theoretical approaches introduced to study the Kondo model attest to the importance and difficulty of this problem. These include perturbative renormalization group~\cite{Anderson1970PoorMansScaling}, Bethe ansatz~\cite{Andrei1980BetheAnsatzKondo, Vigman1980ExactKondo, Andrei.1983, zhou1999algebraic}, 
large-$N$~\cite{coleman1983KondoLattice, bickers1987review} and non-crossing approximations~\cite{keiter1971non-crossing, grewe1981non-crossing, kuramoto1983non-crossing, muller1984non-crossing}, as well as numerical studies utilizing the numerical renormalization group (NRG)~\cite{wilson1975renormalization}, \corr{density matrix renormalization group (DMRG)~\cite{White1992DMRG}, 
density matrix numerical renormalization group (DM-NRG)}~\cite{hofstetter2000generalized} approaches and the flow equation method~\cite{Wegner1994FlowEquationMethod}. \corr{The Kondo effect has been one of the most fruitful areas of condensed matter theory, with many techniques developed in this field subsequently being extended to other systems.}

Most of the earlier theoretical work on Kondo systems focused on equilibrium properties on the linear response to external perturbations. In the last few years, out-of-equilibrium properties of the system have also become the subject of active research. This analysis is motivated by experimental studies of transport through quantum dots at finite bias voltage~\cite{yacoby1995coherence, roch2008quantum, roch2009observation}, 
optical spectroscopy~\cite{Latta2011NonEqKondoOpticallyExcited, tureci2011many}, as well as pump and probe experiments~\cite{Bascov2011Electrodynamics}. Theoretical analysis of the out-of-equilibrium dynamics of Kondo systems is particularly challenging due to the interplay of degrees of freedom at different energies. 

\corr{We have recently} seen considerable progress in realizing a new experimental platform for studying strongly correlated many-body systems, using systems of ultracold atoms as a quantum simulator~\cite{Bloch2008Review}. Recent experiments demonstrated the fermionic Mott state with long-range antiferromagnetic order~\cite{Mazurenko2016AFM}, observed spin-charge separation in the one-dimensional Fermi-Hubbard model~\cite{hilker2017revealing}, studied BCS to BEC crossover in the vicinity of the Feshbach resonance~\cite{vanHoucke2011DiagrammaticMC, ku2014motion}, \corr{and observed long-lived}  prethermalized state in one-dimensional Bose systems~\cite{Schmiedmayer2012Prethermalization}, just to name a few. In this paper, we consider a system of $^{173}{\rm Yb}$ atoms in an optical lattice, such as the one studied recently in experiments by S. F\"olling and collaborators~\cite{riegger2017localized}. We show how by adding Floquet-type control of interactions one can realize a particularly intriguing regime of the Kondo model: ferromagnetic (FM) interactions with tunable exchange anisotropy \corrnew{between the Kondo couplings $J_z$ and $J_\perp$ corresponding to the $z$ and the $(x, y)$ directions, respectively}. While in the commonly studied antiferromagnetic (AFM) Kondo model spin anisotropy is irrelevant, systems with ferromagnetic easy-axis and easy-plane couplings behave in a very different way. Systems with easy-axis anisotropy (\corrnew{$|J_z| > |J_\perp|$}) and those with $SU(2)$ symmetry flow to weak coupling, so that at low temperatures the impurity spin becomes effectively decoupled from the conduction electrons. Easy-plane systems (\corrnew{$|J_z| < |J_\perp|$}), by contrast, have a non-trivial renormalization group flow, which first goes in the direction of decreasing ferromagnetic coupling, but later crosses over to the antiferromagnetic regime and flows toward the strong coupling fixed point. Therefore, at the lowest temperatures, the impurity spin acquires a screening cloud~\cite{barzykin1996kondo, simon2003kondo,  
holzner2009kondo}, 
although the original microscopic model had ferromagnetic interactions with \corr{easy-plane} anisotropy. We consider several types of experiments that can probe this exotic regime. We also consider quench-type experiments, such as analysis of the formation of the Kondo cloud in time. This can not be handled with the NRG approach on the ferromagnetic easy-plane side since it requires analyzing long-time dynamics of the low-temperature system. We introduce a new nonpertubative variational approach to describe the time-evolution of anisotropic Kondo systems across the phase diagram.

\begin{figure}
\includegraphics[width = 7.5cm]{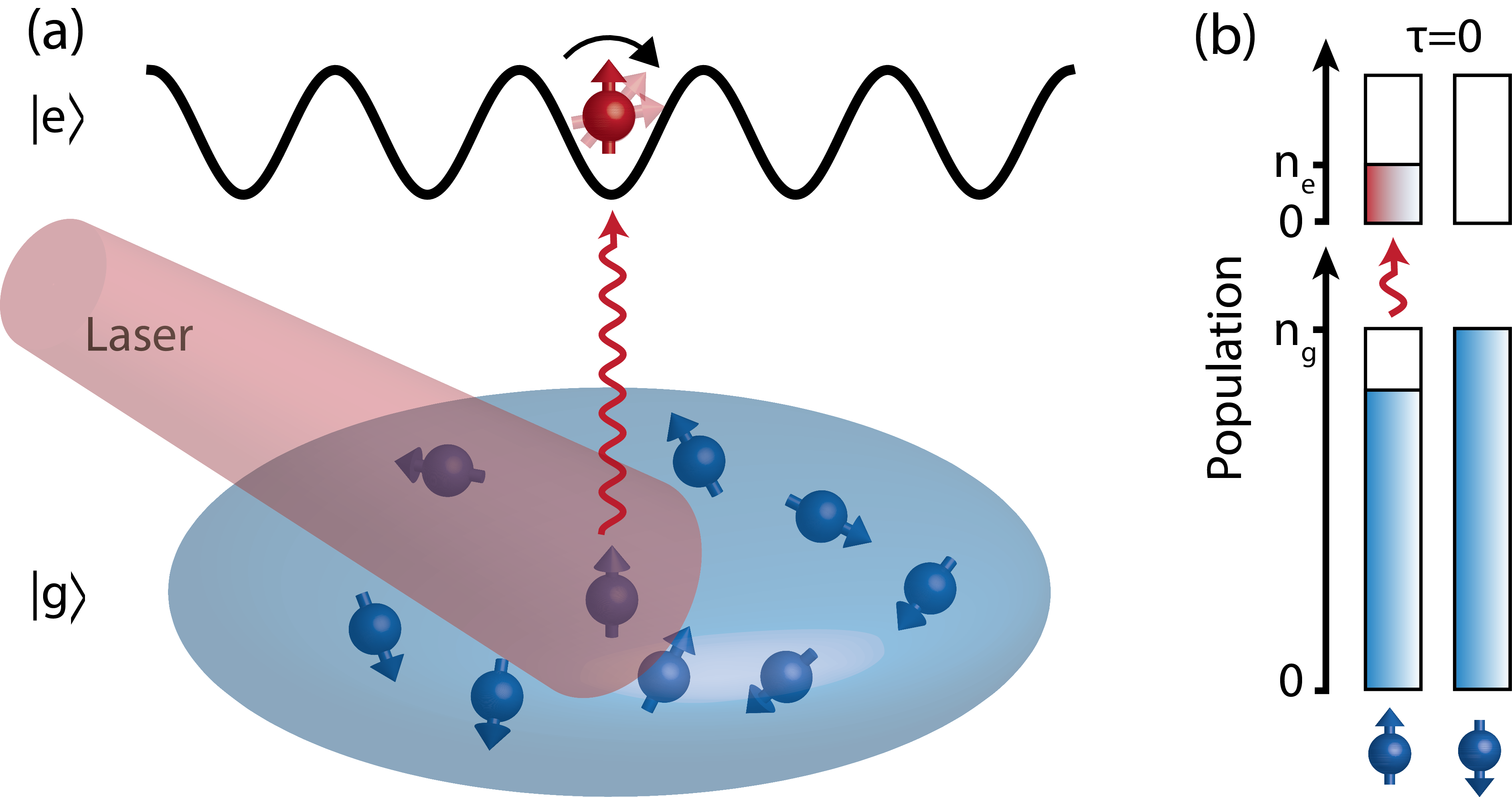}
\caption{ (Color.) Experimental realization. (a) Blue (red) atoms denote the $\ket{g}$ ($\ket{e}$) states of alkaline-earth atoms. Only two of the $2 I + 1$ nuclear spin states of $\ket{g}$ atoms are populated initially. A dim laser pulse excites a small fraction of $\ket{g \upa}$ atoms into the $\ket{e \Uparrow}$ state, whereas the  $\ket{g \downarrow}$ atoms are left unaltered. The $\ket{e}$ atoms are anchored by a deep optical lattice, acting as impurities that interact with the itinerant $\ket{g}$ atoms through strong on-site interaction. (b) In the quench experiment discussed in Sec.~\ref{sec:zero_temperature}, the $\ket{e}$ atoms are excited at time \corr{$\tau=0$} into the $\ket{\Upa}$ state, during a time that can be considered  instantaneous on the timescales of the Kondo dynamics. They gradually lose their spin orientation due to the spin exchange with the $\ket{g}$ atoms. The magnetization of the impurity $\langle S_e^z(\corr{\tau})\rangle$ can be measured after an evolution time $\corr{\tau}$.
}
\label{fig:proposal}
\end{figure}

\section{Kondo model} 

\subsection{\corr{Formulation of the model}} \label{sec:KondoIntro}


 \begin{figure}
 \includegraphics[width = 6.5cm]{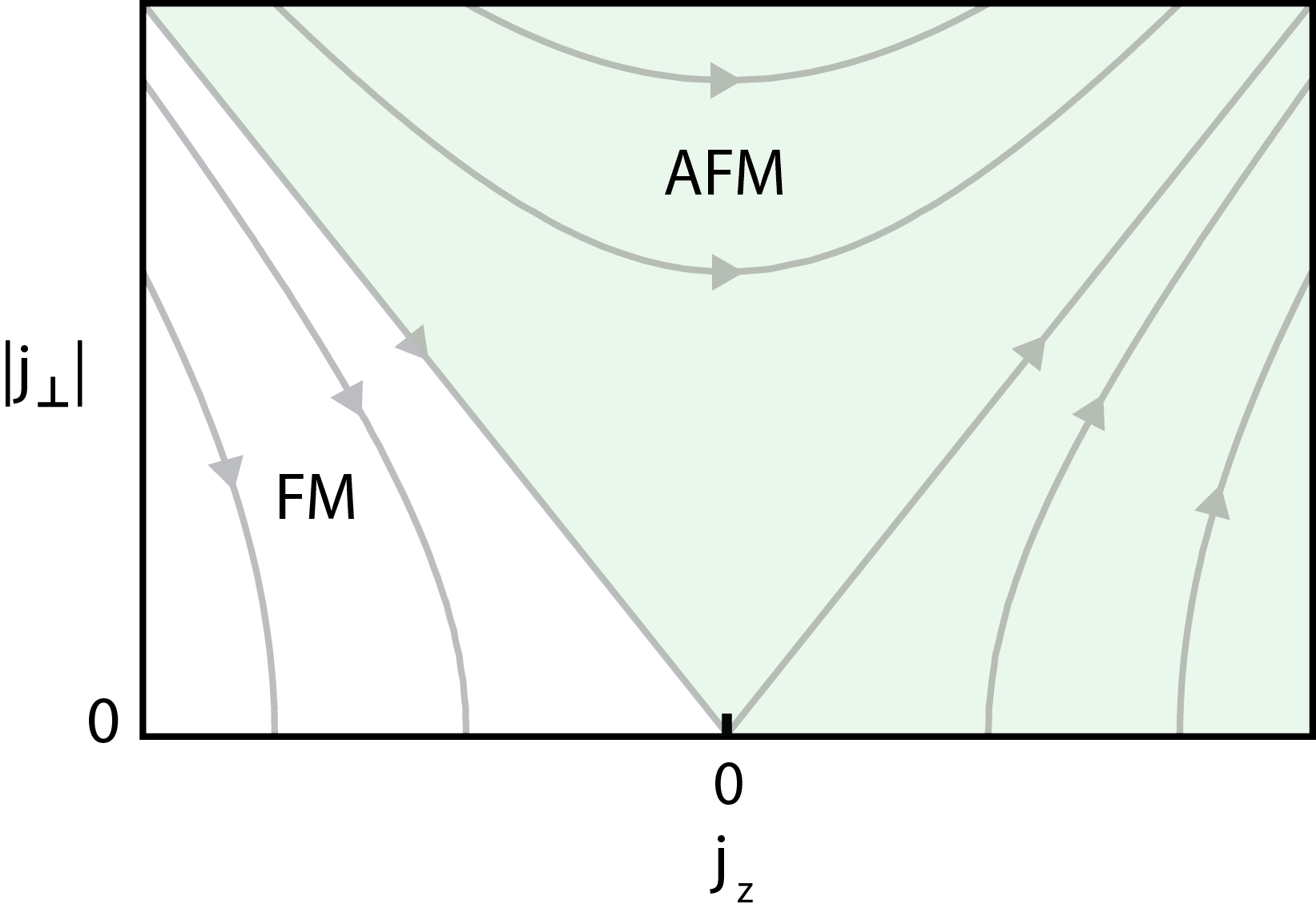}
 \caption{(Color.) Anisotropic Kondo phase diagram in the absence of an effective magnetic field. $j_z$ is the dimensionless coupling between the $z$ components of the impurity and the bath atoms, whereas $j_\perp$ is associated with spin-flip processes. Gray arrows indicate the poor man's scaling renormalization group flows, arising from the second order perturbation theory~\cite{Anderson1970PoorMansScaling}. Couplings in the white region flow to the line of ferromagnetic fixed points $(j_\perp = 0$, $j_z < 0)$, where the Kondo impurity remains unscreened. In contrast, all points in the dark (green) region flow into the antiferromagnetic fixed point, where the Kondo impurity forms a singlet with the surrounding cloud of atoms.
}
\label{fig:Kondo_phdiag}
\end{figure}

\corr{To establish notations, we begin by introducing the Kondo model including some additional terms, which will be used in subsequent discussions. We will also remind the readers of some basic facts about the Kondo model that will lay the foundation for analysis in the next sections.}
The Kondo Hamiltonian $H_{\rm Kondo} = H^K_{\rm int} + H^K_{\rm bath} + H^K_{\rm m}$  describes the interaction of a localized impurity spin with the surrounding fermionic bath. The dynamics of the bath is governed by the Hamiltonian
\beq
H^{K}_{\rm bath} =\sum_{\alpha\sigma} \epsilon_{\alpha\sigma} \; g_{\alpha\sigma}^\dagger g_{\alpha\sigma},
\label{eq:KondoBathHamiltonian}
\eeq
where the operators $g_{\alpha \sigma}$ annihilate a bath atom with spin $\sigma$ in the bath eigenmode $\alpha$. Here, the indices $\alpha$ run over those eigenmodes of the bath around the impurity which couple to the impurity. The single particle energies $\epsilon_{\alpha\sigma}$ also include a Zeeman splitting between the $\ket{\upa}$ and $\ket{\downa}$ bath fermions. \corr{The density of states of these modes $\corr{\varrho}(\epsilon)$ factors in the strength of the coupling of the  bath to the impurity (see~\ref{app:hybridization} for the full definition).} In a three dimensional system and at half-filling it is given by $\corr{\varrho}(0) = 0.118 / t$.  
Interaction between the impurity spin ${\bf S}_e = (S_e^x, S_e^y, S_e^z)$ and the surrounding bath of fermions is given by the anisotropic Kondo interaction Hamiltonian 
\bea
H^K_{{\rm int}} = \frac{1}{M} & \sum\limits_{\alpha \beta\, \sigma\sigma^\prime} &
\left( \, (J_z/2)  \, S^z_e \, \sigma^z_{\sigma\sigma^\prime} \, g_{\alpha \sigma}^\dagger \; g_{\beta \sigma^\prime} \right. \label{eq:KondoInteractionHamiltonian} \\
&+& (J_\perp/2)  \left(S^x_e \, \sigma^x_{\sigma\sigma^\prime} +  S^y_e \, \sigma^y_{\sigma\sigma^\prime}\right) \; g_{\alpha \sigma}^\dagger \, g_{\beta \sigma^\prime} \nonumber \\
&+& \left. K \, \delta_{\sigma\sigma^\prime} \; g_{\alpha \sigma}^\dagger \, g_{\beta \sigma^\prime} \, \right), \nonumber
\eea
where the Pauli matrices are denoted by $(\sigma^x, \sigma^y, \sigma^z)$.
$M$ stands for the number of lattice sites of the system, whereas $J_z$ and $J_\perp$ denote the longitudinal and transverse Kondo couplings. $S_e^\pm = S_e^x \pm i S_e^y$ are spin-flip operators acting on the impurity. The associated dimensionless couplings 
$j_z = J_z \corr{\varrho}$ and  $j_\perp = J_\perp \corr{\varrho}$
characterize the coupling strength between the impurity and the bath, and determine the temperature scale of the onset of the Kondo effect~\cite{HewsonBook}.
\corr{In addition to the spin-dependent scattering, the impurity also gives rise to the potential scattering term $K$ which is the last term in Eq.~\eqref{eq:KondoInteractionHamiltonian}.}
This term has no significant effect on the low energy Kondo dynamics of the system and can be eliminated using a basis transformation~\cite{HewsonBook}. 

Using a static external Zeeman field in the experiment breaks the $SU(2)$ symmetry of the low energy Kondo model to a $U(1)$ symmetry, \corrnew{associated with the conserved spin in the $z$ direction}. This leads to anisotropy between the Kondo couplings $J_z \neq J_\perp$. In addition, the field also breaks the $\pi$ rotation symmetry along the $x$ or $y$ axis, and therefore allows for the appearance of additional effective magnetic couplings
\cpm{
\beq
H^K_{\rm m} =  -m_e \, S^z_e - \frac{1}{2} \, \frac{m_{g}}{M} \sum_{\alpha \, \beta \, \sigma \, \sigma^\prime} \sigma^z_{\sigma\sigma^\prime} \; g_{\alpha \sigma}^\dagger \, g_{\beta \sigma^\prime}.
\label{eq:KondoZeemanHamiltonian}
\eeq
}
Whereas $m_e$ acts as an external local magnetic field for the impurity atom, the coupling $m_g$ creates magnetic scattering for the bath atoms. 
\corr{Note that this scattering occurs only at the position of the impurity, however, it does not involve the spin of the impurity.}
As we discuss in Sec.~\ref{sec:finiteTemperatureEquilibrium}, large enough values of these magnetic terms can be detrimental to the formation of the screening cloud in the antiferromagnetic model. However, modulating the external field restores the $\pi$ rotation symmetry of the low energy Floquet Hamiltonian, and the magnetic couplings vanish (see Sec.~\ref{sec:SchriefferWolff}). We show that a combination of static and modulated external fields can be used to control the magnetic terms $m_e$ and $m_g$ independently from the anisotropy $a = J_\perp - J_z$.

\subsection{Phase diagram and the relevant energy scales}\label{sec:pd_es}

When the effective local magnetic fields $m_e$ and $m_g$ are zero, the Kondo model is described by the phase diagram shown in Fig.~\ref{fig:Kondo_phdiag}. The universal equilibrium behavior is determined by the dimensionless Kondo parameters defined as $j_z = J_z \, \corr{\varrho}$ and $j_\perp = J_\perp \, \corr{\varrho}$. The gray lines in the phase diagram denote the renormalization group flows of the dimensionless Kondo couplings $j_z$ and $j_\perp$ under the poor's man scaling flow~\cite{HewsonBook, Anderson1970PoorMansScaling}. Note that the sign of $j_\perp$ is \corr{not relevant} as it can be changed by a $\pi$ rotation of the spins in the $x$-$y$ plane, \cpm{but  the sign of $j_z$ is important.}

\cpm{
In the shaded region of the phase diagram in Fig.~\ref{fig:Kondo_phdiag} the parameters flow towards the strong coupling AFM fixed point, $(j_z ,j_{\perp}) \to \infty$. At zero temperature, the impurity spin is completely screened by a cloud of itinerant atoms, whose total spin forms a singlet with the impurity.  The Kondo screening survives  as long as the temperature is below a fundamental energy scale, called the Kondo temperature, $T_K$ (see \ref{app:DifferentDOS}). A hallmark of the Kondo regime is that every physical quantity depends on the microscopic model parameters solely through $T_K$, so determining $T_K$ precisely is essential.  To better understand  how we define $T_K$ it is useful to discuss the isotropic situation $j_z = j_\perp \equiv j_{\rm eff}$ first. The isotropic coupling $j_{\rm eff}$ is always positive in the AFM region. Then $T_K$ is found to depend on this dimensionless parameter and is regularized by an energy cut-off of the order of bandwidth $\cal D$ \cite{HewsonBook}
\begin{equation}
T_K\simeq  {\cal D}\sqrt{j_{\rm eff}} \exp(-1/j_{\rm eff}),\, \;\;\;\;\; j_{\rm eff}>0\, .
\label{eq:TK}
\end{equation}
In the case of anisotropic coupling, $T_K$ is associated with the infrared divergence in the poor's man scaling equations \corr{(discussed in~\ref{app:DifferentDOS})}. As a rule of thumb, in the limit when $j_{\rm eff}\to 0$, the Kondo temperature $T_K$, vanishes exponentially.

The formation of the Kondo cloud does not survive in the \corr{easy-axis} FM (white) region of the phase diagram in Fig.~\ref{fig:Kondo_phdiag}, where the physics is completely different: At low temperatures, 
apart from some logarithmic corrections, the impurity spin behaves essentially as a free local moment.
Here the fixed point Hamiltonian corresponds to free fermions with an additional degeneracy due to 
the uncoupled spin. In this limit, the effective coupling $j_{\rm eff}$ is always negative, 
and represents a marginally irrelevant interaction. Solving the same scaling equation \corr{(see~~\eqref{eq:poor_man} in \ref{app:DifferentDOS})} allows us to introduce another characteristic energy scale~\cite{Koller.2005}
\begin{equation}
E_0\simeq {\cal D}\sqrt{|j_{\rm eff}|} \exp(-1/j_{\rm eff})\, , \;\;\;\;\; j_{\rm eff}<0.
\end{equation}
\corrnew{This expression is formally similar to that of $T_K$} and, furthermore, can be associated with the 
ultraviolet divergence in the scaling equations. Therefore, we expect $E_0$, in general, to be larger than the bandwidth $\cal D$ itself and to diverge in the limit when $j_{\rm eff}\to 0$. 
Numerical results for $T_K$ and $E_0$ are presented in Sec.~\ref{sec:lm}.
}

Discovering the crossover between these two regions, understanding the effect of magnetic fields and following the low energy non-equilibrium dynamics of the impurity constitutes \corr{both a challenge and an opportunity for experiments with ultracold atoms.}

\section{Experimental realization of the Kondo Hamiltonian} \label{sec:Proposal}

In this section, we present our proposal for creating a tunable version of the anisotropic Kondo model Eqs.~(\ref{eq:KondoInteractionHamiltonian}-\ref{eq:KondoZeemanHamiltonian}) using alkaline earth atoms. These species have been widely used both in atomic clocks~\cite{Takamoto2005, Hinkley2013, Bloom2014} and in quantum emulation experiments recently~\cite{ye2008review, krauser2012coherent, scazza2014observation, krauser2014giant, zhang2014spectroscopic, mancini2015observation, hofer2015observation, taie20126}. 
Their special properties arise from their closed outer electron shell, making the total electronic angular momentum zero. Their nuclear spin thus decouples from the electronic degrees of freedom, and it is not affected by ultracold collisions~\cite{Zhang2014}. Fermionic isotopes with nuclear spin $I$  realize systems with $SU(N)$ symmetric interactions~\cite{gorshkov2010two, Zhang2014}. By populating only $N$ spin components, the symmetry group of the model is tunable from $N = 1$ to its maximal value of $2 I + 1$. This can be as large as $N=6$ and $N=10$ for $^{173}{\rm Yb}$ and $^{87}{\rm Sr}$, respectively.
Furthermore, besides their electronic ground states $^{1}S_0 = \g$, these atoms exhibit an excited clock-state $^{3}P_0 = \e$ of exceptionally long lifetime~\cite{Takamoto2005,Daley2008,Hinkley2013,Bloom2014}. The ultranarrow linewidth of the $\e$ state is the basis of the significantly increased precision of recent atomic clocks  based on these species. As the $\e$ state also has a closed outer shell, the interaction is SU($N$) symmetric in all channels, $\g - \g$, $\e - \g$ and $\e - \e$~\cite{gorshkov2010two,scazza2014observation, Zhang2014}.
In quantum emulation experiments, this makes it possible to realize higher symmetry analogs of several impurity models, where the role of the impurity is played by atoms in the excited state~\cite{gorshkov2010two}. 

\corr{Our starting point is the Hubbard-Anderson model of the $\g$ and $\e$ states of alkaline-earth atoms. (For a detailed discussion of the microscopic model of alkaline-earth atoms in optical lattices we refer the readers to Refs.~\onlinecite{gorshkov2010two, Daley2008, taie2010realization, stellmer2011detection, 
pagano2014one}.) 
The key element of our setup is the state-dependent optical lattice, which allows to strongly localize $\e$ fermions while keeping the $\g$ atoms highly mobile~\cite{riegger2017localized}. We use a time-dependent Schrieffer-Wolff transformation to show that the low energy properties of this system can be described by the Kondo Hamiltonian. Our analysis extends earlier work on the subject (see e.g.~\cite{gorshkov2010two}) by including both static and modulated Zeeman fields, which leads to a much broader class of anisotropic Kondo Hamiltonians.}

\subsection{Hubbard-Anderson model} \label{sec:HubbardAnderson}

\corr{We now outline the steps needed to realize the spin-$1/2$ anisotropic Kondo model. Two nuclear spin components provide the analog of electron spin in electron systems.}
Atoms are initialized in the $\g$ state in a three-dimensional optical trap. A weak $\pi$-polarized laser pulse is then used  to excite a small fraction of one of the nuclear spin components into state $\e$ (see Fig.~\ref{fig:proposal}~(a)). Different polarizability of $\g$ and $\e$ states makes it possible to create an optical lattice that anchors the atoms in the clock-state but creates only a weak lattice potential for those in the ground state~\cite{Daley2008}. \corr{Thus, the few impurities created by the laser pulse are} coupled to the Fermi sea of mobile ground state atoms~\cite{gorshkov2010two}.

\begin{figure}
\includegraphics[width = 6.5cm]{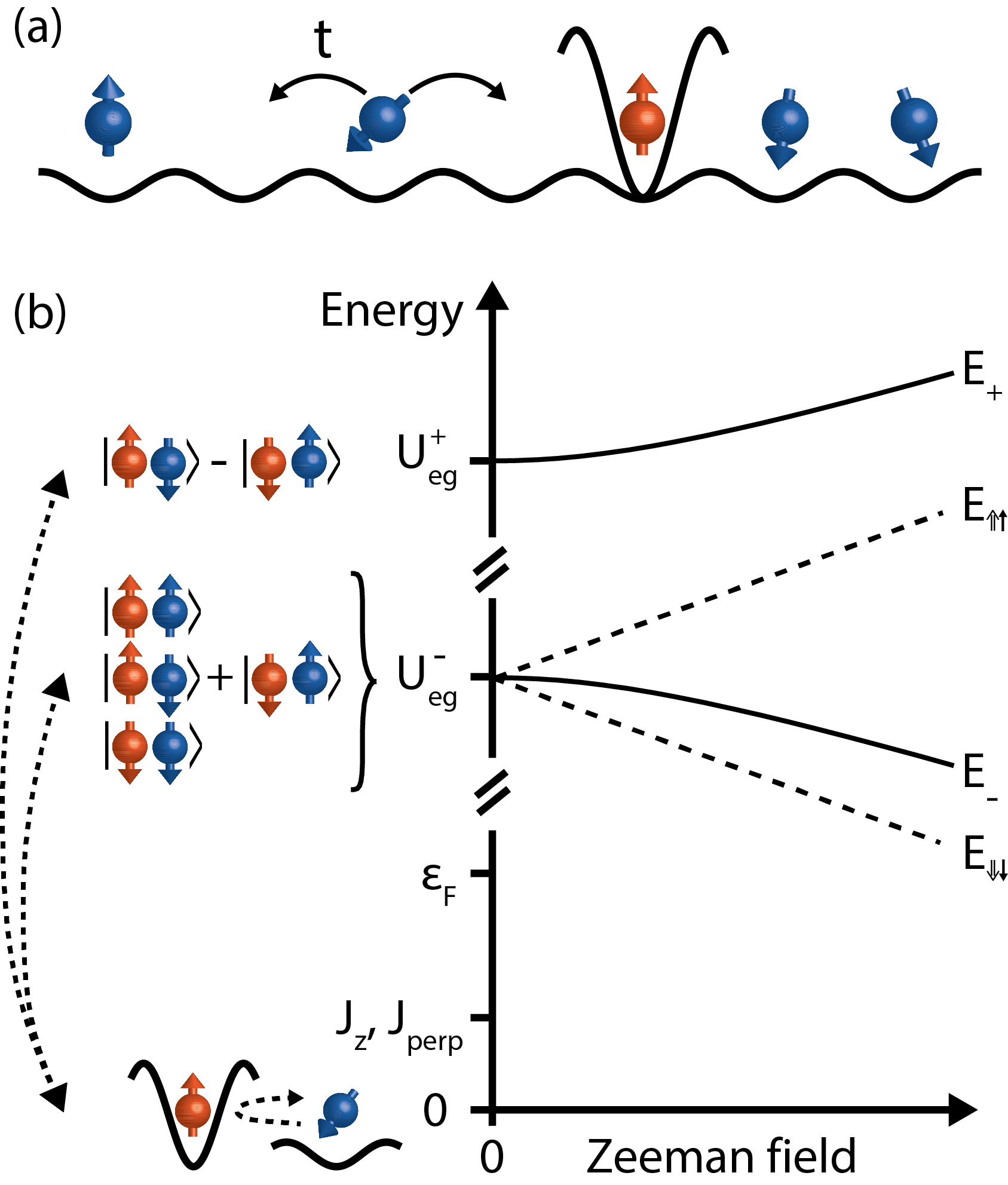}
\caption{ (Color.) Low energy spin dynamics in the Hubbard-Anderson model. (a) Whereas the impurity atom $\e$ is localized by a strong optical potential, the bath of $\g$ atoms is itinerant, with a hopping energy $t$. Energy scales of the system are shown in (b), with $\epsilon_F$ denoting the Fermi energy. The impurity interacts with the bath through on-site interactions $U_{eg}^-$ and $U_{eg}^+$, corresponding to the triplet and singlet spin channels, respectively. The on-site interactions are much larger than the tunneling matrix element to the impurity site. Interactions with the impurity, therefore, happen only virtually through second-order processes. Since $U_{eg}^-\ll U_{eg}^+$, the virtual state is dominated by the spin triplet channel, which leads to FM Kondo couplings $J_z = J_\perp < 0$ in the low energy effective Hamiltonian. An external Zeeman field creates $\corrnew{\Delta}_e$ and $\corrnew{\Delta}_g$ Zeeman splittings, acting on the $\e$ and $\g$ atoms, respectively.\corr{The Zeeman splitting $\corrnew{\Delta} = \corrnew{\Delta}_e - \corrnew{\Delta}_g$ leads to level repulsion and mixing between the singlet and the triplet channels. As a result, the Kondo parameters become anisotropic and a finite magnetic term appears in the Kondo Hamiltonian, see} Eqs.~(\ref{eq:KondoInteractionHamiltonian}-\ref{eq:KondoZeemanHamiltonian}).
}
\label{fig:AndersonModel}
\end{figure}

The bath atoms interact with each other through the nuclear-spin-independent scattering length $a_{gg}$. In case of $^{173}{\rm Yb}$, this is given by $a_{gg} = 199.4 \, a_0$, where $a_0$ denotes the Bohr radius. As the optical lattice is shallow for the bath atoms, we assume that they are in the Fermi liquid phase and the $\g - \g$ interaction only renormalizes the Fermi liquid parameters. Interaction between the impurities and the gas is characterized by two scattering lengths $a_{eg}^{\pm}$, corresponding to symmetric and antisymmetric combinations of their \corr{orbital} wave functions $(| ge \rangle \pm | eg \rangle)/\sqrt{2}$, as shown in Fig.~\ref{fig:AndersonModel}. \corr{Due to the Pauli principle, the nuclear spins are thus in a singlet and triplet state, respectively.}

\corr{Assuming that both the $\g$ and $\e$ atoms occupy the lowest vibrational state on each lattice site, their on-site repulsion is given by
$U_{eg}^\pm = \frac{4\pi \hbar^2}{m} \, a_{eg}^\pm \, \int d^3 r \; |w_g({\bf r})|^2 \, |w_e({\bf r})|^2$. Here, $w_g$ and $w_e$ denote the Wannier orbitals of $\g$ and $\e$ atoms, respectively.
This expression holds as long as the oscillator frequency of the local potential is much larger than the on-site repulsion.
When the scattering length of an interaction channel becomes large, the band gap created by the harmonic oscillator potential effectively limits the interaction energy~\cite{busch_two_1998}. In particular for $^{173}$Yb, this is the case for the symmetric interaction channel scattering length $a_{eg}^+$ close to 2000\,$a_0$~\cite{hofer2015observation}. In addition, bound states can strongly influence the interaction in case of certain trap configurations \cite{riegger2017localized}.
The antisymmetric scattering channel is also repulsive, with $a_{eg}^- = 219.5 \, a_0$. Therefore, the on-site repulsion in the absence of bound state  resonances is much stronger in the symmetric than in the antisymmetric channel. For the computations in this paper, we use the constant ratio
\beq
\frac{U_{eg}^+}{U_{eg}^-} \approx 15. \nonumber
\eeq

In the case of a two-component gas, the interaction decouples in the triplet 
\bea
\ket{\Uparrow\uparrow} &\equiv& \frac{1}{\sqrt 2} \left( \ket{ge} - \ket{eg} \right) \, \ket{\Uparrow\upa}, \nonumber \\
\ket{\Downarrow\downarrow} &\equiv& \frac{1}{\sqrt 2} \left( \ket{ge} - \ket{eg} \right) \,\ket{\Downarrow\downa}, \nonumber \\
\ket{-} &\equiv& \frac{1}{2} (\ket{ge} - \ket{eg}) \, (\ket{\Uparrow\downa} + \ket{\Downarrow\upa}), \nonumber 
\eea
and the singlet channels
\beq
\ket{+} \equiv \frac{1}{2} (\ket{ge} + \ket{eg}) (\ket{\Uparrow \downa} - \ket{\Downarrow \upa}), \nonumber
\eeq
where $\ket{\Uparrow}$ and $\ket{\Downarrow}$ denote the spin states of the impurity and $\ket{\upa}$ and $\ket{\downa}$ are those of the bath atom.
The (nuclear) spin singlet configuration, which interacts with the symmetric molecular potential, therefore experiences stronger repulsive interaction than the triplet configuration.}

The dynamics of the impurity at the origin $i=0$ and the surrounding gas is thus governed by the Hamiltonian 
\beq
H = H_{{\rm kin}}^{(0)} + H_{{\rm imp}}^{(0)}. \nonumber
\eeq 
The kinetic and impurity parts of $H$ are given by 
\bea
H_{\rm kin}^{(0)} &=&  -t \sum_{\langle i, j \rangle, \sigma} g_{i\sigma}^\dagger \, g_{j\sigma}, \label{eq:AndersonHamiltonianKin}\\
H_{\rm imp}^{(0)} &=& U \, (n_{g 0\upa} + n_{g 0\downa}) (n_{e 0 \Uparrow} + n_{e 0 \Downarrow}) \label{eq:AndersonHamiltonianImp} \\
&+& U_{\rm ex} \, \sum_{\sigma \sigma^\prime} g_{0\sigma^\prime}^\dagger e_{0\sigma}^\dagger \, e_{0 \sigma^\prime} g_{0\sigma}, \nonumber
\eea
where $U = (U_{eg}^- + U_{eg}^+)/2$ and $U_{\rm ex} = (U_{eg}^- -  U_{eg}^+)/2 < 0$ are the on-site charge and spin exchange interactions. The operator $g_{i\sigma}$ annihilates a bath atom of spin $\sigma$ on site $i$, whereas $e_{0 \sigma}$ is the annihilation operator of the impurity on site $i=0$ with spin $\sigma$.
At the impurity site, the number operator of the impurity and bath atoms are given by $n_{g 0 \sigma} = g_{0\sigma}^\dagger \, g_{0\sigma}$ and $n_{e 0 \sigma} = e_{0 \sigma}^\dagger e_{0 \sigma}$, respectively.

Due to the large scattering lengths $a_{eg}^\pm$ and the strong confinement of the impurities, the system is in the regime $U_{eg}^\pm \gg t$ \corrnew{where the Schrieffer-Wolff transformation can be applied}. The impurity site is filled by the $\e$ atom only, and bath atoms  interact with the impurity through virtual tunneling. This leads to the spin interactions of the Kondo model, shown in Eq.~\eqref{eq:KondoInteractionHamiltonian}. Since the interaction $U_{eg}^-$ in the triplet channel is much weaker, virtual tunneling into these states has a higher amplitude. This leads to ferromagnetic isotropic couplings $J_z = J_\perp < 0$ between the impurity and the bath atoms (see Sec.~\ref{sec:SchriefferWolff}).

We mention that earlier proposals discussed the opposite regime of weak to intermediate interactions $t \corr{\gtrsim} U_{eg}^-$~\cite{zhang2016kondo, nakagawa2015LaserInducedKondo}. A caveat of this regime is that the \corr{fast formation of a weakly bound state may change the on-site interaction and break down the Kondo dynamics at long times (see e.g. the discussion in Ref.~\onlinecite{Pekker2011Competition}).}

\subsubsection{Artificial Zeeman fields}

\corr{The $SU(2)$ symmetry of the Kondo model can be broken using an external effective magnetic field. Ultracold experiments with alkaline-earth atoms have used various ways to create different (effective) Zeeman fields for the $\e$ and $\g$ atoms. A well-established approach is to create a state-dependent optical Stark shift, which has been routinely used for optical Stern-Gerlach separation of the nuclear spin components~\cite{taie2010realization, stellmer2011detection, zhang2014spectroscopic, pagano2014one}. This allows one to create both static and modulated Zeeman fields for the atoms. By modulating the intensities or detunings of the lasers, time-dependent Zeeman fields can be created (see \ref{app:opticalStark}). Effective Zeeman fields can also be created using a large external magnetic field, as has been demonstrated in the recent realization of orbital Feshbach resonances of alkaline-earth atoms~\cite{hofer2015observation}. This technique relies on the slightly different Land\'e $g$-factors of the bath and impurity atoms~\cite{Zhang2015}. It works well in case of static Zeeman shifts, requiring external magnetic fields of the order of $50 \, {\rm G}$~\cite{hofer2015observation}.  Modulating such large magnetic fields at radio frequencies can, however, be challenging experimentally. 

In an external effective magnetic field, the Zeeman shifts $\corrnew{\Delta}_e$ and $\corrnew{\Delta}_g$ are slightly different in the Hamiltonian}
\bea
H_{\rm kin} &=& H_{\rm kin}^{(0)} -\frac{\corrnew{\Delta}_g}{2} \sum_{i\neq 0} \corrnew{(g_{i\uparrow}^\dagger g_{i\uparrow} - g_{i\downarrow}^\dagger g_{i\downarrow})}, \label{eq:HZeemanKin}\\
H_{\rm imp} &=& H_{\rm imp}^{(0)} - \frac{\corrnew{\Delta}_e}{2} \left(\ket{\Upa} \bra{\Upa} - \ket{\Downa}\bra{\Downa}\right) \label{eq:HZeemanImp} \\
&-& \frac{\corrnew{\Delta}_g}{2} (n_{g 0\upa} - n_{g 0\downa}). \nonumber
\eea
In the subspace of the $\{ \ket{\Upa\downa}, \ket{\Downa\upa}\}$ states, the interaction Hamiltonian of the impurity site with a single $\g$ atom reads~\cite{gorshkov2010two}
\beq 
{\bf H}_{\rm imp}^{\rm ex} = 
\begin{pmatrix}
U - \corrnew{\Delta}/2 & U_{\rm ex}\\
U_{\rm ex} & U + \corrnew{\Delta}/2 
\end{pmatrix}.
\label{eq:Himpex}
\eeq
Here, $\corrnew{\Delta} = \corrnew{\Delta}_e - \corrnew{\Delta}_g$ denotes the difference between the Zeeman splittings of $\e$ and $\g$ atoms. This magnetic coupling mixes the singlet $\ket{+}$ and triplet $\ket{-}$ states and breaks the $SU(2)$ symmetry of the model.
As we show in Fig.~\ref{fig:AndersonModel}~(b), this leads to on-site energies  $E_{\pm} = U \pm \sqrt{U_{\rm ex}^2 + (\corrnew{\Delta}/2)^2}$~\cite{gorshkov2010two}. In contrast, the energies of the states $\ket{\Upa \upa}$ and $\ket{\Downa \downa}$ simply get shifted by $\pm (\corrnew{\Delta}_e + \corrnew{\Delta}_g)$. 

The breakdown of the spin rotation symmetry in the Hubbard-Anderson Hamiltonian leads to an anisotropy in the corresponding low energy Kondo model, as we discuss in Sec.~\ref{sec:SchriefferWolff}. This anisotropy allows us to realize a large fraction of the Kondo phase diagram in Fig.~\ref{fig:Kondo_phdiag}. The additional magnetic terms in Eq.~\eqref{eq:KondoZeemanHamiltonian} can be used to mimic the effect of an external magnetic field $m_e$ acting on the Kondo impurity as well as the magnetic scattering term $m_g$. Oscillating Zeeman fields on the other hand average the magnetic terms out, while they preserve the anisotropy of the model (see Sec.~\ref{subsec:DrivenField}).

In order to reach sensitive control of the Kondo parameters, the driving frequency often needs to be in the range of $U_{eg}^\pm$ (see Sec.~\ref{sec:SchriefferWolff}). This means that the driving is usually in the $\rm 1-10 \, kHz$ regime, and it is much faster than the dynamical timescales of the system
\beq
J_z, J_\perp \ll t \ll \omega \sim U_{eg}^\pm.
\label{eq:energy_scales}
\eeq
Therefore, on the timescales of Kondo dynamics the modulation averages out, and we can use an effective Floquet description to model the system, as we show in \ref{app:SchriefferWolffDynamic}.

\subsection{Kondo parameters of the driven model} \label{sec:SchriefferWolff}

\corr{In this subsection, we derive the Kondo Hamiltonian governing the low energy impurity-bath dynamics.}
Due to the strong confinement and interaction between the impurity and the bath atoms, tunneling to the impurity site by bath atoms is strongly suppressed, $t \ll U_{eg}^-, U_{eg}^+$.  \corrnew{This is the regime where} the coupling between the impurity and the bath arises from virtual tunneling to the impurity site. The impurity's on-site interaction is described by $H_{\rm int}$ in Eq.~\eqref{eq:HZeemanImp}, where the periodically modulated Zeeman splittings depend on time $\tau$. The bath Hamiltonian 
\beq
H_{\rm bath} =  \sum_\alpha \left( \epsilon_\alpha - \sigma \corrnew{\Delta}_g(\tau) / 2 \right) \, g_{\alpha\sigma}^\dagger g_{\alpha\sigma} \nonumber 
\eeq
also has time-dependent energies.
The coupling between the impurity and the bath modes is given by the Hamiltonian~\cite{HewsonBook}
\beq
H_{\rm mix} = \sum_{\alpha \sigma}\frac{V}{\sqrt{M}} \; g_{0\sigma}^\dagger\,  g_{{\alpha} \sigma} + {\rm h.c.},
\label{eq:Hmix}
\eeq
with the hybridization matrix element $V = \sqrt{z} t$, where $z$ denotes the coordination number of the optical lattice. The origin of the mixing term as well as the calculation of the density of states $\corr{\varrho}$ of bath eigenmodes is discussed in \ref{app:hybridization}.

\begin{figure}
\includegraphics[width = 7.0cm]{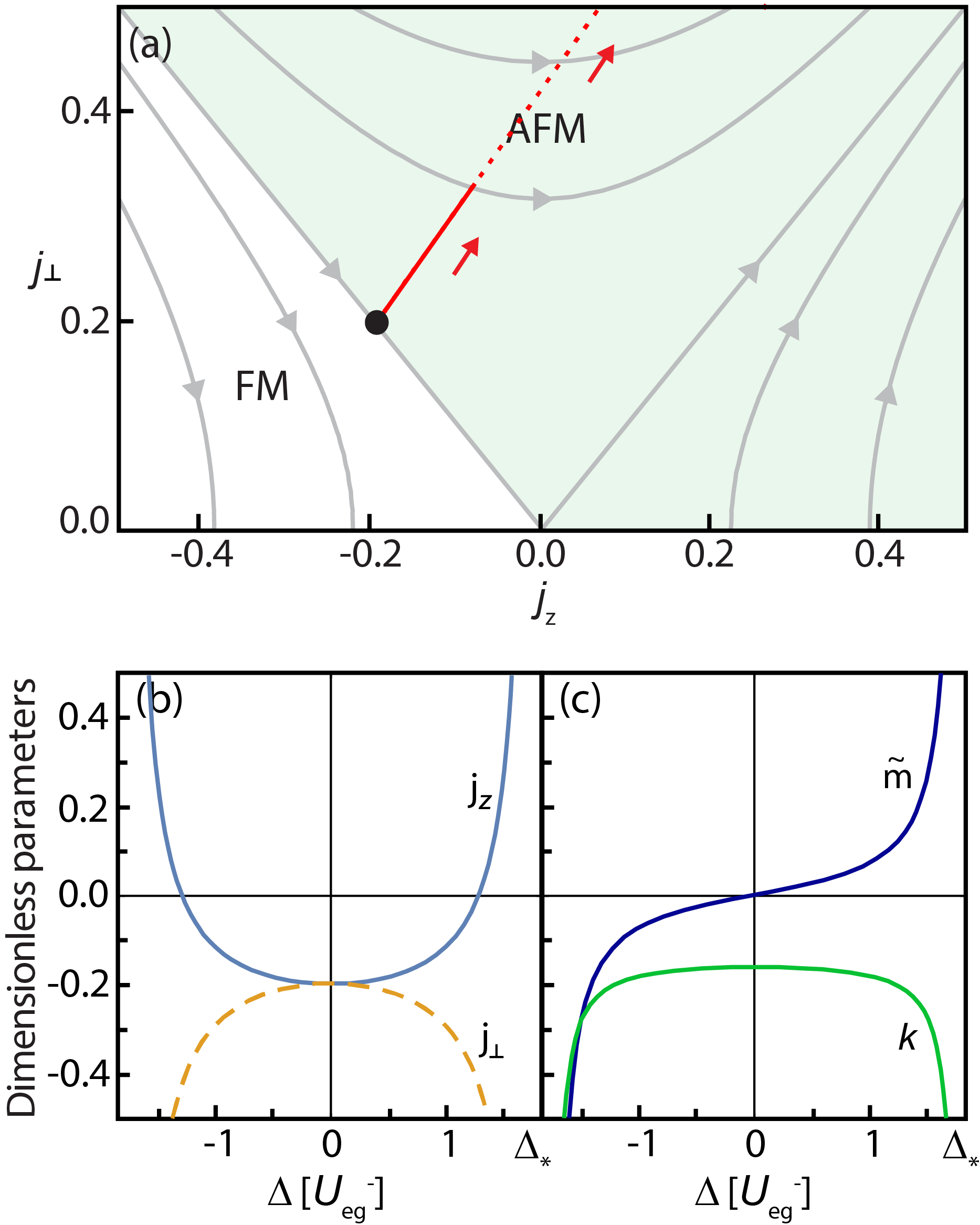}
\caption{ (Color.) (a) Dependence of the dimensionless Kondo parameters $j_z$ and $j_\perp$ in Eqs.~(\ref{eq:Jz},\ref{eq:Jperp}) on a static Zeeman splitting $\corrnew{\Delta}$ in the anisotropic Kondo phase diagram. \corr{Red line shows the effect of increasing $\corrnew{\Delta}$ on the isotropic system $j_z = - j_\perp = 0.2$ (black dot).}
Gray lines denote the directions of the poor man's scaling flow~\cite{Anderson1970PoorMansScaling} in the absence of magnetic terms.
(b, c) Dependence of Kondo model parameters of Eqs.~(\ref{eq:KondoInteractionHamiltonian}-\ref{eq:KondoZeemanHamiltonian}) \corr{on $\corrnew{\Delta}$.} Here $k = K \corr{\varrho}$ denotes the dimensionless potential scattering term and $\tilde{m} = m_e \corr{\varrho} = -m_g \corr{\varrho}$ corresponds to the dimensionless magnetic couplings. These parameters become resonant at the Zeeman field $\corrnew{\Delta}_*$ at the edges of the plots. The Schrieffer-Wolff transformation and the \corrnew{Kondo} description is not valid anymore in the vicinity of $\corrnew{\Delta}_*$, as indicated by the dotted line in (a).
[Parameters of the plot: $U_{eg}^+ = 15 \, U_{eg}^-$, $t = 0.35 \, U_{eg}^-$, and $\epsilon_{F} = 0$.] 
}
\label{fig:Kondo_parameters}
\end{figure}

\subsubsection{Static Zeeman field} \label{sec:SchriefferWolffStatic}

Here, we discuss how static Zeeman fields can be used to control the Kondo parameters in Eqs.~(\ref{eq:KondoBathHamiltonian}-\ref{eq:KondoZeemanHamiltonian}). We derive the effective Hamiltonian of the system using the Schrieffer-Wolff transformation $S = -S^\dagger$~\cite{Schrieffer1966} and obtain
\beq
H_{\rm eff} = \mathbb{P}_0 \, e^S \, \left(H_{\rm bath} + H_{\rm imp} + H_{\rm mix} \right) \, e^{-S} \mathbb{P}_0, \label{eq:HeffStatic1}
\eeq
where the projector $\mathbb{P}_0$ maps onto the subspace of excitations with no $\g$ atoms at the impurity site. \corr{States with a single $(\mathbb{P}_1)$ and two $(\mathbb{P}_2)$ $\g$ atoms can be neglected from the low energy description of the system, as they are separated by an energy $U_{eg}^\pm$ from the $\mathbb{P}_0$ sector.} The transformation $S$ is chosen such that it cancels the coupling between the bath and the impurity at first order
\beq
\mathbb{P}_1 \, H_{\rm mix} \, \mathbb{P}_0 =\mathbb{P}_1 \left[H_{\rm bath} + H_{\rm imp}, \, S \right] \mathbb{P}_0. \label{eq:SchriefferWolffStatic}
\eeq
The resulting effective Hamiltonian at second order in $S$ then becomes
\beq
H_{\rm eff} =  \mathbb{P}_0 \left( H_{\rm bath} +  H_{\rm imp} + \frac{1}{2} \left[ S, H_{\rm mix} \right] \right)  \mathbb{P}_0
\label{eq:HeffStatic2}
\eeq

We solve the Schrieffer-Wolff equation Eq.~\eqref{eq:SchriefferWolffStatic} using the ansatz
\beq
S = \frac{1}{\sqrt{M}} \sum_{{\bf k}, \sigma \sigma^\prime \tilde{\sigma} \tilde{\sigma}^\prime} \Gamma_{\tilde{\sigma} \tilde{\sigma}^\prime}^{\sigma \sigma^\prime}({\bf k}) \; g_{0 \sigma}^\dagger e_{0 \tilde{\sigma}}^\dagger \, e_{0 \tilde{\sigma}^\prime} g_{{\bf k} \sigma^\prime} - {\rm h.c.} \label{eq:SchriefferWolffAnsatz}
\eeq
In order to assure spin conservation, the amplitudes $\Gamma_{\tilde{\sigma} \tilde{\sigma}^\prime}^{\sigma \sigma^\prime}({\bf k})$ should be non-zero only in case when $\sigma + \tilde{\sigma} = \sigma^\prime + \tilde{\sigma}^\prime$. Using the ansatz in the last equation, we obtain the Schrieffer-Wolff parameters in the $\ket{\Upa\upa} $ and $\ket{\Downa\downa}$ channels
\beq
\left(U + U_{\rm ex} - \epsilon_{{\bf k}} \right) \Gamma_{\sigma\sigma}^{\sigma\sigma}({\bf k}) = V.
\label{eq:Gamma_no_spin_flip}
\eeq
In the $\{ \ket{\Upa\downa}, \ket{\Downa\upa} \}$ basis, the on-site energy $U_{\rm ex}$ mixes spin channels, and the Schrieffer-Wolff coefficients
\beq
{\bf \Gamma}(k) = 
\begin{pmatrix}
\Gamma_{\Upa\Upa}^{\downa\downa}({\bf k}) & \Gamma_{\Upa\Downa}^{\downa\upa}({\bf k}) \\
\Gamma_{\Downa\Upa}^{\upa\downa}({\bf k}) & \Gamma_{\Downa\Downa}^{\upa\upa}({\bf k})
\end{pmatrix}
\eeq
obey a matrix equation
\beq
\left[
\begin{pmatrix}
{\bf H}_{\rm imp}^{\rm ex} & 0 \\
0 & {\bf H}_{\rm bath}^{\rm ex}(k)
\end{pmatrix} 
,
\begin{pmatrix}
0 & {\bf \Gamma}^{\rm ex}({\bf k}) \\ 
-\left( {\bf \Gamma}^{{\rm ex}}({\bf k}) \right)^\dagger & 0
\end{pmatrix} \right]
= 
V.
\nonumber
\eeq
The Hamiltonian matrix ${\bf H}_{\rm imp}^{\rm ex}$ is defined in Eq.~\eqref{eq:Himpex}, whereas
\beq
{\bf H}_{\rm bath}^{\rm ex}(k) = 
\begin{pmatrix}
\epsilon_{\bf k} - \corrnew{\Delta}/2 & 0 \\
0 & \epsilon_{\bf k} + \corrnew{\Delta}/2
\end{pmatrix}
\eeq
describes the energies of the incoming modes.
The Kondo parameters only depend on the difference $\corrnew{\Delta}$ between the fields $\corrnew{\Delta}_e$ and $\corrnew{\Delta}_g$ but not on their average (see also \ref{app:SchriefferWolffDynamic}), which can be removed using a unitary transformation~\cite{Zhang2015}.

The effective Hamiltonian Eq.~\eqref{eq:HeffStatic2} takes on the same form as the Kondo model in Eqs.~(\ref{eq:KondoInteractionHamiltonian} - \ref{eq:KondoZeemanHamiltonian}). We note that during spin-flip processes $\ket{\Upa \downa} \leftrightarrow \ket{\Downa\upa}$, the scattered $\g$ atom changes its energy with $\pm \corrnew{\Delta}$. In order to make sure that the scattering exchanges particles between the Fermi levels of $\ket{g  \upa}$ and $\ket{g \downa}$ atoms, the Fermi energies also need to be separated with this energy $\epsilon_{F \upa} - \epsilon_{F \downa} = \corrnew{\Delta}$. 
The imbalance of Fermi energies can lead to differences in the density of states for the two components $\corr{\varrho}_{\upa} \neq \corr{\varrho}_{\downa}$. The Kondo scaling equations, and thus the low energy properties of the model, are determined by the dimensionless product of the couplings and the densities of states, as we discuss in \ref{app:DifferentDOS}. Therefore, the anisotropies of the dimensionless couplings might be different from that of $J_z$ and $J_\perp$.

The Zeeman field dependence of the Kondo parameters at the Fermi energy is given by 
\bea
J_z(\corrnew{\Delta}) &=& 2 V^2 \; 
\frac{
U_{\rm ex} \left( U_{\bf k}^2 - U_{\rm ex}^2 - \frac{U_{\bf k}}{U_{\bf k}+U_{\rm ex}} \, \corrnew{\Delta}^2 \right)
}{
\left( U_{\bf k}^2 - U_{\rm ex}^2 \right)^2 - U_{\bf k}^2 \, \corrnew{\Delta}^2
},  \label{eq:Jz} \\
J_\perp(\corrnew{\Delta}) &=& 2 V^2 \; 
\frac{ 
U_{\rm ex} \left( U_{\bf k}^2 - U_{\rm ex}^2 \right)
}{
\left( U_{\bf k}^2 - U_{\rm ex}^2 \right)^2 - U_{\bf k}^2 \, \corrnew{\Delta}^2
}, \label{eq:Jperp} 
\eea
where we introduced the notation $U_{\bf k} \equiv U - \epsilon_{\bf k}$ for brevity.
The potential scattering $K$ and the magnetic terms $m_e = -m_g = m$ become
\bea
K(\corrnew{\Delta}) &=&  - \frac{V^2}{2}\; \frac{2 \, U_{\bf k} - U_{\rm ex}}{U_{\bf k}^2 - U_{\rm ex}^2}
\, - \,
\frac{\corrnew{\Delta}^2}{4} \, \frac{U_{\bf k} \, U_{\rm ex}}{(U_{\bf k}^2 - U_{\rm ex}^2)^2} \; J_{\perp}(\corrnew{\Delta})
 \nonumber \\
m(\corrnew{\Delta}) &=&  \corrnew{\Delta} \, \left( \frac{1}{2} - V^2 \, \frac{U_{\rm ex}^2}{(U_{\bf k}^2 - U_{\rm ex}^2)^2 - U_{\bf k}^2 \, \corrnew{\Delta}^2} \right). \label{eq:m}
\eea
The dependence of these parameters on the Zeeman splitting $\corrnew{\Delta}$ is shown in Fig.~\ref{fig:Kondo_parameters}.  
In the absence of magnetic field, the interaction is $SU(2)$ symmetric $J_{z}(0) = J_{\perp}(0)= J$, with
\beq
J =  - 2 V^2 \; \frac{U_{\rm ex}}{U_{\rm ex}^2 - U_{\bf k}^2} = -\left( \frac{V^2}{U_{eg}^-} - \frac{V^2}{U_{eg}^+} \right), \label{eq:J_isotropic}
\eeq
whereas the dimensionless Kondo parameters equal $j \equiv J \, \corr{\varrho}(0)$. The applicability of the Schrieffer-Wolff transformation requires that the broadening parameter over the on-site interaction $\tilde{\Gamma} / U_{eg}^- = \pi \, j$ be smaller than unity. \corr{Larger couplings could be possible to achieve, however, our calculations for the Kondo couplings are not reliable in that regime.}

\corr{At increasing Zeeman splittings, the couplings go into the $|J_\perp| > |J_z|$ easy plane regime. However, the anisotropy is not sufficient to reach the anisotropic AFM Kondo fixed point since static Zeeman fields also lead to the appearance of the finite effective magnetic term $m(\corrnew{\Delta})$. 
Due to this term, the RG flow no longer flows into the AFM fixed point, and the Kondo screening breaks down.
We will show in the next subsection that this term can be canceled by periodically modulating effective Zeeman fields.}

\corr{The primary effect of static external Zeeman fields is that they create magnetic terms $m_e$ and $m_g$, which grow linearly at small values of $\corrnew{\Delta}$ (see Fig.~\ref{fig:Kondo_parameters}). These terms can substantially change the spin susceptibility of the impurity (see Sec.~\ref{sec:finiteTemperatureEquilibrium}).}
Our calculations are reliable at small and intermediate Zeeman splittings, but they break down near $\corrnew{\Delta}_* = \pm (U_{\bf k}^2 - U_{\rm ex}^2) / U_{\bf k}$, where the Kondo parameters become resonant. At this point, the on-site energy $E_-$ turns negative (see Fig.~\ref{fig:AndersonModel}) and double occupancy of the impurity site becomes energetically favorable, therefore our \corrnew{Kondo} description can no longer be applied. Furthermore, in the vicinity of $\corrnew{\Delta}_*$, our assumption that higher order terms in the Schrieffer-Wolff transformation are negligible starts to break down.
However, such large values of the Zeeman field should not be reached in the Kondo regime $t \ll U_{eg}^\pm$. Since the chemical potential difference needs to be comparable to the Zeeman splitting, $\corrnew{\Delta}$ cannot be larger than the bandwidth $z \, t$. Therefore, we can always assume that the static Zeeman field $\corrnew{\Delta}$ remains  smaller than $U_{eg}^-$.
 Since the anisotropy of the Kondo couplings $J_z$ and $J_\perp$ grows quadratically with $\corrnew{\Delta}$, the anisotropy remains small at such small values of the Zeeman energy. As we show in the next subsection, modulated Zeeman fields can reach much larger anisotropies between the Kondo parameters, at large driving amplitudes.

\begin{figure}
\includegraphics[width=8.5cm,clip=true]{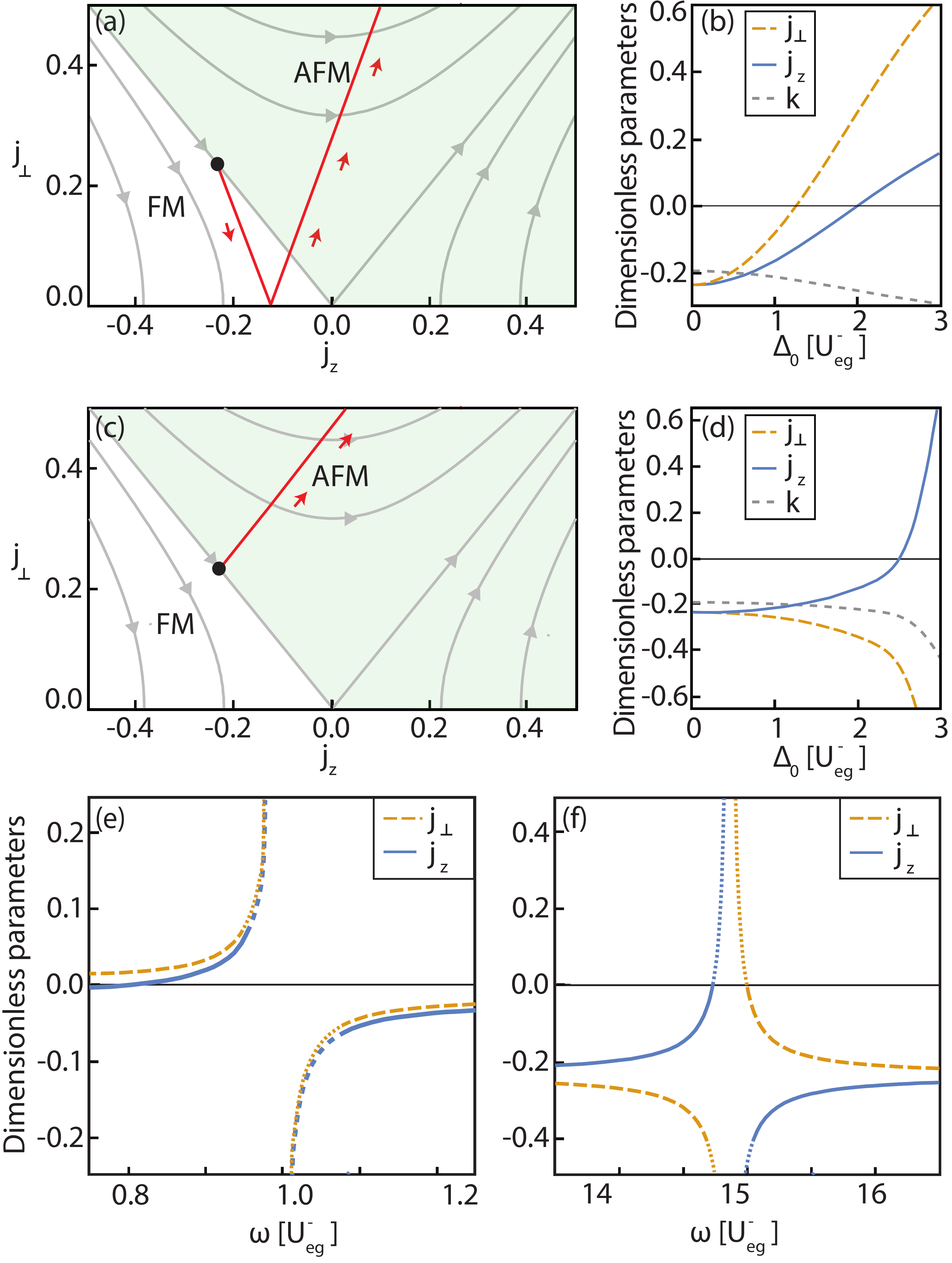}
\caption{ (Color.) Kondo model parameters in an oscillating Zeeman field $\corrnew{\Delta} (\corr{\tau}) = \corrnew{\Delta}_0 \, \cos(\omega \corr{\tau})$. The dimensionless couplings $j_z$ and $j_\perp$ as well as the potential scattering term $k = K \, \varrho(0)$ are determined at driving frequencies (a, b) $\omega = 0.8 \, U_{eg}^-$ and (c, d) $\omega = 0.95 \, U_{eg}^+$ as the Zeeman energy $\corrnew{\Delta}_0$ increases. The panels on the right show these parameters as functions of $\corrnew{\Delta}_0$. As the field has no static component, the effective magnetic couplings $m_e$ and $m_g$ vanish.  (e, f) Resonant behavior of the Kondo parameters at frequencies around (e) $U^-_{eg}$, with a driving amplitude $\corrnew{\Delta}_0 = 0.6 \, U_{eg}^-$, and (f) near $U^+_{eg}$, with an amplitude $\corrnew{\Delta}_0 = 1.5 \, U_{eg}^-$. Close to the resonance (dotted part of the curve) the Schrieffer-Wolff transformation becomes unreliable. [Parameters of the plot: $t=0.35 \, U_{eg}^-$, $U_{eg}^+ = 15  \, U_{eg}^-$.]
}
\label{fig:KondoParametersDriving}
\end{figure}

\subsubsection{Driven Zeeman field} \label{subsec:DrivenField}

In order to obtain full control of the Kondo model, it is important to find a way to tune the anisotropy of the Kondo parameters independently from the magnetic terms $m_e$ and $m_g$ in Eq.~\eqref{eq:KondoZeemanHamiltonian}. This can be achieved using a periodically modulated Zeeman field $\corrnew{\Delta}(\tau)$. The main insight is that the magnetic term in Eq.~\eqref{eq:m} is an odd function of the static Zeeman field $\corrnew{\Delta}$. It is therefore expected to average out to zero when the field is oscillating. In contrast, we expect that the anisotropy will remain finite, since it is an even function of the driving (see Fig.~\ref{fig:Kondo_parameters} and Eqs.~(\ref{eq:Jz}, \ref{eq:Jperp})). Furthermore, by combining a static and an oscillating Zeeman field components, both the anisotropy and the magnetic terms can be controlled individually. 

Engineering of driven Floquet Hamiltonians has been successfully applied in a wide variety of ultracold atomic systems. This technique has been used broadly to create synthetic gauge fields, topological bands~\cite{struck2011quantum, aidelsburger2011experimental, struck2012tunable, hauke2012non, aidelsburger2013realization, celi2014synthetic, aidelsburger2014measuring, jotzu2014experimental} and artificial spin-orbit coupling~\cite{Garcia2015SpinOrbit}. Driving has also been used in interacting systems to tune the superfluid to Mott insulator transition in bosonic systems~\cite{eckardt2005superfluid} 
as well as to control the interaction between atoms~\cite{kaufman2009radio, tscherbul2010rf}. 
Although one could naively expect that driving interacting systems could lead to heating, these experiments \corr{have demonstrated that excessive heating can be avoided by choosing the driving frequency far from the system's many-body excitations. We achieve this by choosing the driving frequency to be larger than the bandwidth, as} we show in Eq.~\eqref{eq:energy_scales}. 

Similarly to the static case, we obtain the low energy Kondo parameters using a Schrieffer-Wolff transformation, that decouples the high energy and the low energy subspace of the Hubbard-Anderson Hamiltonian. Since the bath and on-site Hamiltonians contain oscillating terms, the transformation needs to be time-dependent, and it is chosen to have the same periodicity as the driving field. The low energy sector contains terms that are much smaller than the driving frequency $\omega$. This allows us to perform a Floquet expansion in the transformed basis ~\cite{Shirley1965Floquet, Sambe1973Floquet, Rahav2003Floquet} in powers of $1/\omega$, and thereby derive the static Kondo parameters $J_z$ and $J_\perp$. Since the driving is much faster than the Kondo dynamics, we can stop at the lowest order Floquet term, which is simply the time average of the Hamiltonian. The details of this calculation can be found in \ref{app:SchriefferWolffDynamic}, we only present the results here. 

Figure~\ref{fig:KondoParametersDriving} shows how the Kondo couplings \corrnew{$j_z$ and $j_\perp$} depend on the amplitude of the oscillating field $\corrnew{\Delta}(\tau) = \corrnew{\Delta}_0 \, \cos(\omega \tau)$. Depending on the frequency of the driving, the couplings can show very different anisotropies. In Fig.~\ref{fig:KondoParametersDriving}~(a-b), the driving $\omega$ tuned below $U_{eg}^+$ creates a FM anisotropy at weak Zeeman fields. After an initial decrease where the couplings reach the line of FM fixed points $J_\perp = 0$, they grow again as the driving amplitude increases. Eventually, the couplings go from the FM to the AFM phase, allowing the experimental study of the phase transition. The experimental signatures of this transition, specific to cold atoms, \corrnew{are} discussed in Sec.~\ref{sec:finiteTemperatureEquilibrium}. 
The couplings exhibit the opposite behavior when the driving frequency is tuned below the $U_{eg}^+$ on-site energy of the singlet spin state, see Fig.~\ref{fig:KondoParametersDriving}~(c-d). As the driving amplitude increases, the system goes into the AFM phase already at weak driving.  We confirmed numerically that the magnetic couplings $m_e$ and $m_g$ vanish when the driving field's static component is zero.
In both of the above cases, the driving is red detuned from the on-site interactions $U_{eg}^\pm$. \corr{By suppressing the phase space available for particle-hole excitations, this reduces heating of the bath arising from the optical driving.} Since the driving frequency is below the on-site interaction energies, the excitation processes need to borrow an energy equal to the detuning $\delta\omega = U_{eg}^- - \omega$ from the bath. When the temperature is much smaller than the detuning $T\ll \delta\omega$, the probability of available quasi-particle hole excitations are exponentially reduced, leading to suppressed heating effects.

We find that the Kondo parameters depend resonantly on the driving frequency as it approaches the on-site interactions $U_{eg}^\pm$, shown in Fig.~\ref{fig:KondoParametersDriving}~(e). The driving field dresses the atoms entering the impurity site with multiples of the frequency $\omega$. When the dressed incoming energy approaches one of the on-site energies, we expect a resonant interaction between the impurity and the bath atoms, similarly to traditional Feshbach resonances. \corr{Our second order Schrieffer-Wolff results become unreliable close to the resonance when the Floquet energies become of the order of the coupling $V$, as indicated by the dotted parts of the curves.} In this regime, the higher order terms in the expansion can become non-negligible and more accurate calculations are needed to characterize the Kondo parameters' dependence on the driving.

We finally mention that the Kondo parameters can also be tuned by modulating the optical lattice amplitude, as we discuss in \ref{app:on-site_driving}.  \corr{Similar driving has recently been used to turn antiferromagnetic into ferromagnetic correlations the interactions in the Fermi-Hubbard model~\cite{Esslinger2017Floquet}.}
\corrnew{This type of driving} preserves the $SU(2)$ symmetry of the effective spin-$1/2$ Hubbard-Anderson Hamiltonian. The Kondo parameters, therefore, remain isotropic. \corr{In addition, this method allows for extending the physics to $SU(N>2)$-symmetric systems, as the $SU(N)$-symmetry of the underlying atoms is not broken.}

%
%

\begin{figure}[tbh!]
\includegraphics[width = 8.5cm]{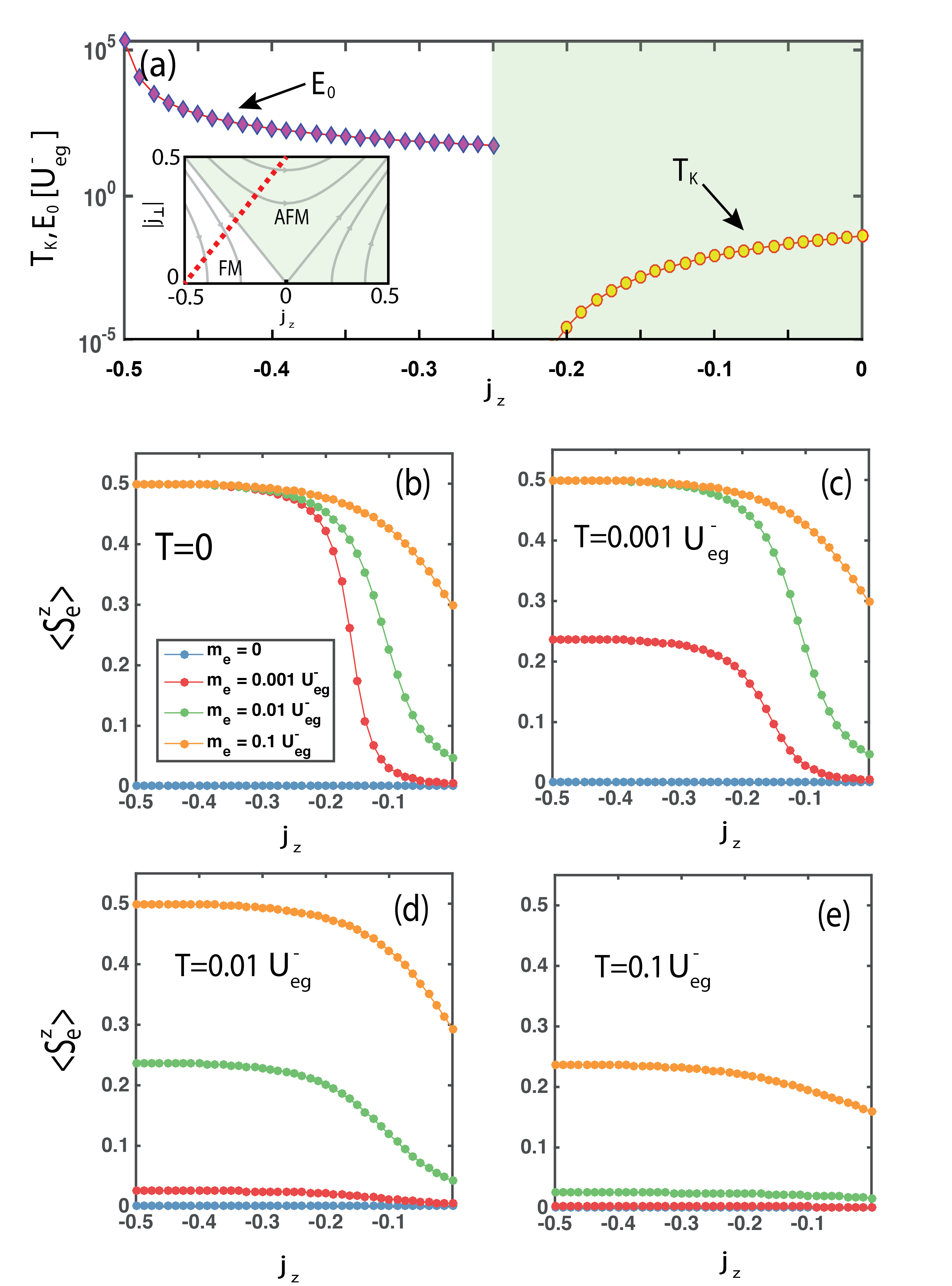}
\caption{ (Color.)  (a) The evolution of the Kondo temperature $T_K$ in the AF regime and the 
characteristic temperature $E_0$ in the FM domain.
 The parameters ($j_z$, $j_\perp$) are continuously tuned along the red (dotted) line in the inset
from (-0.5, 0) towards (0, 0.5). When $-0.5<j_z< -0.25$ the system displays the FM behavior and when 
$-0.25<j_z<0$ the system is in the AFM state.
(b)-(e) Zero and finite temperature equilibrium magnetization 
of the impurity $\langle S^z_e \rangle$  across the phase transition. Different line colors 
corresponds to different magnetic fields, as indicated in panel (b).
}
\label{fig:phaseTransition}
\end{figure}

\section{Ferromagnetic to antiferromagnetic phase transition} \label{sec:finiteTemperatureEquilibrium}

In this section, we discuss the experimental signatures of the phase transition between the easy-axis and easy-plane ferromagnetic Kondo interactions. As Fig.~\ref{fig:KondoParametersDriving}~(a, c) shows, periodically modulated  Zeeman fields allow one to tune the anisotropy of the exchange couplings and cross the phase boundary that separates the two regimes. 

\subsection{Local magnetization}\label{sec:lm}

Here, we consider the particular protocol in which dimensionless couplings ($j_z$, $j_\perp$) are linearly tuned from (-0.5, 0) to (0, 0.5), as indicated by the dashed line in the inset of Fig.~\ref{fig:phaseTransition}(a). The evolution of the characteristic energy scales $T_K$ and $E_0$ is displayed in Fig.~\ref{fig:phaseTransition}(a). \corr{These characteristic energy scales were defined in Sec.~\ref{sec:pd_es}.}
When ($j_z$, $j_\perp$) = (0, 0.5) the Kondo temperature is maximum $T_K\approx 0.05\, U_{eg}^-$. Moving towards the FM-AFM boundary, $T_K$ decreases exponentially and vanishes \corr{at the phase boundary} ($j_z$, $j_\perp$) = (-0.25, 0.25). On the FM side of the phase boundary, $E_0$ is order of magnitudes larger, $E_0\approx 10^2\, U_{eg}^-$ and increases towards the (-0.5, 0) point. Although $T_K$ and $E_0$ are the essential energy scales that characterize the two regimes, measuring them is a difficult task in general. 

A more useful way to visualize the transition between the FM and AFM regimes is to consider the temperature and magnetic field dependence of the magnetization of the impurity $\langle S_e^z\rangle$, shown in Fig.~\ref{fig:phaseTransition}~(b-e). The finite temperature  magnetization  was determined using numerical renormalization group calculations~\cite{toth2008BudapestNRG}.
In the  low temperature AFM regime ($T\ll T_K$), the many-body ground state (GS) is a Kondo singlet, $\langle S_e^z\rangle\approx 0$. In contrast, the ground state becomes a doublet in the FM regime. In the AFM phase, applying an effective magnetic field $m_e\lesssim T_K$ does not break up the singlet state. \corr{On the other hand, in the FM state,} even a small $m_e$ is sufficient to lift the degeneracy of the GS and polarize the local moment. This induces a finite local magnetization $\langle S_e^z\rangle\approx 1/2$. This behavior is clearly captured in Fig.~\ref{fig:phaseTransition}(b) where  results for the local magnetization at $T=0$ are presented. At finite temperature $T>0$, thermal fluctuations suppress the impurity magnetization, therefore larger effective magnetic fields $m_e\gtrsim T$ are required to fully polarize the local spin on the FM side of the transition.

\begin{figure}[tbhp!]
\includegraphics[width = 8.5cm]{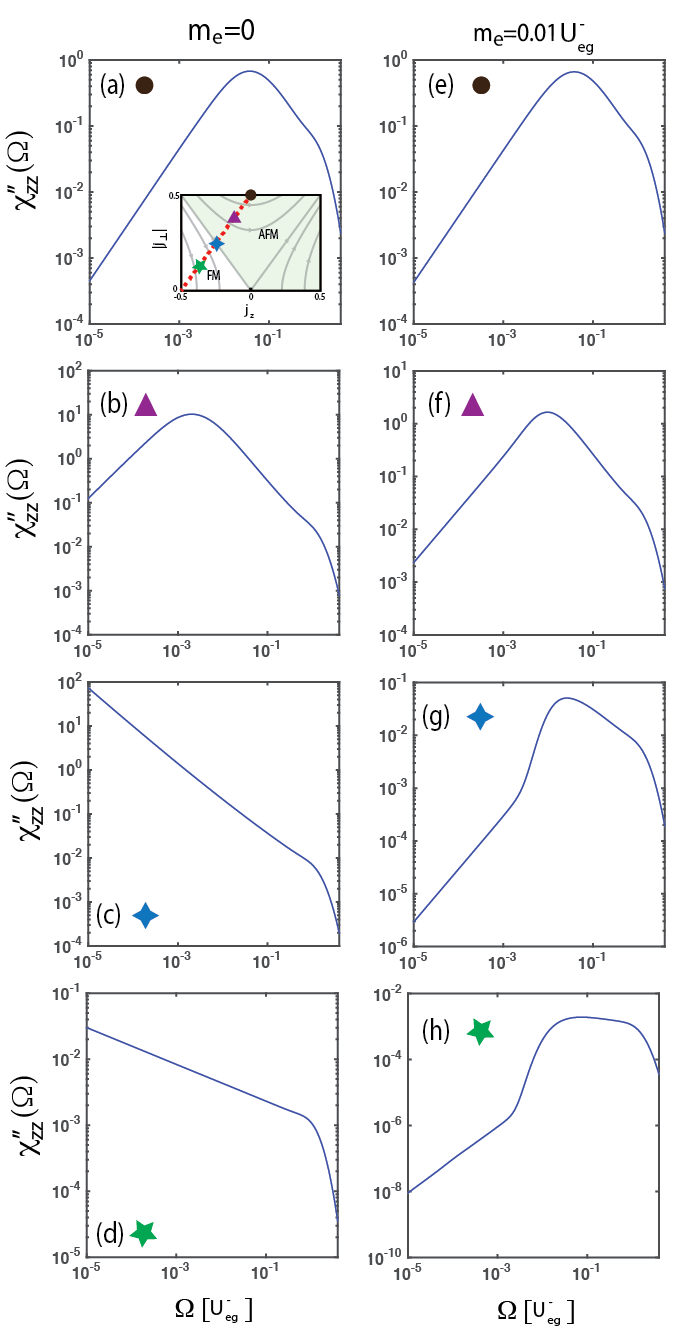}
\caption{ Magnetic susceptibility of the anisotropic Kondo impurity across the FM to AFM phase transition for $T=0$. Figures on the left (right) show results at zero (finite) magnetic fields. At zero magnetic field in the AF regime (a, b), the susceptibility depends linearly on the driving frequency $\chi_{zz}''(\Omega) \sim \Omega$, indicating AFM Kondo screening. This  behavior changes on the other side of the phase transition, where the FM ground state exhibits $\chi_{zz}''(\Omega) \sim 1/\Omega$ scaling. In case of finite magnetic fields, the low-frequency behavior of the imaginary part of the susceptibility always shows $\sim \Omega$ scaling \corr{at} frequencies $\Omega \lesssim B$. The symbols in the inset in panel (a) indicate the points ($j_z$, $j_\perp$) in the phase diagram where the susceptibility has been computed in each panel.
}
\label{fig:Susceptibilities}
\end{figure}

\subsection{Magnetic susceptibility of the impurity}
Whereas the ferromagnetic Kondo behavior can be investigated by measuring the impurity magnetization, this probe does not tell much about the AFM part of the phase diagram. $\langle S_e^z \rangle$ can be suppressed both by temperature fluctuations and by Kondo screening, therefore the on-set of the Kondo effect cannot be determined by looking at this observable alone. In this section, we show, however, that the dynamical spin susceptibility of the impurity can be used to directly detect Kondo screening. It is, therefore, a useful probe to determine the phase transition between the AFM and the FM phase.

\cpm{We investigate the time-dependent correlation of the local spin $S_e^z(\corr{\tau})$, 
\begin{equation}
\chi_{zz}(\corr{\tau-\tau^\prime}) = i\langle [S_e^z(\corr{\tau}), S_e^z(\corr{\tau}^\prime)] \rangle\theta(\corr{\tau}-\corr{\tau}^\prime)
\end{equation}
and determine the corresponding spin susceptibility spectrum $\chi''_{zz}(\Omega) = {\rm Im}\, \chi_{zz}(\Omega)$, which can be measured in an ultracold system using a Ramsey protocol, as \corrnew{described} in Ref.~\onlinecite{Knap2013CorrelationFunctionsRamsey}. \corrnew{$\chi_{zz}(\Omega)$ can be determined numerically using the numerical renormalization group. Analytically, however, it is} easier to compute another response function,  $\tilde \chi_{zz}(\corr{\tau}-\corr{\tau}^\prime) = i\langle [\dot S_e^z(\corr{\tau}), \dot S_e^z(\corr{\tau}^\prime)] \rangle\theta(\corr{\tau}-\corr{\tau}^\prime)$ instead, using perturbation theory. \corr{In frequency space, the spectral functions of these response functions are closely related,}
\begin{equation}
\chi_{zz}''(\Omega) = \frac{\tilde \chi_{zz}''(\Omega)}{\Omega ^2}.
\end{equation}
We rewrite the interaction term in the Hamiltonian by introducing the field $\psi _\sigma = \sum_\alpha g_{\alpha\sigma}$ which annihilates atoms with spin $\sigma$ in the bath. Using the equation of motion we obtain
\begin{equation}
\dot S_e^z(\corr{\tau}) ={j_\perp\over 2}\sum_{\sigma \sigma^\prime} \corrnew{\psi_\sigma^{\dagger} \Big( S_e^x  \sigma^y_{\sigma\sigma^\prime} - S_e^y \sigma^x_{\sigma\sigma^\prime}\Big)  \psi_{\sigma^\prime}}.
\end{equation}
We evaluate $\tilde \chi_{zz}(\corr{\tau})$ perturbatively order by order in $j_\perp$ and $j_z$. 
The 0th order gives
\begin{equation}
\chi_{zz}(\Omega) = {\pi j_\perp^2\over 4}  {1\over \Omega}.
\label{eq:chi}
\end{equation}
 As it is derived, Eq.~\eqref{eq:chi} is valid in both the FM as well as in the AFM regime, irrespective of the sign of the exchange coupling.
  
\subsubsection{AFM regime}

When the system is in the AFM regime, one obtains logarithmic corrections to the exchange coupling at a higher order. These contributions can be summed up by perturbative renormalization group proceduce~\cite{HewsonBook}. This amounts to replacing the bare coupling $j$ with its renormalized counterpart $j\to j(\Omega) \equiv 1/\ln(\Omega/T_K)$. We then find that \begin{equation}
\chi''_{zz}(\Omega) \approx {\pi\over 4} {1\over \corr{\Omega} \ln^2(\Omega/T_K)};\;\;\;\; |\Omega|\gg T_K\,.
\label{eq:chi_decay}
\end{equation}
In the Kondo limit $|\Omega|\ll T_K$ on the other hand, the spin spectral function takes on a  universal form,  $\chi_{zz}(\Omega) = f(\Omega/T_K)/T_K$. Here, $f(x)$ is a universal function that can be determined numerically. Its imaginary part can be approximated as $f''(x)\corrnew{\sim} x$, implying 
\begin{equation}
\chi''_{zz}(\Omega) \corrnew{\sim} {\Omega\over T_K^2};\;\;\;\; |\Omega|\ll T_K\,.
\end{equation}
a behavior which is characteristic for a Fermi liquid~\cite{nozieres1974fermi, mora2015fermi}. This leads to a finite spin susceptibility $\chi_{zz}\corrnew{\sim} 1/T_K$, in agreement with Bethe ansatz results~\cite{Andrei.1983}.
Figures~\ref{fig:Susceptibilities}~(a-b, e-f) show results for $\chi''_{zz}(\Omega)$ obtained using the NRG approach in the AFM regime.
The figures display the Fermi liquid properties described above: a linear increase, $\chi''_{zz}(\Omega)\sim \Omega$ at small frequencies $\Omega\ll T_K$, followed by a broad resonance at $\Omega \approx T_K$, and the decay predicted by Eq.~\eqref{eq:chi_decay} at large frequencies. This behavior survives in the presence of a small magnetic term, as can be seen in Fig.~\ref{fig:Susceptibilities}~(e,f). The Kondo state is affected only by a relatively large magnetic field $m_e\gtrsim T_K$.

\subsubsection{FM regime}
\corrnew{At the transition point}, the logarithmic correction is asymptotically exact down to frequencies $\Omega \to 0$. Using the same procedure, the renormalized coupling becomes $ j(\Omega) = 1/\ln(\Omega/E_0)$. In this regime, the spectral function is then given by
\begin{equation}
\chi''_{zz}(\Omega) \approx {\pi \over 4} {1\over \Omega \ln^2 (\Omega/E_0)};\;\;\;\; |\Omega|\to 0\,
\end{equation}
and it diverges in the $\Omega\to 0$ limit. 
\corrnew{In the ferromagnetic phase, however, the effective coupling $j_\perp(\Omega)$ scales to $0$ as $j_\perp(\Omega)\sim \Omega^\zeta$. As a consequence, here $\chi^”_{zz}$ displays a power law behavior, $\chi^”_{zz}\sim \Omega^{2\zeta -1}$, as can be seen in Fig.~\ref{fig:Susceptibilities}~(d).}
This singular behavior is also supported by the NRG results (see panels (c-d) in Fig.~\ref{fig:Susceptibilities}). In this respect, the FM side of the transition shows a singular Fermi liquid behavior~\cite{Zarand.2005SFL}, as the ferromagnetic coupling  tends to zero very slowly. The  presence of a finite magnetic field introduces a new energy scale, given by the Zeeman energy. This can be associated with the Fermi liquid scale $T_{FL}$ below which the regular nature of the Fermi liquid is restored and the $\chi''_{zz}(\Omega)\sim \Omega$ behavior is recovered. 
}

\section{Non-equilibrium dynamics}

With \corr{several} orders of magnitude slower dynamics than electronic systems, ultracold atoms provide an ideal setup to test the non-equilibrium dynamics of many-body dynamics~\cite{Eisert2014QuantumManyBodyOutOfEquilibrium}. In addition to their good time resolution, quantum gas microscopes allow for spatially resolved imaging of both the Kondo impurity and the bath in real time~\cite{miranda2015site, yamamoto2016ytterbium}.
After the creation of the impurity in the $\ket{\Uparrow}$ spin state (see Fig.~\ref{fig:proposal}b), both the impurity's and the bath's dynamics can be studied. Thereby the dynamical formation of the Kondo screening cloud could be measured. The short time dynamics is governed by the high energy excitations, whereas the long-time behavior is determined by the low energy degrees of freedom. Therefore, the system's behavior mimics that of the RG flows, where the RG parameter's role is played by the time. Depending on the value of the bare couplings, the dynamics of the system either leads to a ferromagnetic or antiferromagnetic behavior at long times.

\subsection{Relaxation in the \corr{easy-axis} ferromagnetic regime at finite temperature} \label{sec:finiteTemperatureRelaxation}

We start this section by focusing on the long-time exponential relaxation of the impurity in the \corr{easy-axis} ferromagnetic regime, as characterized by the Korringa law~\cite{Korringa1950, gupta2012KorringaExpt}. 
Whereas quantum corrections to the Korringa relaxation have been predicted early on~\cite{Zwerger1983}, these corrections have not been observed experimentally so far. We argue that at sufficiently low temperatures, these corrections should be measurable in ultracold atomic experiments.

\corr{In the easy axis FM regime, $|J_\perp| < |J_z|$, the zero temperature behavior of the system is dominated by the spin-dependent scattering term $J_z$}. This is the white region below the isotropic line on the phase diagram Fig.~\ref{fig:Kondo_phdiag}. Here, the RG flow brings the couplings into the line of ferromagnetic fixed points, with vanishing spin-flip terms $J_\perp = 0$. Thus, the ground state of the system is purely ferromagnetic, and bath atoms only participate in Ising type spin scattering. Based on the poor man's scaling equations, one would expect that the impurity spin freezes in this regime.

At finite temperature, the RG flow does not take its full course, and \corr{it is stopped} when the energy cut-off reaches the range of the temperature~\cite{HewsonBook}. At this point, the effective spin-flip term remains finite but suppressed compared to its bare value. Due to the thermal excitations from the bath, the impurity relaxes to its equilibrium value with a \corrnew{rate} $\nu(T)$. The temperature-dependence of the relaxation \corrnew{rate} has been estimated by Korringa based on the phase space available to thermal excitations in Fermi's golden rule. The Korringa law states that the \corrnew{relaxation rate} shall depend linearly on the temperature, and the impurity freezes at zero temperature. This result has been confirmed in a number of NMR measurements in solid state systems. 

However, as has been pointed out early on~\cite{Zwerger1983}, quantum corrections  lead to a power law temperature-dependence of the relaxation \corrnew{rate},
\beq
\nu(T) \sim T^{1+\eta}. \nonumber
\eeq
This work obtained quantum corrections originally within the spin-boson model, \corrnew{describing the relaxation of a spin in a decohering many-body bosonic environment. This model} is intimately related to the Kondo problem through bosonization of the bath~\cite{Leggett1987DissipativeTwoStateSystem}. Quantum corrections obtained in Ref.~\onlinecite{Zwerger1983} also describe the relaxation of the Kondo impurity close to the line of FM fixed points. Making use of the connection between the Kondo and spin-boson model parameters, the \corrnew{relaxation rate} can be expressed as
\beq
\eta = -j_z + \frac{j_z^2}{4} \corrnew{+ \dots}
\nonumber
\eeq
As Fig.~\ref{fig:KondoParametersDriving} shows, the dimensionless coupling can be as large as $j_z \sim 0.1 - 0.2$. The resulting correction to the Korringa law is of the order of $\eta \sim 0.1 - 0.2$, \corr{which could be measurable in the ultracold setup.}

\begin{figure*}
\includegraphics[width = 17cm]{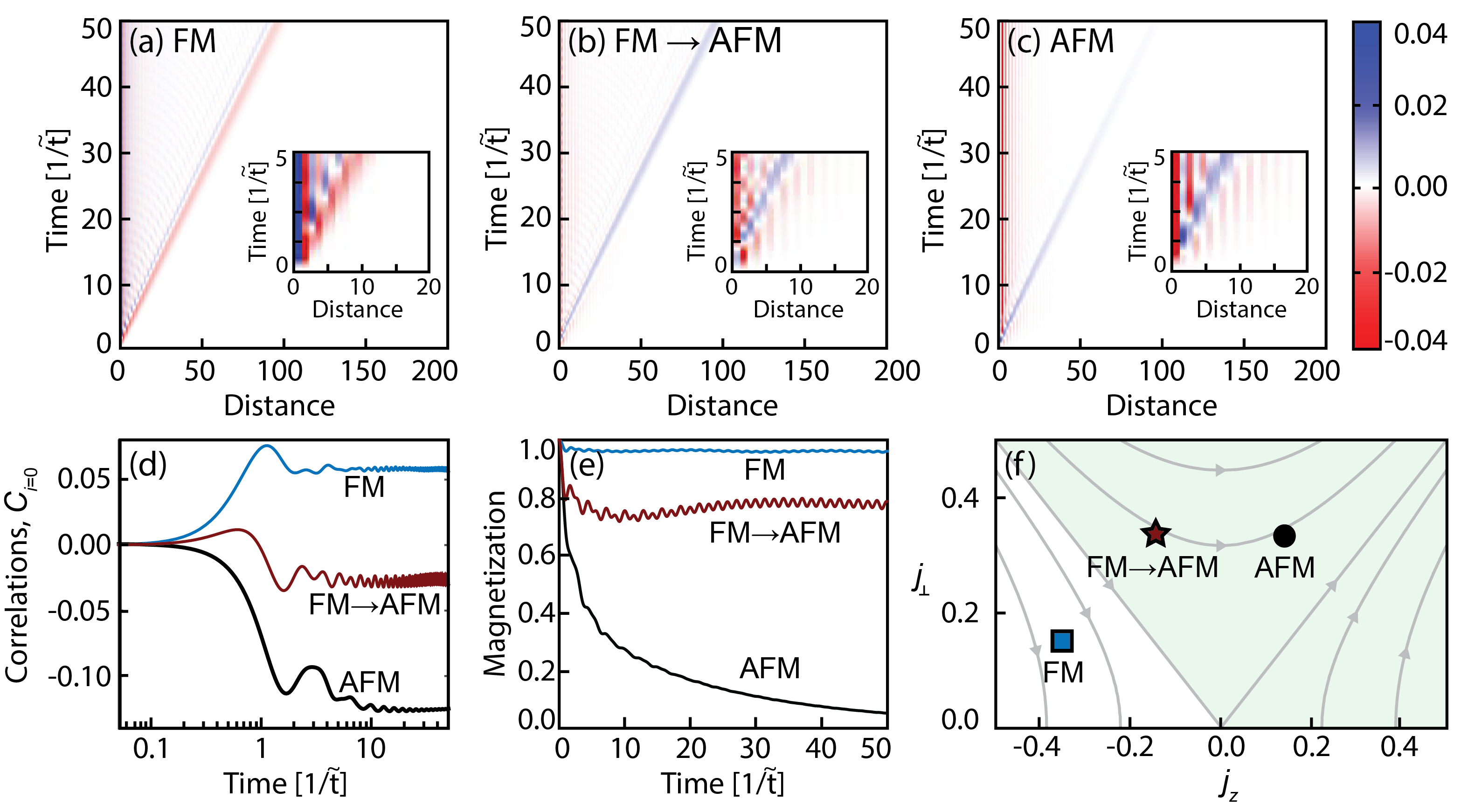}
\caption{ (Color.) 
\corr{Time-evolution after the creation of the impurity. (a-c) show the time-resolved impurity-bath spin correlations $C_i(\corr{\tau})$ (shown in Eq.~\eqref{eq:impurity_bath_spin_corr}). The excess polarization is emitted ballistically after the Kondo cloud forms around the impurity. The time-evolution of the correlations $C_{i=0}(\corr{\tau})$ at the impurity site and that of the impurity magnetization are shown in (d) and (e), respectively. The dynamics of these observables follow our expectations based on the RG flow shown in (f). The bare couplings corresponding to figures (a) $j_z = -0.35$, $j_\perp=0.15$, (b) $j_z = -0.15$, $j_\perp = 0.35$ and (c) $j_z = 0.15$, $j_\perp = 0.35$ are denoted as FM (rectangle), FM$\rightarrow$AFM (star) and AFM (circle). The easy-axis couplings in (a) flow to the ferromagnetic line of fixed points ($j_\perp = 0$) the dynamics remains ferromagnetic, as the impurity becomes ferromagnetically correlated with the surrounding bath of atoms. In contrast, the bare couplings, shown in (b) and (c), flow into the antiferromagnetic fixed point. After the formation of the Kondo screening cloud, the impurity magnetization decays and the impurity becomes antiferromagnetically aligned with the surrounding bath atoms. The bare ferromagnetic couplings in (b) determine the initially ferromagnetic dynamics. However, this quickly crosses over to antiferromagnetic behavior (see also (d)). [The calculation was done for a 1D open chain of range $[-L, L]$ with $L=200$ sites, and with the impurity at the origin. \corrnew{$\tilde{t}$ denotes the tunneling matrix element along the chain.]}}
}
\label{fig:zeroTemperatureBathDynamics}
\end{figure*}

\subsection{\corr{Quench} dynamics at zero temperature} \label{sec:zero_temperature}

\corr{Ultracold atoms not only make it possible to explore the  equilibrium properties of the screening cloud in the ground state; they also allow one to study how it is formed starting from an initial non-equilibrium state. Our discussion will emphasize new aspects of the Kondo dynamics that can be analyzed using quantum gas microscopes. This includes, for example, time-dependent spin correlations between the impurity and spins of the bath atoms.  

We now consider quench dynamics of the anisotropic Kondo model. For $\corr{\tau}<0$ the impurity spin is completely decoupled from the fermionic bath, and this coupling is switched on abruptly at $\corr{\tau}=0$. This protocol is closely related to the optical spectroscopy performed in electron systems in experiments by Tureci et al.~\cite{tureci2011many}. The most intriguing aspect of the Kondo system that we aim to explore is the formation of the screening cloud around the impurity spin. This effect is particularly striking in the ferromagnetic easy-plane regime of the model: the impurity spin gets screened even though interactions are ferromagnetic, to begin with.

\corr{We note that the Kondo model is integrable (when the density of states can be assumed to be constant)~\cite{andrei1981Kondo, wiegmann1981exact, wigman1982exact, andrei1995integrable}. Hence, the dynamics should contain signatures of the conservation laws of the system.} We will not discuss integrability aspects of the problem in the current paper (see Refs.~\onlinecite{wiegmann1981exact, andrei1981Kondo, wigman1982exact,
tsvelick1984solution, affleck1993exact, fendley1996unified, leclair1999minimal, zhou1999algebraic} for a discussion of some of these issues.)}

Arguably the most interesting possibility of the ultracold atomic realization of the Kondo model is the opportunity to measure its non-equilibrium dynamics in real time. We discuss smoking gun experimental signatures of this process specific to cold atomic experiments, at several parts of the phase diagram. Among other observables, we discuss how quantum gas microscopy can be used to measure the screening and the bath's spin dynamics. 

We \corr{point out} that the time-dependent and spatially resolved Kondo dynamics is still an area of active theoretical research, with many open questions. The ultracold atomic toolbox could provide enormous insight into testing theoretical predictions.  Despite the wide variety of methods used to solve this problem, current techniques are often limited to certain parts of the phase diagram or they can only determine the dynamics of the impurity but not that of the bath degrees of freedom. Earlier works have relied on non-equilibrium Monte Carlo~\cite{Schmidt2008MonteCarlo}, 
DMRG~\cite{White2004DMRG, Schmitteckert2004DMRG}, 
TD-NRG~\cite{Anders2005TDNRG, anders2006spin}, 
the flow equation method~\cite{lobaskin2005CrossoverFromNonequlibriumToEquilibrium_FlowEquation, 
Hackl2009FM_Kondo_FlowEquation}, 
time-evolving block decimation (TEBD)~\cite{nuss2015nonequilibriumKondoTEBD, dora2017information},
\corr{as well as analytical solutions}~\cite{guinea1985bosonization, Affleck1991KondoConformalFieldTheory, Affleck1992KondoConformalFieldTheory, Lesagne1996TimeCorrelationsExact, Lesage1998BoundaryExact, medvedyeva2013spatiotemporal}.
Techniques such as perturbative renormalization group methods~\cite{Abrikosov1965PerturbativeKondo, Fowler1971ScalingKondo} have been mainly limited to the regime of weak coupling between the quantum dot and the reservoirs~\cite{nordlander1999HowLongDoesItTakeForTheKondoEffectToDevelop, Keil2001PerturbativeRGSpinBoson, Lobaskin2005PerturbativeRG_FM_Kondo, Hackl2008PerturbativeRGSpinBoson,  pletyukhov2010RelaxationPerturbativeRG_WeakCoupling}.

Our predictions are obtained using a non-Gaussian variational method at zero temperature covering all parts of the phase diagram~\cite{TaoPaper, yuto2017ToBePublished}. \corr{The dynamics of the system is governed by the Kondo Hamiltonian Eqs.~(\ref{eq:KondoBathHamiltonian}, \ref{eq:KondoInteractionHamiltonian}) with a vanishing potential scattering term. In order to make the calculations numerically tractable, we model the bath by a one-dimensional chain, with a tunneling $\tilde{t}$. The density of states of the chain is set to $\varrho_{1D}(0) = 1/(2\pi \tilde{t})$.}

We consider the quench dynamics starting from the decoupled initial state $|\Psi_{0}\rangle=|\Uparrow\rangle|{\rm FS}\rangle$, where $|\Uparrow\rangle$ is the impurity spin in a positive spin-$z$ direction and $|{\rm FS}\rangle$ denotes the Fermi sea of bath fermions, i.e., the ground state without the Kondo coupling. 
\corr{Using the experimental procedure outlined in Sec.~\ref{sec:introduction}, we refer the reader to~\ref{app:ImpurityInitialization} for the details. At time $\corr{\tau}=0$, a $\pi$ pulse of a weak laser excites a small number of bath fermions from the $\ket{g \uparrow}$ state into the $\ket{e \Uparrow}$ state. 
We determine the time-evolution of the coupled bath-impurity system.} 
In Fig.~\ref{fig:zeroTemperatureBathDynamics}~(a-c), we plot the impurity-bath spin correlations 
\beq
\corrnew{C_{i}(\corr{\tau})= \sum_{\sigma \sigma^\prime}\langle S_{e}^{z} \; \frac{1}{2} (g^{\dagger}_{i\uparrow} g_{i\uparrow} - g^{\dagger}_{i\downarrow} g_{i\downarrow})\rangle_{\corr{\tau}}}
\label{eq:impurity_bath_spin_corr}
\eeq
in FM (a) and AFM (b,c) phases, where $i$ labels a lattice site and $\langle\cdots\rangle_{\corr{\tau}}$ denotes an expectation value with respect to the time-evolving state $|\Psi_{\corr{\tau}}\rangle$. (d) shows the correlation $C_{i=0}(\corr{\tau})$ at the impurity site.  Note that our calculations can be done without relying on the bosonization, in which one assumes a strictly linear dispersion of the bath, see e.g. the TD-NRG method.  \corr{This allows us to  analyze an experimentally relevant situation of  fermions on a lattice, where a cut-off scale is naturally given by the lattice bandwidth and the energy dispersion is nonlinear in general.}

Figure~\ref{fig:zeroTemperatureBathDynamics}~(a) corresponds to the easy-axis ferromagnetic part of the phase diagram and demonstrates the formation of the ferromagnetic correlations. This indicates the triplet state of the impurity spin and the co-aligned spin cloud in the bath. The excess local polarization is emitted and propagates ballistically. Inside the light cone of the spin polarization, the ferromagnetic correlations develop at the timescale of the Fermi energy, as a result of the fast response from the Fermi sea.

Above the line of isotropic couplings in Fig.~\ref{fig:Kondo_phdiag}, the parameters flow into the AFM fixed point. Mimicking this RG flow, the dynamics shows \corrnew{shows a cross-over from the \corr{easy-plane} FM regime to the AFM phase,} as shown in Fig.~\ref{fig:zeroTemperatureBathDynamics}~(b). In the short time evolution, the corresponding correlation $C_{i=0}(\corr{\tau})$ increases and becomes positive (see Fig.~\ref{fig:zeroTemperatureBathDynamics} (d)). This indicates the formation of the triplet state between the impurity spin and the surrounding fermions. \corr{At later times, however, \corrnew{bath fermions near the impurity change their spin polarization abruptly,} and become antiferromagnetically aligned, as signified by the antiferromagnetic correlation $C_{i=0}(\corr{\tau})<0$.} This contrasts with the dynamics in the \corr{intrinsically} AFM regime with a coupling $j_z>0$ (Fig.~\ref{fig:zeroTemperatureBathDynamics}~(c)), where the parameters monotonically flow into the AFM fixed point (Fig.~\ref{fig:Kondo_phdiag}) and thus the localized fermions exhibit the antiferromagnetic spin correlation  $C_{i=0}(\corr{\tau})<0$ at all times. After the emission of \corr{the ferromagnetic excess spin polarization, correlations} between the impurity and the surrounding spins quickly become antiferromagnetic (see Fig.~\ref{fig:zeroTemperatureBathDynamics}~(b,c)). The nonvanishing correlations outside the light cone can be attributed to the initial entanglement in the bath Fermi sea in  coordinate space \cite{medvedyeva2013spatiotemporal}.

\begin{figure}
\includegraphics[width=7.5cm]{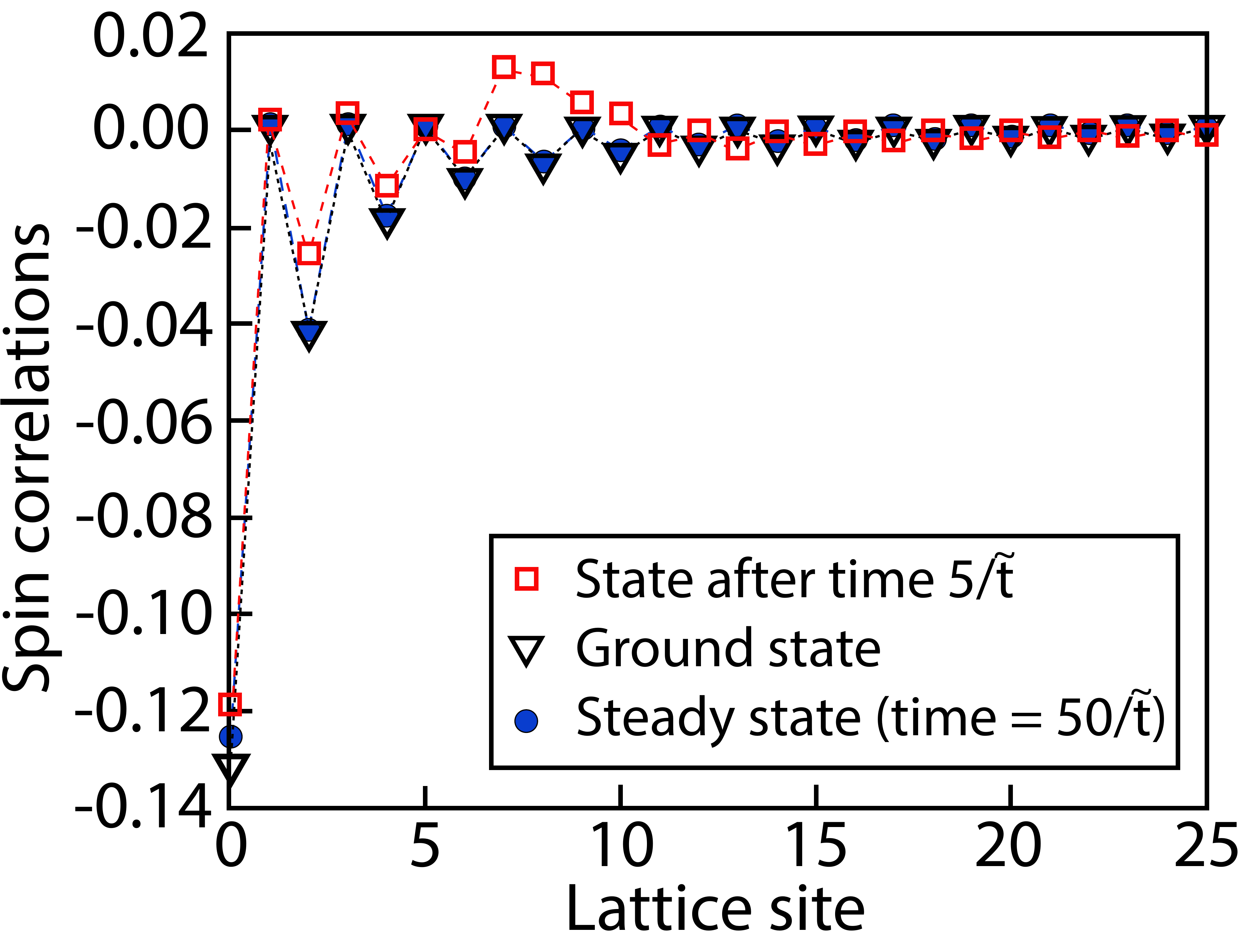}
\caption{ (Color.)  Non-equilibrium and equilibrium impurity-bath spin correlations. The correlation $C_{i}(\corr{\tau})$ \corrnew{(defined in Eq.~\eqref{eq:impurity_bath_spin_corr})} in the steady-state regime $\corr{\tau}=50 / \corrnew{\tilde{t}}$ agrees with the equilibrium values of the corresponding ground state obtained by the imaginary-time evolution. At the intermediate time $\corr{\tau}=5 / \corrnew{\tilde{t}}$, the emitted spin polarization forms an effective light cone in which the AFM correlations are partially developed. All the parameters are set to be the same as in Fig.~\ref{fig:zeroTemperatureBathDynamics}~(c).
}
\label{fig:zeroTemperatureSpin}
\end{figure}

We show the dynamics of the impurity magnetization $\langle S_{e}^{z}\rangle_{\corr{\tau}}$ in Fig.~\ref{fig:zeroTemperatureBathDynamics}~(e) in the corresponding regimes. In the AFM phase with  positive $j_z>0$ (Fig.~\ref{fig:zeroTemperatureBathDynamics}~(f)), the impurity spin monotonically relaxes to zero, indicating the formation of the Kondo singlet. This is consistent with the results \cite{anders2006spin} obtained in the spin-boson model, which is equivalent to the bosonized, low-energy effective theory of the anisotropic Kondo model \cite{Leggett1987DissipativeTwoStateSystem}. We find the oscillations with period $2\pi \hbar / D = \pi \hbar / 2 \tilde{t}$, as characterized by the bandwidth $\mathcal{D}= 4 \tilde{t}$. These are associated with a high-energy excitation of a particle from the bottom of the band to the Fermi level \cite{Knap2012} and were absent in the bosonized treatments. Correspondingly, the long-lasting oscillations with the same period can also be found in the impurity-bath spin correlations, see also Fig.~\ref{fig:zeroTemperatureBathDynamics}~(a,b) and (d). In the AFM phase, the couplings flow into the infinite AFM fixed point, which should make the magnetization ultimately relax to an equilibrium value close to zero (see Fig.~\ref{fig:phaseTransition}). Such an ultimate relaxation is hampered in the plotted timescale due to the small Kondo temperature of the parameters, leading to an exponentially slow decay \corrnew{during the FM to AFM crossover}.

\corrnew{Figure~\ref{fig:zeroTemperatureSpin} shows the spin correlations $C_i(\corr{\tau})$ between the impurity and the bath atoms. After the formation of the Kondo singlet in the long-time regime, the correlations reach the equilibrium values of the ground state, with spatially dependent AFM correlations. These correlations are only formed within a finite light cone at intermediate times. At the edge of the cone, the propagation of the excess spin of the Kondo impurity leads to FM correlations between the bath and the impurity (see also Ref.~\onlinecite{nuss2015nonequilibriumKondoTEBD}).}

\section{Outlook}

\corr{
Alkaline-earth atoms allow the realization of a wide variety of Kondo systems that are beyond the scope of this work. Whereas we considered localized impurities, \corrnew{mobile} heavy impurities are expected to show even more complex behavior. Such impurities can be realized by introducing shallower lattice potentials~\cite{zhang2016kondo, nakagawa2015LaserInducedKondo} as well as by using atomic mixtures~\cite{recati2005atomic, Bauer2013}. In one-dimensional systems, this \corrnew{may} lead to the realization of a two-channel Kondo model, as was shown in Ref.~\onlinecite{Lamacraft2008KondoIn1D}. In higher dimensions, the recoil energy of the collision between the impurity and the bath atoms could suppress low-energy spin exchange processes. We, therefore, expect that Kondo screening will appear only at finite values of the coupling $J$. This behavior is also characteristic of narrow-gap semiconductors and semi-metals such as graphene: since the density of states is suppressed at the Fermi energy, a magnetic impurity only shows a Kondo effect
if the strength of the coupling is strong enough~\cite{withoff1990phase, WithoffFradkin1990Kondo, borkowski1992kondo, chen1995kondo,  ingersent1996gaplessFermi, hentschel2007orthogonality, sengupta2008tuning}. These band structures can be realized using honeycomb and optical superlattices, which allow one to control the density of states at the Fermi level. 

Quantum gas microscopy could provide a completely new experimental perspective on the interplay of two Kondo impurities. This system has been studied early on \cite{JonesVarma1987TwoKondoImpurities, JonesVarma1988TwoKondoImpurities, Affleck1992TwoKondoImpurities, Affleck1995TwoKondoImpurities}:
In the $SU(2)$ symmetric case, its equilibrium properties depend non-universally on the dimensionless ratio of the RKKY interaction strength and the Kondo temperature. This ratio can be controlled by changing the spatial separation of the impurities as well as by modifying the filling of the band~\corrnew{\cite{Affleck1995TwoKondoImpurities, zarand2006quantum}}. Ultracold experiments could study the screening process in these phases in a spatially resolved way. Further intriguing questions arise in the case of quench dynamics, which is exceptionally hard to investigate theoretically, especially in the case of anisotropic interactions, made possible by optical driving.

Creating a Kondo impurity at each site of the optical lattice realizes the Kondo lattice, the paradigmatic model of heavy-fermion materials~\cite{Graebner1975OriginalHeavyFermionPaper, doniach1977kondo, si1999quantum, coleman2001fermi, si2001locally}.
These systems exhibit enormous quasi-particle masses as compared to that of the bath fermions. This mass renormalization should be measurable in transport and Bloch oscillation measurements. These systems also exhibit quantum critical behavior, topological and exotic superconducting orders. Using the periodically modulated  optical fields discussed in this work, one could also realize the anisotropic Kondo lattice model and study its complex phases.

Further interesting questions arise about the effect of disorder on the Kondo dynamics. Optical speckle potentials have been used extensively to create Anderson localized and diffusive phases in cold atomic baths~\cite{billy2008direct, roati2008anderson, jendrzejewski2012three}. Since disorder leads to local changes in the density of states, the Kondo energy scales will also become randomly distributed. The disordered Kondo model still shows non-Fermi-liquid behavior in the AFM phase~\cite{Dobrosavljeviifmmode1992DisorderedKondo}. In quench experiments performed in the localized phase, spin \corr{polarization emitted by the impurity} is expected to show revivals, that might be detrimental to the formation of the Kondo singlet. Three dimensional disordered systems show diffusive behavior below the mobility edge~\cite{jendrzejewski2012three}. Instead of ballistic propagation, the spin \corr{polarization}  emitted by the Kondo impurity will propagate diffusively and will likely lead to a very different time evolution of the impurity-bath correlations as compared to the disorder free case.

We finally mention that by populating $N>2$ spin components of alkaline-earth atoms, one can naturally create an $SU(N)$ symmetric version of the FM Kondo model. As we discuss in Sec.~\ref{app:opticalStark}, the optical driving suggested in this paper can break this symmetry down to a product of $U(1)$ symmetries. In quench experiments, we expect that the anisotropy of the Kondo coupling terms will lead to several different dynamical time-scales. Ultracold experiments would allow studying the effect of this symmetry breaking on the Korringa relaxation and on the cross-over from the FM to the screened phase. 
}

\section{Acknowledgements}
\corrnew{We thank I. Bloch for a stimulating discussion that motivated this project and for many valuable comments and suggestions. Enlightening discussions with D. Greif, R. Schmidt, A. Rosch, J. von Delft, D. Abanin, W. Hofstetter, S. Kehrein, V. Gritsev, N. Andrei, E. Andrei, W. Zwerger, T. Giamarchi, M. Nakagawa and S. Furukawa are gratefully acknowledged. M. K.-N. and E. A. D. acknowledge support from the Harvard-MIT CUA, NSF Grant No. DMR-1308435, AFOSR Quantum Simulation MURI and AFOSR grant number FA9550-16-1-0323. T. S. and J. I. C. were partially funded by the ERC grant QENOCOBA (no. 742102). T. S. acknowledges support from the EUproject SIQS and the Thousand-Youth-Talent Program of China. C. P. M. and G. Z. were partially supported by the National Research Development and Innovation Office of Hungary under Project No. 2017-1.2.1-NKP-2017-00001. C. P. M. acknowledges support from the UEFISCDI Romanian Grant No. PN-III-P4-ID-PCE-2016-0032. Y. A. and T. N. I. acknowledge the Japan Society for the Promotion of Science through Program for Leading Graduate Schools (ALPS) and Grant Nos. JP16J03613 and JP16H06718 as well as Harvard University for hospitality.}

\newpage

\appendix

 \begin{center}
    {\bf APPENDIX}
  \end{center}

\section{Derivation of the hybridization} \label{app:hybridization}

In this section, we discuss the derivation of the hybridization $V$ in Eq.~\eqref{eq:Hmix} and the density of states $\corr{\varrho}$ of bath eigenmodes. The calculation below applies to cubic lattices in any dimension $d$, as well as to arbitrary fillings. 

The hybridization couples the impurity site to the bath modes $|{\alpha}, \sigma \rangle = g^\dagger_{\alpha \sigma} |0\rangle$. This coupling arises from the tunneling Hamiltonian between the impurity and the surrounding sites 
\beq
H_{\rm tun} = -t \, \sum_{\langle \delta, 0 \rangle, \, \sigma} g_{\delta \sigma}^\dagger \, g_{0 \sigma} + {\rm h.c.} = -V \sum_\sigma (g_{h\sigma}^\dagger \, g_{0 \sigma} + {\rm h.c.}),
\nonumber
\eeq 
where $V = \sqrt{z} t$ is the hybridization. In the last equation, we introduced the creation operator of the hybridizing orbit $| h, \sigma \rangle  = g_{h \sigma}^\dagger |0 \rangle$, which is the equal superposition of states on sites neighboring the impurity,
\beq
g_{h \sigma}^\dagger = \frac{1}{\sqrt{z}}\sum_{\langle \delta, 0 \rangle} g_{\delta \sigma}^\dagger.
\nonumber
\eeq

The bath's dynamics is described by the hopping Hamiltonian on the remaining sites, $H_{\rm bath} = H_{\rm kin} - H_{\rm tun}$. This operator obeys $d$-dimensional cubic symmetries. Since the hybridizing orbit transforms trivially under these symmetry group, its overlap $\Lambda_{\alpha} = \langle \alpha, \sigma   | h, \sigma \rangle$ with bath modes is non-zero only for modes with the same symmetry~\cite{JonesBookSymmetries}. These overlaps are non-trivial due to the boundary conditions the bath eigenmodes obey at the impurity site. Since $H_{\rm bath}$ does not contain the tunnel coupling between the impurity and its neighbors, its eigenmodes need to vanish at the impurity site. The  density of states $\varrho(\omega)$ is defined as
\beq
\corr{\varrho}(\omega) = \sum_{\alpha} \left| \Lambda_{\alpha} \right|^2 \, \delta(\omega - \epsilon_{\alpha}) = M \, |\Lambda(\omega)|^2 \, \rho_{{\rm bath}}(\omega).
\nonumber
\eeq
incorporates the density of states of these modes, together with their coupling to the hybridizing orbit.
Here, $\Lambda(\omega)$ denotes the average matrix element of the hybridizing orbit with states at energy $\omega$, and $\rho_{{\rm bath}}(\omega) = \frac{1}{M} \sum_\alpha \delta(\omega - \epsilon_\alpha)$ is the density of states of bath atoms. 

We determine $\corr{\varrho}(\epsilon_F) = -\frac{1}{\pi} \, {\rm Im} \, G^R_h (\epsilon_F)$ using the retarded Green's function $G^R_h(\corr{\tau}) = \langle {\rm FS} | \{g_h(\corr{\tau}), g_h^\dagger(0)\} | {\rm FS}\rangle$. The time evolution of the operator $g_h(\corr{\tau}) = \exp(i H_{\rm bath} \corr{\tau}) \, g_h \, \exp(-i H_{\rm bath} \corr{\tau})$ is generated by the bath Hamiltonian.
We calculate $G_{h}(\omega)$ by introducing an auxiliary lattice Hamiltonian 
\beq
\tilde{H}_{\rm bath}(\lambda) = -t \sum_{\langle i, j \rangle,  \sigma} g^\dagger_{i \sigma} g_{j \sigma} + \lambda \, g_{0\sigma}^\dagger g_{0\sigma}
\nonumber
\eeq
with the potential $\lambda$ at the impurity site. For $\lambda \to \infty$, $\tilde{H}_{\rm bath}(\lambda)$ is equivalent to the bath Hamiltonian. Therefore, the Green's function $\tilde{G}^R_{h, \lambda} (\omega)$ generated by $\tilde{H}_{\rm bath}(\lambda)$ also becomes identical to $G^R_{h} (\omega)$ in this limit.
We determine $\tilde{G}^R_{h, \lambda} (\omega)$ by expanding it in terms of $\lambda$ to infinite order. The lowest order term is given by
\bea
\tilde{G}^R_{h, \lambda = 0} (\omega) &=& \frac{1}{z M} \sum_{\bf k} \frac{(\epsilon_{\bf k} / t)^2}{\omega - \epsilon_{\bf k} + i 0^+} \nonumber 
\\
&=& \frac{1}{z} \int d\epsilon \, (\epsilon/t)^2 \,  \frac{\varrho_{{\rm bath}}(\epsilon)}{\omega - \epsilon + i 0^+},
\nonumber
\eea
with $0^+$ denoting an infinitesimally small positive constant. A straightforward calculation leads to the higher order terms in the Lippmann-Schwinger equation
\beq
\tilde{G}^R_{h, \lambda} (\omega) = \tilde{G}^R_{h, \lambda = 0} (\omega)  
+ \frac{\Lambda^2(\omega)}{z} \,\sum_{n=1}^\infty \lambda^n \, \Pi^{n-1}(\omega).
\label{eq:LippmanSchwinger}
\eeq
The local Green's function $\Pi(\omega)$ in the last equation is defined as
\beq
\Pi(\omega) = \frac{1}{M} \sum_{\bf k}  \frac{1}{\omega - \epsilon_{\bf k} + i 0^+} = \int d\epsilon \frac{\varrho_{{\rm bath}}(\epsilon)}{\omega - \epsilon + i 0^+},
\label{eq:local_Greens_fn}
\eeq
whereas $\Lambda(\omega) \equiv (\omega\, \Pi(\omega) - 1) / t$.

\begin{figure}
\includegraphics[width=7.0cm,clip=true]{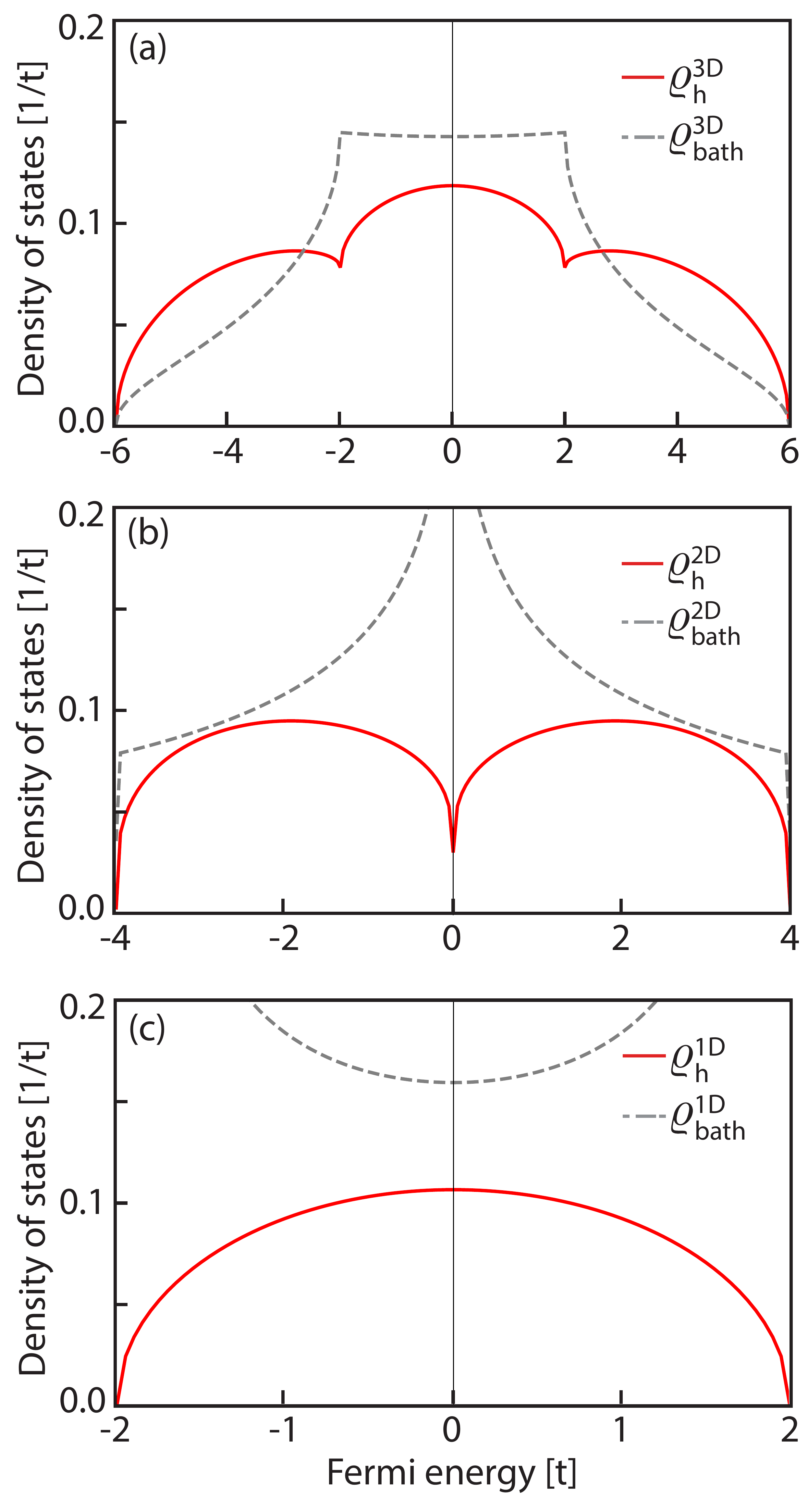}
\caption{
(Color) Density of states $\corr{\varrho}(\epsilon_F)$ in a (a) three (b) two and (c) one-dimensional optical lattice (solid line), as a function of the Fermi energy $\epsilon_F$. The $\rho_{{\rm bath}}$ density of states of the bath (dashed line) is also shown. In three dimensions, $\corr{\varrho}(\epsilon_F)$ is slightly suppressed near half-filling as compared to $\rho_{{\rm bath}}$, whereas it is enhanced towards the band edges. In contrast, the density of states is always smaller than $\rho_{{\rm bath}}$ in lower dimensions.
}
\label{fig:DOS}
\end{figure}

The Green's function in Eq.~\eqref{eq:LippmanSchwinger} can be summed up as a geometric series. After taking the $\lambda \to \infty$ limit, we find that 
\beq
G^R_{h} (\omega) = \tilde{G}^R_{h, \lambda = 0} (\omega) - \frac{1}{z} \, \frac{\Lambda^2(\omega)}{\Pi(\omega)}.
\nonumber
\eeq
As a final step, we determine the tunneling density of states from the imaginary part of the last equation,
\beq
\corr{\varrho}(\omega) = \frac{1}{\pi} \frac{1}{z t^2} \, {\rm Im}\frac{1}{\Pi(\omega)}.
\nonumber
\eeq
Fig.~\ref{fig:DOS} shows $\corr{\varrho}(\epsilon_F)$ together with $\rho_{{\rm bath}}(\epsilon_F)$ in spatial dimensions $d=1,2$ and $3$. At half-filling, the real-part of $\Pi(\omega)$ vanishes due to particle-hole symmetry, and we find $\corr{\varrho}(0) = 1/(\pi^2 \, z t \, \rho_{{\rm bath}}(0))$. In three spatial dimensions, we get $\varrho_h(0) = 0.118/t$, which is slightly suppressed as compared to $\rho_{{\rm bath}}(0) = 0.143/t$. In contrast,  $\corr{\varrho}$ is enhanced significantly towards the band edges.

We mention that, for a numerical evaluation, it is useful to express the local Green's function in Eq.~\eqref{eq:local_Greens_fn} as an integral. We rewrite the first denominator as an exponential integral and make use of the  integral representation of Bessel functions $J_0(x) = \int_{-\pi}^\pi \frac{dk}{2\pi}\exp(i  x \cos(k))$. The Green's function in $d$ dimensions is thus given by 
\beq
\Pi(\omega) = -\frac{i}{2 t} \int_0^\infty dx \, e^{i x (\omega + i 0^+) / 2 t} \, \left(J_0(x)\right)^d. \nonumber
\eeq

\section{Dimensionless parameters at anisotropic density of states} \label{app:DifferentDOS}

In this appendix, we outline Anderson's poor man's scaling equations in the case when the density of states of the fermionic degrees of freedom is different for the two spin components $\corr{\varrho}_{\upa} \neq \corr{\varrho}_{\downa}$. We illustrate how the anisotropy of the dimensionless Kondo couplings can be different from those of $J_z$ and $J_\perp$. \corr{We discuss a simplified case when the magnetic terms in Eq.~\eqref{eq:KondoZeemanHamiltonian} are neglected and the density of states is constant within the bandwidth $[-{\cal D}, {\cal D}]$ of the bath. Scaling in the more general case (with energy-dependent density of states and magnetic terms) more detailed numerical calculations. This can be done using  numerical renormalization group methods~\cite{toth2008BudapestNRG}.}

As a first step, we represent the Kondo interaction Hamiltonian in a vectorial form
\beq
H^K_{\rm int} = \frac{1}{M}\, \sum_{{\alpha} {\beta} \sigma \sigma^\prime} \sum_{a = x, y, z}
J_a \; S_e^a \, s^a_{g \sigma \sigma^\prime} \; c_{{\alpha} \sigma}^\dagger c_{{\beta} \sigma^\prime} \nonumber,
\eeq
where the bath spins are represented by the spin matrices ${\bf s}_g = ({s}_g^x, {s}_g^y, {s}_g^z)$. The couplings are given by $(J_x, J_y, J_z) = (J_\perp, J_\perp, J_z)$. In this representation, Anderson's poor man's scaling relations become~\cite{VladarZawadowski1983theory1, VladarZawadowski1983theory2, VladarZarand1996Low}
\beq
\frac{\delta J_a \, { s}^a_{g \sigma\sigma^\prime}}{\delta \log {\cal D}} = 2 i \sum_{a^\prime a^{\prime\prime} \sigma^{\prime\prime}} \varepsilon^{a a^\prime a^{\prime\prime}} \, \left( J_{a^\prime} {s}^{a^\prime}_{g \sigma\sigma^{\prime\prime}} \right) \, \corr{\varrho}_{\sigma^{\prime\prime}} \, \left( J_{a^{\prime\prime}} s^{a^{\prime\prime}}_{g\sigma^{\prime\prime} \sigma^\prime} \right) \nonumber
\eeq
The dimensionless Kondo couplings are most naturally chosen as
\bea
j_z & \equiv & J_z \, \frac{\corr{\varrho}_{\upa} + \corr{\varrho}_{\downa}}{2} \nonumber \\
j_\perp & \equiv & J_\perp \, \sqrt{\corr{\varrho}_{\upa} \, \corr{\varrho}_{\downa}}. \nonumber
\eea
With this choice, the couplings follow the usual poor man's scaling equations that also arise in the case of equal density of states~\cite{Anderson1970PoorMansScaling},
\bea
\frac{\delta j_z}{\delta \log {\cal D}} &=& -  \, j_{\perp}^2, \nonumber \\
\frac{\delta j_{\perp}}{\delta \log {\cal D}} &=& -  \,  j_z \, j_{\perp}. 
\label{eq:poor_man}
\eea
Thus, the renormalization group flow of these couplings will be identical to the ones shown in Fig.~\ref{fig:Kondo_phdiag}.

\section{Optical Stark shift} \label{app:opticalStark}

The effective Kondo Hamiltonian only depends on the difference between the Zeeman shifts of the impurity and bath atoms. Therefore, it is sufficient to address the impurities that are in the $\ket{^3 P_0} = \e$ electronic state to realize the required driving. \corr{We refer the reader to  Ref.~\onlinecite{stellmer2011detection} for the details of how to realize the optical Stark effect in alkaline-earth atoms. Here, we only summarize the details specific to our proposal. The optical setup requires circularly polarized light, coupling the $\ket{e}$ state to an excited state such as $\ket{6s5d\, ^3 D_1}$.} Since the external electron shell in this state is not closed, the hyperfine coupling can mix the electronic and nuclear spins. Thus, nuclear spins can be addressed by optically exciting the electronic degrees of freedom. 

\begin{figure}[h]
\includegraphics[width = 8.5cm]{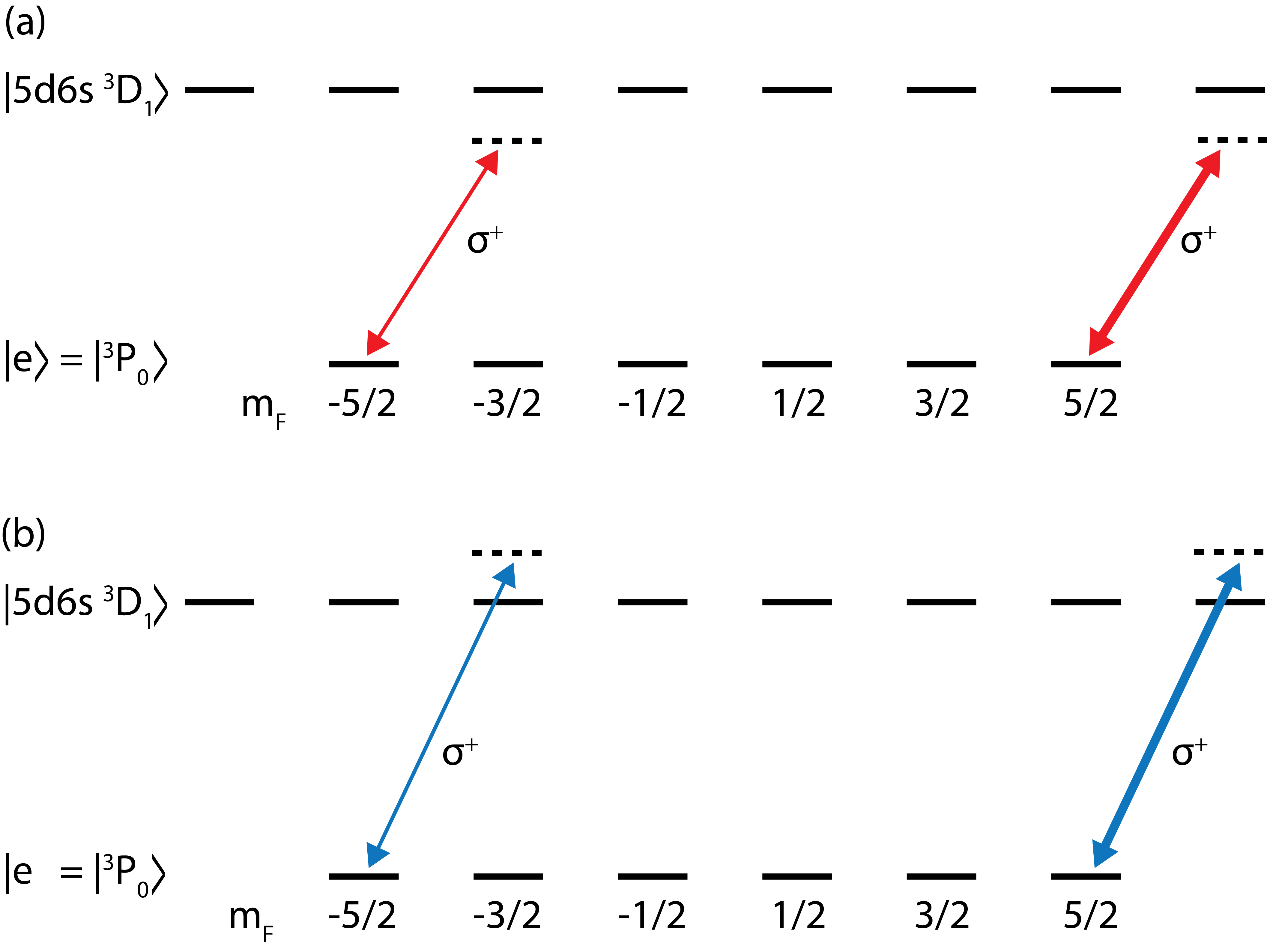}
\caption{ (Color.) \corr{Schematics of the laser configurations realizing effective Zeeman shifts for the $\e$ atoms, using circularly polarized $\sigma^+$ laser fields.}
Only the couplings to the $m_I = \pm 5/2$ fields are shown. The laser frequencies are \corr{detuned relative to the $\e \to \ket{^3 D_1}$-transition with a detuning comparable to the hyperfine splitting. Due to their different matrix elements, the $\sigma^+$ lasers couple differently to the two nuclear spin states.} Thick arrows denote large matrix elements whereas narrow ones correspond to weak couplings. \corr{By choosing appropriate detunings of the lasers, a Zeeman shift (a) $\corrnew{\Delta}_e < 0$ and (b) $\corrnew{\Delta}_e > 0$ can be realized.} Modulated Zeeman fields can be realized by modulating the amplitudes of the laser configurations in (a) and (b).
}
\label{fig:opticalStark}
\end{figure}

In our proposal, we assume that only the smallest and the largest nuclear spin states $m_I = \pm I$ are populated. \corr{Due to its Clebsch-Gordan coefficients, the circularly polarized $\sigma^+$ laser couples stronger to the nuclear spin states with positive $m_I$. As Fig.~\ref{fig:opticalStark} shows, by red (blue) detuning the $\sigma^+$ mode, a negative (positive) Zeeman shift $\corrnew{\Delta}_e$ can be realized.}  A time-dependent effective Zeeman field can thus be created by modulating the intensities of the red- and blue-detuned lasers. The required modulation frequencies are in the $\rm kHz$ regime, which is easily accessible in current experiments.

When we populate all spin states, each pair of states with nuclear spin $\pm m_I$ experience different Zeeman shifts. These two-dimensional subspaces each obey a $U(1)$ spin rotation symmetry. The $SU(N)$ symmetry of the model is thus broken down to $U(1)^{N/2}$. This symmetry breaking could be used in future works to realize anisotropy in Kondo models of higher spin in alkaline-earth atomic systems.

\section{Time-dependent Schrieffer-Wolff transformation} \label{app:SchriefferWolffDynamic}

In this appendix, we derive the Kondo parameters of the periodically modulated model, assuming that the Zeeman fields $\corrnew{\Delta}_e(\tau)$ and $\corrnew{\Delta}_g(\tau)$ have both static as well as an oscillating component
\bea
\corrnew{\Delta}_{e}(\tau) = \corrnew{\Delta}_{e 0} \cos(\omega \tau) + \corrnew{\Delta}_{e 1}, \nonumber \\
\corrnew{\Delta}_{g}(\tau) = \corrnew{\Delta}_{g 0} \cos(\omega \tau) + \corrnew{\Delta}_{g 1}. \nonumber
\eea
The driving frequencies are the same for both states since we assume that the Zeeman fields are generated by the same laser field.
As a first step, we perform a unitary transformation on the Hamiltonian that removes the oscillating part of the Zeeman energy 
\bea
H_{{\rm Z} 0}(\tau) &=& 
-\frac{\corrnew{\Delta}_{e0}}{2} \cos(\omega \tau) \left( \ket{\Upa}\bra{\Upa} - \ket{\Downa}\bra{\Downa} \right) 
\nonumber \\ 
&-& \frac{\corrnew{\Delta}_{g0}}{2} \cos(\omega \tau) (n_{g 0\upa} - n_{g 0\downa}) \nonumber \\
&-& \frac{\corrnew{\Delta}_{g0}}{2} \cos(\omega \tau) \, \sum_{{\bf k} \sigma} \sigma \, g_{{\bf k}\sigma}^\dagger \, g_{{\bf k}\sigma} \nonumber
\eea
by using the unitary transformation
$W(\tau) = \exp\left(- i \int^\tau d\tau^\prime \, H_{\rm Z 0}(\tau^\prime) \right)$ that brings the system into the rotating frame. The Hamiltonian then becomes
\bea
\tilde{H}(\tau) &=& i (\partial_\tau W^\dagger) \, W + W^\dagger H W \nonumber \\
&=& \tilde{H}_{\rm bath} + \tilde{H}_{\rm imp} (\tau) + H_{\rm mix}. \nonumber
\eea
The transformation does not affect the mixing term and the transformed bath Hamiltonian only contains the static part of the Zeeman field $\tilde{H}_{\rm bath} = \sum_{{\bf k}\sigma} (\epsilon_{{\bf k}} -\sigma \, \corrnew{\Delta}_{g1}/2) \, g_{{\bf k}\sigma}^\dagger g_{{\bf k}\sigma}$.
The exchange term in the impurity Hamiltonian, however, depends on the oscillating part of the Zeeman energy,
\bea
\tilde{H}_{\rm imp}(\tau) &=& U \, (n_{g 0\upa} + n_{g 0\downa}) (n_{e 0 \Uparrow} + n_{e 0 \Downarrow}) \nonumber \\ 
&-& \frac{\corrnew{\Delta}_{e1}}{2} \, (n_{g 0\upa} - n_{g 0\downa}) \nonumber \\
&+& U_{\rm ex} \, \sum_{\sigma \sigma^\prime} g_{0\sigma^\prime}^\dagger e_{0\sigma}^\dagger \, e_{0\sigma^\prime} g_{0 \sigma} \, e^{-i (\sigma - \sigma^\prime) \corrnew{\Phi}_0(\tau)/2}.
\nonumber
\eea
The phase factor $\corrnew{\Phi}_0(\tau) = \int^\tau d\tau^\prime \corrnew{\Delta}_0 \cos(\omega \tau^\prime)$ in the last equation is the anti-derivative of the Zeeman splitting $\corrnew{\Delta}_0(\tau) = (\corrnew{\Delta}_{e 0}- \corrnew{\Delta}_{g 0}) \cos(\omega \tau)$, and it arises from the Zeeman energy gains from spin-exchanging collisions with the impurity.

We derive the low energy effective Hamiltonian using a time-dependent Schrieffer-Wolff transformation
\bea
H_{SW}(\tau) &=& \mathbb{P}_0 \left( i(\partial_\tau e^{S(\tau)}) \, e^{-S(\tau)} \right. \label{eq:HeffTimeDependent}  \\
&+& \left. e^{S(\tau)} \, (\tilde{H}_{\rm bath} + \tilde{H}_{\rm imp}(\tau) + H_{\rm mix}) \, e^{-S(\tau)} \right) \mathbb{P}_0. 
\nonumber
\eea
 Similarly to Sec.~\ref{sec:SchriefferWolffStatic}, we choose the transformation $S(\tau)$ such that the first order terms cancel the coupling between the impurity and the bath,
\beq
\mathbb{P}_1 \, \left(i\partial_\tau S(\tau) + H_{\rm mix} \right) \mathbb{P}_0 
= \mathbb{P}_1 \left[ \tilde{H}_{\rm bath} + \tilde{H}_{\rm imp}(\tau), \, S(\tau)\right] \mathbb{P}_0.
\label{eq:TimeDependentSchriefferWolffEq}
\eeq
We use the ansatz Eq.~\eqref{eq:SchriefferWolffAnsatz} to solve the last equation numerically, with time-dependent coefficients. In the $\ket{\Upa\upa}$ and $\ket{\Downa\downa}$ sectors, the time evolution of the coefficients becomes
\beq
\left(-i \partial_\tau + U + U_{\rm ex} - {\epsilon}_{{\bf k}} \right) \Gamma_{\sigma\sigma}^{\sigma\sigma}({\bf k}) = V. \nonumber
\eeq
Therefore, this channel obeys the same solution Eq.~\eqref{eq:Gamma_no_spin_flip} as in the static case. 
In the spin-flip channel, the coefficients obey the equations
\bea
\left[
-i \partial_{\tau} + \begin{pmatrix}
\tilde{\bf H}_{\rm imp}^{\rm ex}(\tau) & 0 \\
0 & \tilde{\bf H}_{\rm bath}^{\rm ex}({\bf k})
\end{pmatrix} 
,
\begin{pmatrix}
0 & {\bf \Gamma}^{\rm ex}({\bf k}) \\ 
-\left( {\bf \Gamma}^{{\rm ex}}({\bf k}) \right)^\dagger & 0
\end{pmatrix} \right] \nonumber \\
=
V,
\nonumber
\eea
with the Hamiltonian matrices
\beq
\tilde{\bf H}_{\rm kin}^{\rm ex}(k) = 
\begin{pmatrix}
\epsilon_{\bf k} - \corrnew{\Delta}_{1}/2 & 0 \\
0 & \epsilon_{\bf k} + \corrnew{\Delta}_{1}/2
\end{pmatrix}, \nonumber
\eeq
and 
\beq
\tilde{\bf H}_{\rm imp}^{\rm ex}(\tau) = 
\begin{pmatrix}
U - \corrnew{\Delta}_1 / 2 & U_{\rm ex} \, e^{i \corrnew{\Phi}_{0}(\tau)} \\
U_{\rm ex} \, e^{-i \corrnew{\Phi}_{0}(\tau)} & U + \corrnew{\Delta}_1 / 2
\end{pmatrix}, \nonumber
\eeq
where we used the notation $\corrnew{\Delta}_1 = \corrnew{\Delta}_{e1} - \corrnew{\Delta}_{g1}$. As the last two equations show, the Schrieffer-Wolff transformation only depends on the difference of the Zeeman energies $\corrnew{\Delta}_0(\tau)$ and $\corrnew{\Delta}_1$ but not on their average value.
In order to ensure the time periodicity of the transformed Hamiltonian, we require the transformation to be periodic $S(\tau + T) = S(\tau)$ as well.

The Schrieffer-Wolff transformation  decouples the low and high energy sectors of the Hamiltonian. The low energy part contains terms of the order of $V^2 / U_{eg}^-$, whereas the high energy sector is of the order of $U_{eg}^\pm$. After the transformation, the effective Hamiltonian reads
\beq
H_{SW}(\tau) = \mathbb{P}_0 \left( \tilde{H}_{\rm bath} + \tilde{H}_{\rm imp}(\tau)+ 
\frac{1}{2} \left[H_{\rm mix}, S(\tau) \right] \right) \mathbb{P}_0 \nonumber
\eeq
up to second order in $S(\tau)$. As the driving frequency $\omega$ is much larger than the energy scale of the Kondo dynamics $V^2 / U_{eg}^-$, we can simply obtain a low energy effective Hamiltonian using the lowest order Floquet term. This is given by the time average of the effective Hamiltonian~\cite{Shirley1965Floquet, Sambe1973Floquet}
\beq
H_{\rm eff} \, = \, \frac{1}{T} \, \int_0^T d\tau \, H_{SW}(\tau).
\label{eq:H_eff_TD}
\eeq
$H_{\rm eff}$ is of the order of $\mathcal{O}(V^2 / U_{eg}^-)$, and the next order correction, of the order $\mathcal{O}\left(\frac{1}{\omega} \, (V^2/U_{eg}^-)^2\right)$, is negligible given that $\omega$ is usually of the order of the on-site energies. 
We obtain the Kondo parameters by comparing $H_{\rm eff}$ in the last equation to Eqs.~(\ref{eq:KondoInteractionHamiltonian} - \ref{eq:KondoZeemanHamiltonian}).

\section{Modulation of the optical lattice} \label{app:on-site_driving}

Varying the optical lattice potential leads to the modulation of both the hopping $t$ and the on-site interactions $U_{eg}^\pm$. The latter depends polynomially on the amplitude of the lattice potential, whereas $t$ is suppressed exponentially~\cite{Bloch2008Review}. The driving therefore leads to the oscillation of the ratios $t/U_{eg}^-$ and $t/U_{eg}^+$. 

In order to discuss how the Kondo couplings get modified by the driving, we first note that the oscillation of an overall energy scale of the Hamiltonian can be removed by a gauge transformation. We can thus choose the transformation such that the tunneling remains constant in the rotating frame, and only the on-site interactions are modulated. As Eq.~\eqref{eq:on-site_driving} shows, the singlet and triplet sectors shall oscillate at the same relative amplitude. Neglecting higher harmonics, we model the driving as
\beq
U^\pm_{eg}(\tau) = U_{eg, 0}^\pm \, \left(1 + \delta u \, \cos(\omega \tau) \right).
\label{eq:on-site_driving}
\eeq

\begin{figure}[h]
\includegraphics[width = 7.5cm]{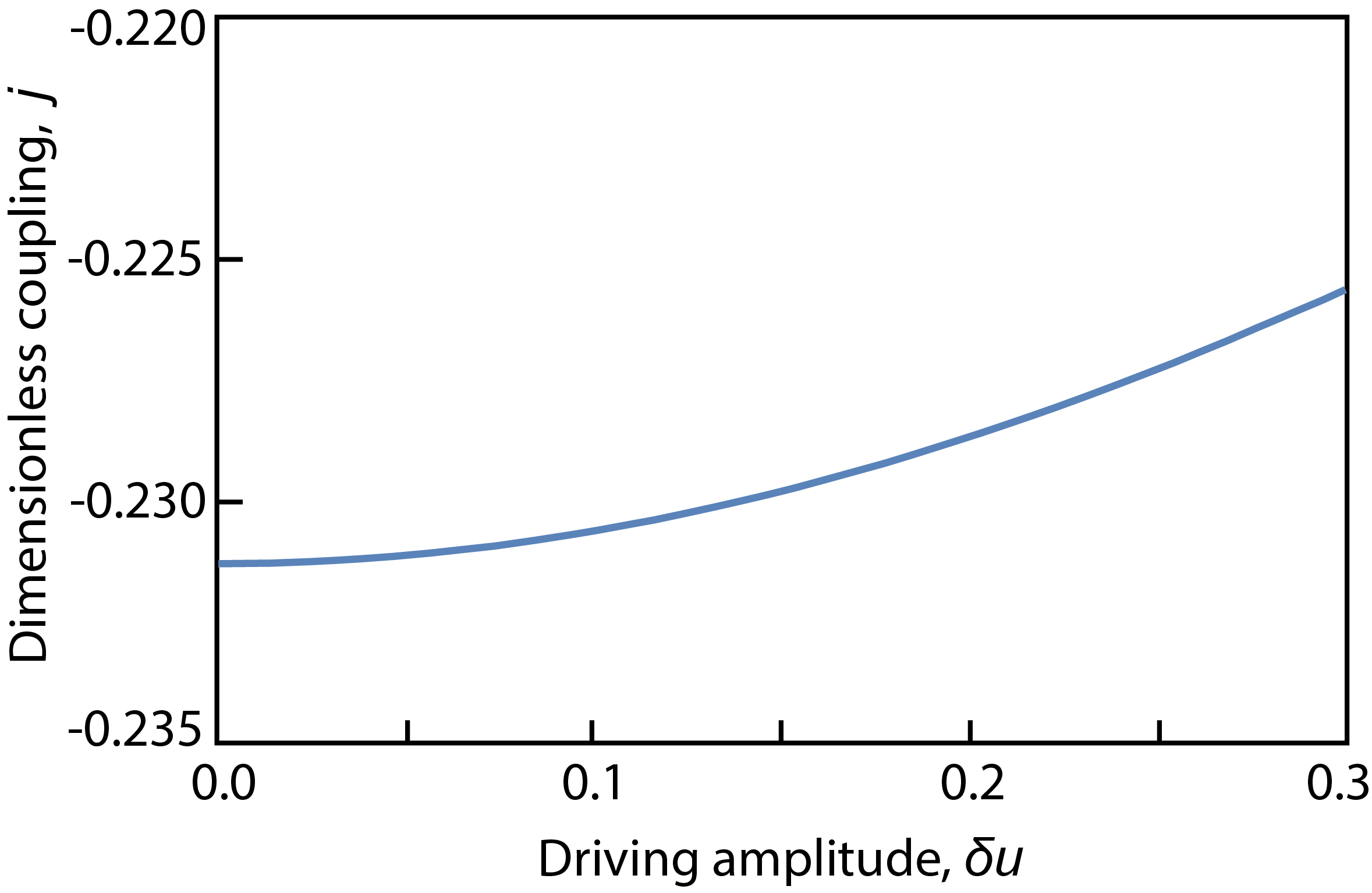}
\caption{ (Color.) Dimensionless isotropic Kondo coupling $j = j_z = j_\perp$ in the presence of periodically modulated optical lattice potentials. The driving frequency $\omega = 14 \, U_{eg, 0}^-$ is red detuned . As the amplitude $\delta u$ in Eq.~\eqref{eq:on-site_driving} grows, the FM coupling $j<0$ initially decreases in amplitude, then it crosses over to the AFM regime $j>0$. [Parameters of the plot: $U_{eg, 0}^+ = 15 \, U_{eg,0 }^-$, $t = 0.35 \, U_{eg}^-$ and $\epsilon_{F} = 0$.]
}
\label{fig:on-site_driving}
\end{figure}

We derive the Kondo parameters using a calculation similar to~\ref{app:SchriefferWolffDynamic}. We use a periodic Schrieffer-Wolff transformation to decouple the low energy sector of the Hubbard-Anderson Hamiltonian. We then determine the couplings by keeping the lowest order Floquet term. Since the driving does not break the $SU(2)$ symmetry of the model, the Schrieffer-Wolff transformation is diagonal in the singlet $(\ket{+})$ and triplet $(\ket{-})$ spin channels
\beq
\left(-i \partial_\tau + U_{eg}^\pm - \epsilon_{\bf k}\right) \Gamma_{\pm}({\bf k}) = V,
\nonumber
\eeq
respectively.
The effective magnetic terms $m_e$ and $m_g$ thus remain zero, whereas the Kondo parameters $j_z = j_\perp = j$ are tunable. Fig.~\ref{fig:on-site_driving} shows how $j$ depends on the driving amplitude $\delta u$ in case of a red-detuned driving $\omega \lesssim U_{eg}^-$. The lowest order Schrieffer-Wolff calculation is most reliable for small driving amplitudes $\delta u \ll 1$, and it will acquire corrections at larger driving amplitudes. In this regime, the isotropic FM coupling $j$ will weaken as $\delta u$ increases.

\section{Initialization of the impurity spin} \label{app:ImpurityInitialization}

In order to study the non-equilibrium dynamics discussed in Sec.~\ref{sec:zero_temperature}, the impurity has to be created instantaneously in the $\ket{\Uparrow}$ state. 
This is possible due to the separation of energy scales in Eq.~\eqref{eq:energy_scales}. The ultranarrow linewidth of the $\g \to \e$ transition (below $1~{\rm Hz}$) allows one to address the spin states of the atoms independently. This can be achieved by introducing an effective Zeeman field optically or using an external magnetic field~\cite{scazza2014observation}. The Zeeman splitting can easily be made larger than the linewidth of the transition. Thus, an appropriately tuned laser pulse can create only $\ket{g\upa} \to \ket{e\Upa}$ transitions, while leaving the $\ket{g\downa}$ atoms unaltered. 

It is also important to ensure that the impurity sites are singly occupied at the beginning of the dynamics. This can be achieved if the laser pulse excites only  $\g$ atoms only on singly occupied sites. Since the on-site repulsions are of the order of $U_{eg}^\pm \sim 1-10 \, {\rm kHz}$, the laser can be tuned such that the doubly occupied sites become off-resonant.
The time $\tau_{\rm ex}$ to create the impurity needs to be instantaneous on the timescales of the Kondo dynamics. The finite time duration of the pulse leads to a frequency broadening of the order of $\gamma \sim 1/\tau_{\rm ex}$. $\tau_{\rm ex}$ can be chosen such that  $\gamma \ll U_{eg}^\pm$ so that the singly occupied sites can be selectively addressed. \corr{By choosing $J \ll \gamma \ll U_{eg}^-$}, the pulse can still be made instantaneous on the timescales of the Kondo dynamics.


\begin{thebibliography}{168}
\expandafter\ifx\csname natexlab\endcsname\relax\def\natexlab#1{#1}\fi
\expandafter\ifx\csname bibnamefont\endcsname\relax
  \def\bibnamefont#1{#1}\fi
\expandafter\ifx\csname bibfnamefont\endcsname\relax
  \def\bibfnamefont#1{#1}\fi
\expandafter\ifx\csname citenamefont\endcsname\relax
  \def\citenamefont#1{#1}\fi
\expandafter\ifx\csname url\endcsname\relax
  \def\url#1{\texttt{#1}}\fi
\expandafter\ifx\csname urlprefix\endcsname\relax\def\urlprefix{URL }\fi
\providecommand{\bibinfo}[2]{#2}
\providecommand{\eprint}[2][]{\url{#2}}

\bibitem[{\citenamefont{Anderson}(1961)}]{anderson1961localized}
\bibinfo{author}{\bibfnamefont{P.~W.} \bibnamefont{Anderson}},
  \bibinfo{journal}{Physical Review} \textbf{\bibinfo{volume}{124}},
  \bibinfo{pages}{41} (\bibinfo{year}{1961}).

\bibitem[{\citenamefont{Kondo}(1964)}]{Kondo1964}
\bibinfo{author}{\bibfnamefont{J.}~\bibnamefont{Kondo}},
  \bibinfo{journal}{Progress of Theoretical Physics}
  \textbf{\bibinfo{volume}{32}}, \bibinfo{pages}{37} (\bibinfo{year}{1964}),
  ISSN \bibinfo{issn}{0033-068X},
  \urlprefix\url{http://ptp.oxfordjournals.org/content/32/1/37.short}.

\bibitem[{\citenamefont{Abrikosov}(1965)}]{Abrikosov1965PerturbativeKondo}
\bibinfo{author}{\bibfnamefont{A.}~\bibnamefont{Abrikosov}},
  \bibinfo{journal}{Phys. Rev} \textbf{\bibinfo{volume}{138}},
  \bibinfo{pages}{A515} (\bibinfo{year}{1965}).

\bibitem[{\citenamefont{Andres et~al.}(1975)\citenamefont{Andres, Graebner, and
  Ott}}]{Graebner1975OriginalHeavyFermionPaper}
\bibinfo{author}{\bibfnamefont{K.}~\bibnamefont{Andres}},
  \bibinfo{author}{\bibfnamefont{J.~E.} \bibnamefont{Graebner}},
  \bibnamefont{and} \bibinfo{author}{\bibfnamefont{H.~R.} \bibnamefont{Ott}},
  \bibinfo{journal}{Phys. Rev. Lett.} \textbf{\bibinfo{volume}{35}},
  \bibinfo{pages}{1779} (\bibinfo{year}{1975}),
  \urlprefix\url{https://link.aps.org/doi/10.1103/PhysRevLett.35.1779}.

\bibitem[{\citenamefont{Doniach}(1977)}]{doniach1977kondo}
\bibinfo{author}{\bibfnamefont{S.}~\bibnamefont{Doniach}},
  \bibinfo{journal}{Physica B+ C} \textbf{\bibinfo{volume}{91}},
  \bibinfo{pages}{231} (\bibinfo{year}{1977}).

\bibitem[{\citenamefont{Si et~al.}(1999)\citenamefont{Si, Smith, and
  Ingersent}}]{si1999quantum}
\bibinfo{author}{\bibfnamefont{Q.}~\bibnamefont{Si}},
  \bibinfo{author}{\bibfnamefont{J.~L.} \bibnamefont{Smith}}, \bibnamefont{and}
  \bibinfo{author}{\bibfnamefont{K.}~\bibnamefont{Ingersent}},
  \bibinfo{journal}{International Journal of Modern Physics B}
  \textbf{\bibinfo{volume}{13}}, \bibinfo{pages}{2331} (\bibinfo{year}{1999}).

\bibitem[{\citenamefont{Coleman
  et~al.}(2001{\natexlab{a}})\citenamefont{Coleman, P{\'e}pin, Si, and
  Ramazashvili}}]{coleman2001fermi}
\bibinfo{author}{\bibfnamefont{P.}~\bibnamefont{Coleman}},
  \bibinfo{author}{\bibfnamefont{C.}~\bibnamefont{P{\'e}pin}},
  \bibinfo{author}{\bibfnamefont{Q.}~\bibnamefont{Si}}, \bibnamefont{and}
  \bibinfo{author}{\bibfnamefont{R.}~\bibnamefont{Ramazashvili}},
  \bibinfo{journal}{Journal of Physics: Condensed Matter}
  \textbf{\bibinfo{volume}{13}}, \bibinfo{pages}{R723}
  (\bibinfo{year}{2001}{\natexlab{a}}).

\bibitem[{\citenamefont{Si et~al.}(2001)\citenamefont{Si, Rabello, Ingersent,
  and Smith}}]{si2001locally}
\bibinfo{author}{\bibfnamefont{Q.}~\bibnamefont{Si}},
  \bibinfo{author}{\bibfnamefont{S.}~\bibnamefont{Rabello}},
  \bibinfo{author}{\bibfnamefont{K.}~\bibnamefont{Ingersent}},
  \bibnamefont{and} \bibinfo{author}{\bibfnamefont{J.~L.} \bibnamefont{Smith}},
  \bibinfo{journal}{Nature} \textbf{\bibinfo{volume}{413}},
  \bibinfo{pages}{804} (\bibinfo{year}{2001}).

\bibitem[{\citenamefont{Nozieres and
  Blandin}(1980)}]{Nozieres1980KondoEffectInRealMetals}
\bibinfo{author}{\bibfnamefont{P.}~\bibnamefont{Nozieres}} \bibnamefont{and}
  \bibinfo{author}{\bibfnamefont{A.}~\bibnamefont{Blandin}},
  \bibinfo{journal}{Journal de Physique} \textbf{\bibinfo{volume}{41}},
  \bibinfo{pages}{193} (\bibinfo{year}{1980}).

\bibitem[{\citenamefont{Tsvelik and
  Reizer}(1993)}]{tsvelik1993HeavyFermionKondo}
\bibinfo{author}{\bibfnamefont{A.~M.} \bibnamefont{Tsvelik}} \bibnamefont{and}
  \bibinfo{author}{\bibfnamefont{M.}~\bibnamefont{Reizer}},
  \bibinfo{journal}{Phys. Rev. B} \textbf{\bibinfo{volume}{48}},
  \bibinfo{pages}{9887} (\bibinfo{year}{1993}),
  \urlprefix\url{https://link.aps.org/doi/10.1103/PhysRevB.48.9887}.

\bibitem[{\citenamefont{Coleman
  et~al.}(2001{\natexlab{b}})\citenamefont{Coleman, P{\'e}pin, Si, and
  Ramazashvili}}]{Coleman2001HeavyFermions}
\bibinfo{author}{\bibfnamefont{P.}~\bibnamefont{Coleman}},
  \bibinfo{author}{\bibfnamefont{C.}~\bibnamefont{P{\'e}pin}},
  \bibinfo{author}{\bibfnamefont{Q.}~\bibnamefont{Si}}, \bibnamefont{and}
  \bibinfo{author}{\bibfnamefont{R.}~\bibnamefont{Ramazashvili}},
  \bibinfo{journal}{Journal of Physics: Condensed Matter}
  \textbf{\bibinfo{volume}{13}}, \bibinfo{pages}{R723}
  (\bibinfo{year}{2001}{\natexlab{b}}).

\bibitem[{\citenamefont{Hewson}(1997)}]{HewsonBook}
\bibinfo{author}{\bibfnamefont{A.~C.} \bibnamefont{Hewson}},
  \emph{\bibinfo{title}{The Kondo problem to heavy fermions}}
  (\bibinfo{publisher}{Cambridge University Press}, \bibinfo{address}{Cambridge
  New York}, \bibinfo{year}{1997}), ISBN \bibinfo{isbn}{0521599474}.

\bibitem[{\citenamefont{L{\"o}hneysen et~al.}(2007)\citenamefont{L{\"o}hneysen,
  Rosch, Vojta, and W{\"o}lfle}}]{Lohneysen2007FermiLiquidInstabilities}
\bibinfo{author}{\bibfnamefont{H.~v.} \bibnamefont{L{\"o}hneysen}},
  \bibinfo{author}{\bibfnamefont{A.}~\bibnamefont{Rosch}},
  \bibinfo{author}{\bibfnamefont{M.}~\bibnamefont{Vojta}}, \bibnamefont{and}
  \bibinfo{author}{\bibfnamefont{P.}~\bibnamefont{W{\"o}lfle}},
  \bibinfo{journal}{Reviews of Modern Physics} \textbf{\bibinfo{volume}{79}},
  \bibinfo{pages}{1015} (\bibinfo{year}{2007}).

\bibitem[{\citenamefont{Glazman and Raikh}(1988)}]{glazman1988resonant}
\bibinfo{author}{\bibfnamefont{L.}~\bibnamefont{Glazman}} \bibnamefont{and}
  \bibinfo{author}{\bibfnamefont{M.}~\bibnamefont{Raikh}},
  \bibinfo{journal}{JETP lett} \textbf{\bibinfo{volume}{47}},
  \bibinfo{pages}{452} (\bibinfo{year}{1988}).

\bibitem[{\citenamefont{Ng and Lee}(1988)}]{ng1988site}
\bibinfo{author}{\bibfnamefont{T.~K.} \bibnamefont{Ng}} \bibnamefont{and}
  \bibinfo{author}{\bibfnamefont{P.~A.} \bibnamefont{Lee}},
  \bibinfo{journal}{Physical review letters} \textbf{\bibinfo{volume}{61}},
  \bibinfo{pages}{1768} (\bibinfo{year}{1988}).

\bibitem[{\citenamefont{Goldhaber-Gordon
  et~al.}(1998)\citenamefont{Goldhaber-Gordon, Shtrikman, Mahalu,
  Abusch-Magder, Meirav, and Kastner}}]{GoldhaberGordon1998KondoExperiment}
\bibinfo{author}{\bibfnamefont{D.}~\bibnamefont{Goldhaber-Gordon}},
  \bibinfo{author}{\bibfnamefont{H.}~\bibnamefont{Shtrikman}},
  \bibinfo{author}{\bibfnamefont{D.}~\bibnamefont{Mahalu}},
  \bibinfo{author}{\bibfnamefont{D.}~\bibnamefont{Abusch-Magder}},
  \bibinfo{author}{\bibfnamefont{U.}~\bibnamefont{Meirav}}, \bibnamefont{and}
  \bibinfo{author}{\bibfnamefont{M.}~\bibnamefont{Kastner}},
  \bibinfo{journal}{Nature} \textbf{\bibinfo{volume}{391}},
  \bibinfo{pages}{156} (\bibinfo{year}{1998}).

\bibitem[{\citenamefont{Cronenwett et~al.}(1998)\citenamefont{Cronenwett,
  Oosterkamp, and Kouwenhoven}}]{Cronenwett1998TunableKondoEffectInQuantumDots}
\bibinfo{author}{\bibfnamefont{S.~M.} \bibnamefont{Cronenwett}},
  \bibinfo{author}{\bibfnamefont{T.~H.} \bibnamefont{Oosterkamp}},
  \bibnamefont{and} \bibinfo{author}{\bibfnamefont{L.~P.}
  \bibnamefont{Kouwenhoven}}, \bibinfo{journal}{Science}
  \textbf{\bibinfo{volume}{281}}, \bibinfo{pages}{540} (\bibinfo{year}{1998}).

\bibitem[{\citenamefont{Kouwenhoven and
  Glazman}(2001)}]{kouwenhoven2001revival}
\bibinfo{author}{\bibfnamefont{L.}~\bibnamefont{Kouwenhoven}} \bibnamefont{and}
  \bibinfo{author}{\bibfnamefont{L.}~\bibnamefont{Glazman}},
  \bibinfo{journal}{Physics world} \textbf{\bibinfo{volume}{14}},
  \bibinfo{pages}{33} (\bibinfo{year}{2001}).

\bibitem[{\citenamefont{Jeong et~al.}(2001)\citenamefont{Jeong, Chang, and
  Melloch}}]{jeong2001kondo}
\bibinfo{author}{\bibfnamefont{H.}~\bibnamefont{Jeong}},
  \bibinfo{author}{\bibfnamefont{A.~M.} \bibnamefont{Chang}}, \bibnamefont{and}
  \bibinfo{author}{\bibfnamefont{M.~R.} \bibnamefont{Melloch}},
  \bibinfo{journal}{Science} \textbf{\bibinfo{volume}{293}},
  \bibinfo{pages}{2221} (\bibinfo{year}{2001}).

\bibitem[{\citenamefont{Mebrahtu et~al.}(2012)\citenamefont{Mebrahtu,
  Borzenets, Liu, Zheng, Bomze, Smirnov, Baranger, and
  Finkelstein}}]{mebrahtu2012quantum}
\bibinfo{author}{\bibfnamefont{H.~T.} \bibnamefont{Mebrahtu}},
  \bibinfo{author}{\bibfnamefont{I.~V.} \bibnamefont{Borzenets}},
  \bibinfo{author}{\bibfnamefont{D.~E.} \bibnamefont{Liu}},
  \bibinfo{author}{\bibfnamefont{H.}~\bibnamefont{Zheng}},
  \bibinfo{author}{\bibfnamefont{Y.~V.} \bibnamefont{Bomze}},
  \bibinfo{author}{\bibfnamefont{A.~I.} \bibnamefont{Smirnov}},
  \bibinfo{author}{\bibfnamefont{H.~U.} \bibnamefont{Baranger}},
  \bibnamefont{and}
  \bibinfo{author}{\bibfnamefont{G.}~\bibnamefont{Finkelstein}},
  \bibinfo{journal}{Nature} \textbf{\bibinfo{volume}{488}}, \bibinfo{pages}{61}
  (\bibinfo{year}{2012}).

\bibitem[{\citenamefont{de~Bruyn~Ouboter and Bol}(1982)}]{deBruyn1982influence}
\bibinfo{author}{\bibfnamefont{R.}~\bibnamefont{de~Bruyn~Ouboter}}
  \bibnamefont{and} \bibinfo{author}{\bibfnamefont{D.}~\bibnamefont{Bol}},
  \bibinfo{journal}{Physica B+ C} \textbf{\bibinfo{volume}{112}},
  \bibinfo{pages}{15} (\bibinfo{year}{1982}).

\bibitem[{\citenamefont{Zwerger}(1983{\natexlab{a}})}]{zwerger1983dynamics}
\bibinfo{author}{\bibfnamefont{W.}~\bibnamefont{Zwerger}},
  \bibinfo{journal}{Zeitschrift f{\"u}r Physik B Condensed Matter}
  \textbf{\bibinfo{volume}{53}}, \bibinfo{pages}{53}
  (\bibinfo{year}{1983}{\natexlab{a}}).

\bibitem[{\citenamefont{Leggett et~al.}(1987)\citenamefont{Leggett,
  Chakravarty, Dorsey, Fisher, Garg, and
  Zwerger}}]{Leggett1987DissipativeTwoStateSystem}
\bibinfo{author}{\bibfnamefont{A.~J.} \bibnamefont{Leggett}},
  \bibinfo{author}{\bibfnamefont{S.}~\bibnamefont{Chakravarty}},
  \bibinfo{author}{\bibfnamefont{A.~T.} \bibnamefont{Dorsey}},
  \bibinfo{author}{\bibfnamefont{M.~P.~A.} \bibnamefont{Fisher}},
  \bibinfo{author}{\bibfnamefont{A.}~\bibnamefont{Garg}}, \bibnamefont{and}
  \bibinfo{author}{\bibfnamefont{W.}~\bibnamefont{Zwerger}},
  \bibinfo{journal}{Reviews of Modern Physics} \textbf{\bibinfo{volume}{59}},
  \bibinfo{pages}{1} (\bibinfo{year}{1987}), ISSN \bibinfo{issn}{0034-6861},
  \urlprefix\url{http://link.aps.org/doi/10.1103/RevModPhys.59.1}.

\bibitem[{\citenamefont{Chakravarty and
  Rudnick}(1995)}]{Chakravarty1995DissipativeDynamics}
\bibinfo{author}{\bibfnamefont{S.}~\bibnamefont{Chakravarty}} \bibnamefont{and}
  \bibinfo{author}{\bibfnamefont{J.}~\bibnamefont{Rudnick}},
  \bibinfo{journal}{Phys. Rev. Lett.} \textbf{\bibinfo{volume}{75}},
  \bibinfo{pages}{501} (\bibinfo{year}{1995}),
  \urlprefix\url{https://link.aps.org/doi/10.1103/PhysRevLett.75.501}.

\bibitem[{\citenamefont{Strong}(1997)}]{Strong1997TransitionBetweenCoherenceAndIncoherence}
\bibinfo{author}{\bibfnamefont{S.~P.} \bibnamefont{Strong}},
  \bibinfo{journal}{Phys. Rev. E} \textbf{\bibinfo{volume}{55}},
  \bibinfo{pages}{6636} (\bibinfo{year}{1997}),
  \urlprefix\url{https://link.aps.org/doi/10.1103/PhysRevE.55.6636}.

\bibitem[{\citenamefont{Costi}(1998)}]{costi1998scaling}
\bibinfo{author}{\bibfnamefont{T.}~\bibnamefont{Costi}},
  \bibinfo{journal}{Physical review letters} \textbf{\bibinfo{volume}{80}},
  \bibinfo{pages}{1038} (\bibinfo{year}{1998}).

\bibitem[{\citenamefont{Voss and
  Webb}(1981)}]{Voss1981MacrosopicQuantumTunneling}
\bibinfo{author}{\bibfnamefont{R.~F.} \bibnamefont{Voss}} \bibnamefont{and}
  \bibinfo{author}{\bibfnamefont{R.~A.} \bibnamefont{Webb}},
  \bibinfo{journal}{Phys. Rev. Lett.} \textbf{\bibinfo{volume}{47}},
  \bibinfo{pages}{265} (\bibinfo{year}{1981}),
  \urlprefix\url{https://link.aps.org/doi/10.1103/PhysRevLett.47.265}.

\bibitem[{\citenamefont{Chakravarty}(1982)}]{Chakravarty1982QuantumTunnelingInSuperconductors}
\bibinfo{author}{\bibfnamefont{S.}~\bibnamefont{Chakravarty}},
  \bibinfo{journal}{Physical Review Letters} \textbf{\bibinfo{volume}{49}},
  \bibinfo{pages}{681} (\bibinfo{year}{1982}), ISSN \bibinfo{issn}{0031-9007},
  \urlprefix\url{http://link.aps.org/doi/10.1103/PhysRevLett.49.681}.

\bibitem[{\citenamefont{Han et~al.}(1991)\citenamefont{Han, Lapointe, and
  Lukens}}]{han1991observation}
\bibinfo{author}{\bibfnamefont{S.}~\bibnamefont{Han}},
  \bibinfo{author}{\bibfnamefont{J.}~\bibnamefont{Lapointe}}, \bibnamefont{and}
  \bibinfo{author}{\bibfnamefont{J.}~\bibnamefont{Lukens}},
  \bibinfo{journal}{Physical review letters} \textbf{\bibinfo{volume}{66}},
  \bibinfo{pages}{810} (\bibinfo{year}{1991}).

\bibitem[{\citenamefont{Friedman et~al.}(2000)\citenamefont{Friedman, Patel,
  Chen, Tolpygo, and Lukens}}]{Friedman2000MacroscopicQuantumSuperposition}
\bibinfo{author}{\bibfnamefont{J.~R.} \bibnamefont{Friedman}},
  \bibinfo{author}{\bibfnamefont{V.}~\bibnamefont{Patel}},
  \bibinfo{author}{\bibfnamefont{W.}~\bibnamefont{Chen}},
  \bibinfo{author}{\bibfnamefont{S.}~\bibnamefont{Tolpygo}}, \bibnamefont{and}
  \bibinfo{author}{\bibfnamefont{J.~E.} \bibnamefont{Lukens}},
  \bibinfo{journal}{nature} \textbf{\bibinfo{volume}{406}}, \bibinfo{pages}{43}
  (\bibinfo{year}{2000}).

\bibitem[{\citenamefont{Van Der~Wal et~al.}(2000)\citenamefont{Van Der~Wal,
  Ter~Haar, Wilhelm, Schouten, Harmans, Orlando, Lloyd, and
  Mooij}}]{VanDerWal2000MacroscopicQuantumSuperposition}
\bibinfo{author}{\bibfnamefont{C.~H.} \bibnamefont{Van Der~Wal}},
  \bibinfo{author}{\bibfnamefont{A.}~\bibnamefont{Ter~Haar}},
  \bibinfo{author}{\bibfnamefont{F.}~\bibnamefont{Wilhelm}},
  \bibinfo{author}{\bibfnamefont{R.}~\bibnamefont{Schouten}},
  \bibinfo{author}{\bibfnamefont{C.}~\bibnamefont{Harmans}},
  \bibinfo{author}{\bibfnamefont{T.}~\bibnamefont{Orlando}},
  \bibinfo{author}{\bibfnamefont{S.}~\bibnamefont{Lloyd}}, \bibnamefont{and}
  \bibinfo{author}{\bibfnamefont{J.}~\bibnamefont{Mooij}},
  \bibinfo{journal}{Science} \textbf{\bibinfo{volume}{290}},
  \bibinfo{pages}{773} (\bibinfo{year}{2000}).

\bibitem[{\citenamefont{Anderson}(1970)}]{Anderson1970PoorMansScaling}
\bibinfo{author}{\bibfnamefont{P.}~\bibnamefont{Anderson}},
  \bibinfo{journal}{Journal of Physics C: Solid State Physics}
  \textbf{\bibinfo{volume}{3}}, \bibinfo{pages}{2436} (\bibinfo{year}{1970}).

\bibitem[{\citenamefont{Nozieres}(1974)}]{nozieres1974fermi}
\bibinfo{author}{\bibfnamefont{P.}~\bibnamefont{Nozieres}},
  \bibinfo{journal}{Journal of low temp{\'e}rature physics}
  \textbf{\bibinfo{volume}{17}}, \bibinfo{pages}{31} (\bibinfo{year}{1974}).

\bibitem[{\citenamefont{Andrei}(1980{\natexlab{a}})}]{Andrei1980BetheAnsatzKondo}
\bibinfo{author}{\bibfnamefont{N.}~\bibnamefont{Andrei}},
  \bibinfo{journal}{Phys. Rev. Lett.} \textbf{\bibinfo{volume}{45}},
  \bibinfo{pages}{379} (\bibinfo{year}{1980}{\natexlab{a}}),
  \urlprefix\url{https://link.aps.org/doi/10.1103/PhysRevLett.45.379}.

\bibitem[{\citenamefont{Vigman}(1980)}]{Vigman1980ExactKondo}
\bibinfo{author}{\bibfnamefont{P.}~\bibnamefont{Vigman}},
  \bibinfo{journal}{JETP Lett} \textbf{\bibinfo{volume}{31}}
  (\bibinfo{year}{1980}).

\bibitem[{\citenamefont{Andrei et~al.}(1983)\citenamefont{Andrei, Furuya, and
  Lowenstein}}]{Andrei.1983}
\bibinfo{author}{\bibfnamefont{N.}~\bibnamefont{Andrei}},
  \bibinfo{author}{\bibfnamefont{K.}~\bibnamefont{Furuya}}, \bibnamefont{and}
  \bibinfo{author}{\bibfnamefont{J.~H.} \bibnamefont{Lowenstein}},
  \bibinfo{journal}{Rev. Mod. Phys.} \textbf{\bibinfo{volume}{55}},
  \bibinfo{pages}{331} (\bibinfo{year}{1983}),
  \urlprefix\url{https://link.aps.org/doi/10.1103/RevModPhys.55.331}.

\bibitem[{\citenamefont{Zhou and Gould}(1999)}]{zhou1999algebraic}
\bibinfo{author}{\bibfnamefont{H.-Q.} \bibnamefont{Zhou}} \bibnamefont{and}
  \bibinfo{author}{\bibfnamefont{M.~D.} \bibnamefont{Gould}},
  \bibinfo{journal}{Physics Letters A} \textbf{\bibinfo{volume}{251}},
  \bibinfo{pages}{279} (\bibinfo{year}{1999}).

\bibitem[{\citenamefont{Coleman}(1983)}]{coleman1983KondoLattice}
\bibinfo{author}{\bibfnamefont{P.}~\bibnamefont{Coleman}},
  \bibinfo{journal}{Physical Review B} \textbf{\bibinfo{volume}{28}},
  \bibinfo{pages}{5255} (\bibinfo{year}{1983}).

\bibitem[{\citenamefont{Bickers}(1987)}]{bickers1987review}
\bibinfo{author}{\bibfnamefont{N.}~\bibnamefont{Bickers}},
  \bibinfo{journal}{Reviews of modern physics} \textbf{\bibinfo{volume}{59}},
  \bibinfo{pages}{845} (\bibinfo{year}{1987}).

\bibitem[{\citenamefont{Keiter and Kimball}(1971)}]{keiter1971non-crossing}
\bibinfo{author}{\bibfnamefont{H.}~\bibnamefont{Keiter}} \bibnamefont{and}
  \bibinfo{author}{\bibfnamefont{J.}~\bibnamefont{Kimball}},
  \bibinfo{journal}{Journal of Applied Physics} \textbf{\bibinfo{volume}{42}},
  \bibinfo{pages}{1460} (\bibinfo{year}{1971}).

\bibitem[{\citenamefont{Grewe and Keiter}(1981)}]{grewe1981non-crossing}
\bibinfo{author}{\bibfnamefont{N.}~\bibnamefont{Grewe}} \bibnamefont{and}
  \bibinfo{author}{\bibfnamefont{H.}~\bibnamefont{Keiter}},
  \bibinfo{journal}{Physical Review B} \textbf{\bibinfo{volume}{24}},
  \bibinfo{pages}{4420} (\bibinfo{year}{1981}).

\bibitem[{\citenamefont{Kuramoto}(1983)}]{kuramoto1983non-crossing}
\bibinfo{author}{\bibfnamefont{Y.}~\bibnamefont{Kuramoto}},
  \bibinfo{journal}{Zeitschrift f{\"u}r Physik B Condensed Matter}
  \textbf{\bibinfo{volume}{53}}, \bibinfo{pages}{37} (\bibinfo{year}{1983}).

\bibitem[{\citenamefont{M{\"u}ller-Hartmann}(1984)}]{muller1984non-crossing}
\bibinfo{author}{\bibfnamefont{E.}~\bibnamefont{M{\"u}ller-Hartmann}},
  \bibinfo{journal}{Zeitschrift f{\"u}r Physik B Condensed Matter}
  \textbf{\bibinfo{volume}{57}}, \bibinfo{pages}{281} (\bibinfo{year}{1984}).

\bibitem[{\citenamefont{Wilson}(1975)}]{wilson1975renormalization}
\bibinfo{author}{\bibfnamefont{K.~G.} \bibnamefont{Wilson}},
  \bibinfo{journal}{Reviews of Modern Physics} \textbf{\bibinfo{volume}{47}},
  \bibinfo{pages}{773} (\bibinfo{year}{1975}).

\bibitem[{\citenamefont{White}(1992)}]{White1992DMRG}
\bibinfo{author}{\bibfnamefont{S.~R.} \bibnamefont{White}},
  \bibinfo{journal}{Phys. Rev. Lett.} \textbf{\bibinfo{volume}{69}},
  \bibinfo{pages}{2863} (\bibinfo{year}{1992}),
  \urlprefix\url{https://link.aps.org/doi/10.1103/PhysRevLett.69.2863}.

\bibitem[{\citenamefont{Hofstetter}(2000)}]{hofstetter2000generalized}
\bibinfo{author}{\bibfnamefont{W.}~\bibnamefont{Hofstetter}},
  \bibinfo{journal}{Physical review letters} \textbf{\bibinfo{volume}{85}},
  \bibinfo{pages}{1508} (\bibinfo{year}{2000}).

\bibitem[{\citenamefont{Wegner}(1994)}]{Wegner1994FlowEquationMethod}
\bibinfo{author}{\bibfnamefont{F.}~\bibnamefont{Wegner}},
  \bibinfo{journal}{Annalen der physik} \textbf{\bibinfo{volume}{506}},
  \bibinfo{pages}{77} (\bibinfo{year}{1994}).

\bibitem[{\citenamefont{Yacoby et~al.}(1995)\citenamefont{Yacoby, Heiblum,
  Mahalu, and Shtrikman}}]{yacoby1995coherence}
\bibinfo{author}{\bibfnamefont{A.}~\bibnamefont{Yacoby}},
  \bibinfo{author}{\bibfnamefont{M.}~\bibnamefont{Heiblum}},
  \bibinfo{author}{\bibfnamefont{D.}~\bibnamefont{Mahalu}}, \bibnamefont{and}
  \bibinfo{author}{\bibfnamefont{H.}~\bibnamefont{Shtrikman}},
  \bibinfo{journal}{Physical review letters} \textbf{\bibinfo{volume}{74}},
  \bibinfo{pages}{4047} (\bibinfo{year}{1995}).

\bibitem[{\citenamefont{Roch et~al.}(2008)\citenamefont{Roch, Florens,
  Bouchiat, Wernsdorfer, and Balestro}}]{roch2008quantum}
\bibinfo{author}{\bibfnamefont{N.}~\bibnamefont{Roch}},
  \bibinfo{author}{\bibfnamefont{S.}~\bibnamefont{Florens}},
  \bibinfo{author}{\bibfnamefont{V.}~\bibnamefont{Bouchiat}},
  \bibinfo{author}{\bibfnamefont{W.}~\bibnamefont{Wernsdorfer}},
  \bibnamefont{and} \bibinfo{author}{\bibfnamefont{F.}~\bibnamefont{Balestro}},
  \bibinfo{journal}{Nature} \textbf{\bibinfo{volume}{453}},
  \bibinfo{pages}{633} (\bibinfo{year}{2008}).

\bibitem[{\citenamefont{Roch et~al.}(2009)\citenamefont{Roch, Florens, Costi,
  Wernsdorfer, and Balestro}}]{roch2009observation}
\bibinfo{author}{\bibfnamefont{N.}~\bibnamefont{Roch}},
  \bibinfo{author}{\bibfnamefont{S.}~\bibnamefont{Florens}},
  \bibinfo{author}{\bibfnamefont{T.~A.} \bibnamefont{Costi}},
  \bibinfo{author}{\bibfnamefont{W.}~\bibnamefont{Wernsdorfer}},
  \bibnamefont{and} \bibinfo{author}{\bibfnamefont{F.}~\bibnamefont{Balestro}},
  \bibinfo{journal}{Physical review letters} \textbf{\bibinfo{volume}{103}},
  \bibinfo{pages}{197202} (\bibinfo{year}{2009}).

\bibitem[{\citenamefont{Latta et~al.}(2011)\citenamefont{Latta, Haupt, Hanl,
  Weichselbaum, Claassen, Wuester, Fallahi, Faelt, Glazman, von Delft
  et~al.}}]{Latta2011NonEqKondoOpticallyExcited}
\bibinfo{author}{\bibfnamefont{C.}~\bibnamefont{Latta}},
  \bibinfo{author}{\bibfnamefont{F.}~\bibnamefont{Haupt}},
  \bibinfo{author}{\bibfnamefont{M.}~\bibnamefont{Hanl}},
  \bibinfo{author}{\bibfnamefont{A.}~\bibnamefont{Weichselbaum}},
  \bibinfo{author}{\bibfnamefont{M.}~\bibnamefont{Claassen}},
  \bibinfo{author}{\bibfnamefont{W.}~\bibnamefont{Wuester}},
  \bibinfo{author}{\bibfnamefont{P.}~\bibnamefont{Fallahi}},
  \bibinfo{author}{\bibfnamefont{S.}~\bibnamefont{Faelt}},
  \bibinfo{author}{\bibfnamefont{L.}~\bibnamefont{Glazman}},
  \bibinfo{author}{\bibfnamefont{J.}~\bibnamefont{von Delft}},
  \bibnamefont{et~al.}, \bibinfo{journal}{Nature}
  \textbf{\bibinfo{volume}{474}}, \bibinfo{pages}{627} (\bibinfo{year}{2011}).

\bibitem[{\citenamefont{T{\"u}reci et~al.}(2011)\citenamefont{T{\"u}reci, Hanl,
  Claassen, Weichselbaum, Hecht, Braunecker, Govorov, Glazman, Imamoglu, and
  von Delft}}]{tureci2011many}
\bibinfo{author}{\bibfnamefont{H.~E.} \bibnamefont{T{\"u}reci}},
  \bibinfo{author}{\bibfnamefont{M.}~\bibnamefont{Hanl}},
  \bibinfo{author}{\bibfnamefont{M.}~\bibnamefont{Claassen}},
  \bibinfo{author}{\bibfnamefont{A.}~\bibnamefont{Weichselbaum}},
  \bibinfo{author}{\bibfnamefont{T.}~\bibnamefont{Hecht}},
  \bibinfo{author}{\bibfnamefont{B.}~\bibnamefont{Braunecker}},
  \bibinfo{author}{\bibfnamefont{A.}~\bibnamefont{Govorov}},
  \bibinfo{author}{\bibfnamefont{L.}~\bibnamefont{Glazman}},
  \bibinfo{author}{\bibfnamefont{A.}~\bibnamefont{Imamoglu}}, \bibnamefont{and}
  \bibinfo{author}{\bibfnamefont{J.}~\bibnamefont{von Delft}},
  \bibinfo{journal}{Physical review letters} \textbf{\bibinfo{volume}{106}},
  \bibinfo{pages}{107402} (\bibinfo{year}{2011}).

\bibitem[{\citenamefont{Basov et~al.}(2011)\citenamefont{Basov, Averitt,
  van~der Marel, Dressel, and Haule}}]{Bascov2011Electrodynamics}
\bibinfo{author}{\bibfnamefont{D.~N.} \bibnamefont{Basov}},
  \bibinfo{author}{\bibfnamefont{R.~D.} \bibnamefont{Averitt}},
  \bibinfo{author}{\bibfnamefont{D.}~\bibnamefont{van~der Marel}},
  \bibinfo{author}{\bibfnamefont{M.}~\bibnamefont{Dressel}}, \bibnamefont{and}
  \bibinfo{author}{\bibfnamefont{K.}~\bibnamefont{Haule}},
  \bibinfo{journal}{Rev. Mod. Phys.} \textbf{\bibinfo{volume}{83}},
  \bibinfo{pages}{471} (\bibinfo{year}{2011}),
  \urlprefix\url{https://link.aps.org/doi/10.1103/RevModPhys.83.471}.

\bibitem[{\citenamefont{Bloch and Zwerger}(2008)}]{Bloch2008Review}
\bibinfo{author}{\bibfnamefont{I.}~\bibnamefont{Bloch}} \bibnamefont{and}
  \bibinfo{author}{\bibfnamefont{W.}~\bibnamefont{Zwerger}},
  \bibinfo{journal}{Reviews of Modern Physics} \textbf{\bibinfo{volume}{80}},
  \bibinfo{pages}{885} (\bibinfo{year}{2008}), ISSN \bibinfo{issn}{0034-6861},
  \urlprefix\url{http://link.aps.org/doi/10.1103/RevModPhys.80.885}.

\bibitem[{\citenamefont{Mazurenko et~al.}(2016)\citenamefont{Mazurenko, Chiu,
  Ji, Parsons, Kan{\'a}sz-Nagy, Schmidt, Grusdt, Demler, Greif, and
  Greiner}}]{Mazurenko2016AFM}
\bibinfo{author}{\bibfnamefont{A.}~\bibnamefont{Mazurenko}},
  \bibinfo{author}{\bibfnamefont{C.~S.} \bibnamefont{Chiu}},
  \bibinfo{author}{\bibfnamefont{G.}~\bibnamefont{Ji}},
  \bibinfo{author}{\bibfnamefont{M.~F.} \bibnamefont{Parsons}},
  \bibinfo{author}{\bibfnamefont{M.}~\bibnamefont{Kan{\'a}sz-Nagy}},
  \bibinfo{author}{\bibfnamefont{R.}~\bibnamefont{Schmidt}},
  \bibinfo{author}{\bibfnamefont{F.}~\bibnamefont{Grusdt}},
  \bibinfo{author}{\bibfnamefont{E.}~\bibnamefont{Demler}},
  \bibinfo{author}{\bibfnamefont{D.}~\bibnamefont{Greif}}, \bibnamefont{and}
  \bibinfo{author}{\bibfnamefont{M.}~\bibnamefont{Greiner}},
  \bibinfo{journal}{arXiv preprint arXiv:1612.08436}  (\bibinfo{year}{2016}).

\bibitem[{\citenamefont{Hilker et~al.}(2017)\citenamefont{Hilker, Salomon,
  Grusdt, Omran, Boll, Demler, Bloch, and Gross}}]{hilker2017revealing}
\bibinfo{author}{\bibfnamefont{T.~A.} \bibnamefont{Hilker}},
  \bibinfo{author}{\bibfnamefont{G.}~\bibnamefont{Salomon}},
  \bibinfo{author}{\bibfnamefont{F.}~\bibnamefont{Grusdt}},
  \bibinfo{author}{\bibfnamefont{A.}~\bibnamefont{Omran}},
  \bibinfo{author}{\bibfnamefont{M.}~\bibnamefont{Boll}},
  \bibinfo{author}{\bibfnamefont{E.}~\bibnamefont{Demler}},
  \bibinfo{author}{\bibfnamefont{I.}~\bibnamefont{Bloch}}, \bibnamefont{and}
  \bibinfo{author}{\bibfnamefont{C.}~\bibnamefont{Gross}},
  \bibinfo{journal}{arXiv preprint arXiv:1702.00642}  (\bibinfo{year}{2017}).

\bibitem[{\citenamefont{Van~Houcke et~al.}(2011)\citenamefont{Van~Houcke,
  Werner, Kozik, Prokofev, Svistunov, Ku, Sommer, Cheuk, Schirotzek, and
  Zwierlein}}]{vanHoucke2011DiagrammaticMC}
\bibinfo{author}{\bibfnamefont{K.}~\bibnamefont{Van~Houcke}},
  \bibinfo{author}{\bibfnamefont{F.}~\bibnamefont{Werner}},
  \bibinfo{author}{\bibfnamefont{E.}~\bibnamefont{Kozik}},
  \bibinfo{author}{\bibfnamefont{N.}~\bibnamefont{Prokofev}},
  \bibinfo{author}{\bibfnamefont{B.}~\bibnamefont{Svistunov}},
  \bibinfo{author}{\bibfnamefont{M.}~\bibnamefont{Ku}},
  \bibinfo{author}{\bibfnamefont{A.}~\bibnamefont{Sommer}},
  \bibinfo{author}{\bibfnamefont{L.}~\bibnamefont{Cheuk}},
  \bibinfo{author}{\bibfnamefont{A.}~\bibnamefont{Schirotzek}},
  \bibnamefont{and}
  \bibinfo{author}{\bibfnamefont{M.}~\bibnamefont{Zwierlein}},
  \bibinfo{journal}{arXiv preprint arXiv:1110.3747}  (\bibinfo{year}{2011}).

\bibitem[{\citenamefont{Ku et~al.}(2014)\citenamefont{Ku, Ji, Mukherjee,
  Guardado-Sanchez, Cheuk, Yefsah, and Zwierlein}}]{ku2014motion}
\bibinfo{author}{\bibfnamefont{M.~J.} \bibnamefont{Ku}},
  \bibinfo{author}{\bibfnamefont{W.}~\bibnamefont{Ji}},
  \bibinfo{author}{\bibfnamefont{B.}~\bibnamefont{Mukherjee}},
  \bibinfo{author}{\bibfnamefont{E.}~\bibnamefont{Guardado-Sanchez}},
  \bibinfo{author}{\bibfnamefont{L.~W.} \bibnamefont{Cheuk}},
  \bibinfo{author}{\bibfnamefont{T.}~\bibnamefont{Yefsah}}, \bibnamefont{and}
  \bibinfo{author}{\bibfnamefont{M.~W.} \bibnamefont{Zwierlein}},
  \bibinfo{journal}{Physical review letters} \textbf{\bibinfo{volume}{113}},
  \bibinfo{pages}{065301} (\bibinfo{year}{2014}).

\bibitem[{\citenamefont{Gring et~al.}(2012)\citenamefont{Gring, Kuhnert,
  Langen, Kitagawa, Rauer, Schreitl, Mazets, Smith, Demler, and
  Schmiedmayer}}]{Schmiedmayer2012Prethermalization}
\bibinfo{author}{\bibfnamefont{M.}~\bibnamefont{Gring}},
  \bibinfo{author}{\bibfnamefont{M.}~\bibnamefont{Kuhnert}},
  \bibinfo{author}{\bibfnamefont{T.}~\bibnamefont{Langen}},
  \bibinfo{author}{\bibfnamefont{T.}~\bibnamefont{Kitagawa}},
  \bibinfo{author}{\bibfnamefont{B.}~\bibnamefont{Rauer}},
  \bibinfo{author}{\bibfnamefont{M.}~\bibnamefont{Schreitl}},
  \bibinfo{author}{\bibfnamefont{I.}~\bibnamefont{Mazets}},
  \bibinfo{author}{\bibfnamefont{D.~A.} \bibnamefont{Smith}},
  \bibinfo{author}{\bibfnamefont{E.}~\bibnamefont{Demler}}, \bibnamefont{and}
  \bibinfo{author}{\bibfnamefont{J.}~\bibnamefont{Schmiedmayer}},
  \bibinfo{journal}{Science} \textbf{\bibinfo{volume}{337}},
  \bibinfo{pages}{1318} (\bibinfo{year}{2012}).

\bibitem[{\citenamefont{Riegger et~al.}(2017)\citenamefont{Riegger, Oppong,
  H{\"o}fer, Fernandes, Bloch, and F{\"o}lling}}]{riegger2017localized}
\bibinfo{author}{\bibfnamefont{L.}~\bibnamefont{Riegger}},
  \bibinfo{author}{\bibfnamefont{N.~D.} \bibnamefont{Oppong}},
  \bibinfo{author}{\bibfnamefont{M.}~\bibnamefont{H{\"o}fer}},
  \bibinfo{author}{\bibfnamefont{D.~R.} \bibnamefont{Fernandes}},
  \bibinfo{author}{\bibfnamefont{I.}~\bibnamefont{Bloch}}, \bibnamefont{and}
  \bibinfo{author}{\bibfnamefont{S.}~\bibnamefont{F{\"o}lling}},
  \bibinfo{journal}{arXiv preprint arXiv:1708.03810}  (\bibinfo{year}{2017}).

\bibitem[{\citenamefont{Barzykin and Affleck}(1996)}]{barzykin1996kondo}
\bibinfo{author}{\bibfnamefont{V.}~\bibnamefont{Barzykin}} \bibnamefont{and}
  \bibinfo{author}{\bibfnamefont{I.}~\bibnamefont{Affleck}},
  \bibinfo{journal}{Physical review letters} \textbf{\bibinfo{volume}{76}},
  \bibinfo{pages}{4959} (\bibinfo{year}{1996}).

\bibitem[{\citenamefont{Simon and Affleck}(2003)}]{simon2003kondo}
\bibinfo{author}{\bibfnamefont{P.}~\bibnamefont{Simon}} \bibnamefont{and}
  \bibinfo{author}{\bibfnamefont{I.}~\bibnamefont{Affleck}},
  \bibinfo{journal}{Physical Review B} \textbf{\bibinfo{volume}{68}},
  \bibinfo{pages}{115304} (\bibinfo{year}{2003}).

\bibitem[{\citenamefont{Holzner et~al.}(2009)\citenamefont{Holzner, McCulloch,
  Schollw{\"o}ck, von Delft, and Heidrich-Meisner}}]{holzner2009kondo}
\bibinfo{author}{\bibfnamefont{A.}~\bibnamefont{Holzner}},
  \bibinfo{author}{\bibfnamefont{I.~P.} \bibnamefont{McCulloch}},
  \bibinfo{author}{\bibfnamefont{U.}~\bibnamefont{Schollw{\"o}ck}},
  \bibinfo{author}{\bibfnamefont{J.}~\bibnamefont{von Delft}},
  \bibnamefont{and}
  \bibinfo{author}{\bibfnamefont{F.}~\bibnamefont{Heidrich-Meisner}},
  \bibinfo{journal}{Physical Review B} \textbf{\bibinfo{volume}{80}},
  \bibinfo{pages}{205114} (\bibinfo{year}{2009}).

\bibitem[{\citenamefont{Koller et~al.}(2005)\citenamefont{Koller, Hewson, and
  Meyer}}]{Koller.2005}
\bibinfo{author}{\bibfnamefont{W.}~\bibnamefont{Koller}},
  \bibinfo{author}{\bibfnamefont{A.~C.} \bibnamefont{Hewson}},
  \bibnamefont{and} \bibinfo{author}{\bibfnamefont{D.}~\bibnamefont{Meyer}},
  \bibinfo{journal}{Phys. Rev. B} \textbf{\bibinfo{volume}{72}},
  \bibinfo{pages}{045117} (\bibinfo{year}{2005}),
  \urlprefix\url{https://link.aps.org/doi/10.1103/PhysRevB.72.045117}.

\bibitem[{\citenamefont{Takamoto et~al.}(2005)\citenamefont{Takamoto, Hong,
  Higashi, and Katori}}]{Takamoto2005}
\bibinfo{author}{\bibfnamefont{M.}~\bibnamefont{Takamoto}},
  \bibinfo{author}{\bibfnamefont{F.-L.} \bibnamefont{Hong}},
  \bibinfo{author}{\bibfnamefont{R.}~\bibnamefont{Higashi}}, \bibnamefont{and}
  \bibinfo{author}{\bibfnamefont{H.}~\bibnamefont{Katori}},
  \bibinfo{journal}{Nature} \textbf{\bibinfo{volume}{435}},
  \bibinfo{pages}{321} (\bibinfo{year}{2005}), ISSN \bibinfo{issn}{1476-4687},
  \urlprefix\url{http://www.nature.com.ezp-prod1.hul.harvard.edu/nature/journal/v435/n7040/full/nature03541.html}.

\bibitem[{\citenamefont{Hinkley et~al.}(2013)\citenamefont{Hinkley, Sherman,
  Phillips, Schioppo, Lemke, Beloy, Pizzocaro, Oates, and
  Ludlow}}]{Hinkley2013}
\bibinfo{author}{\bibfnamefont{N.}~\bibnamefont{Hinkley}},
  \bibinfo{author}{\bibfnamefont{J.~A.} \bibnamefont{Sherman}},
  \bibinfo{author}{\bibfnamefont{N.~B.} \bibnamefont{Phillips}},
  \bibinfo{author}{\bibfnamefont{M.}~\bibnamefont{Schioppo}},
  \bibinfo{author}{\bibfnamefont{N.~D.} \bibnamefont{Lemke}},
  \bibinfo{author}{\bibfnamefont{K.}~\bibnamefont{Beloy}},
  \bibinfo{author}{\bibfnamefont{M.}~\bibnamefont{Pizzocaro}},
  \bibinfo{author}{\bibfnamefont{C.~W.} \bibnamefont{Oates}}, \bibnamefont{and}
  \bibinfo{author}{\bibfnamefont{A.~D.} \bibnamefont{Ludlow}},
  \bibinfo{journal}{Science (New York, N.Y.)} \textbf{\bibinfo{volume}{341}},
  \bibinfo{pages}{1215} (\bibinfo{year}{2013}), ISSN \bibinfo{issn}{1095-9203},
  \urlprefix\url{http://www.sciencemag.org.ezp-prod1.hul.harvard.edu/content/341/6151/1215}.

\bibitem[{\citenamefont{Bloom et~al.}(2014)\citenamefont{Bloom, Nicholson,
  Williams, Campbell, Bishof, Zhang, Zhang, Bromley, and Ye}}]{Bloom2014}
\bibinfo{author}{\bibfnamefont{B.~J.} \bibnamefont{Bloom}},
  \bibinfo{author}{\bibfnamefont{T.~L.} \bibnamefont{Nicholson}},
  \bibinfo{author}{\bibfnamefont{J.~R.} \bibnamefont{Williams}},
  \bibinfo{author}{\bibfnamefont{S.~L.} \bibnamefont{Campbell}},
  \bibinfo{author}{\bibfnamefont{M.}~\bibnamefont{Bishof}},
  \bibinfo{author}{\bibfnamefont{X.}~\bibnamefont{Zhang}},
  \bibinfo{author}{\bibfnamefont{W.}~\bibnamefont{Zhang}},
  \bibinfo{author}{\bibfnamefont{S.~L.} \bibnamefont{Bromley}},
  \bibnamefont{and} \bibinfo{author}{\bibfnamefont{J.}~\bibnamefont{Ye}},
  \bibinfo{journal}{Nature} \textbf{\bibinfo{volume}{506}}, \bibinfo{pages}{71}
  (\bibinfo{year}{2014}), ISSN \bibinfo{issn}{1476-4687},
  \urlprefix\url{http://www.nature.com.ezp-prod1.hul.harvard.edu/nature/journal/v506/n7486/full/nature12941.html}.

\bibitem[{\citenamefont{Ye et~al.}(2008)\citenamefont{Ye, Kimble, and
  Katori}}]{ye2008review}
\bibinfo{author}{\bibfnamefont{J.}~\bibnamefont{Ye}},
  \bibinfo{author}{\bibfnamefont{H.}~\bibnamefont{Kimble}}, \bibnamefont{and}
  \bibinfo{author}{\bibfnamefont{H.}~\bibnamefont{Katori}},
  \bibinfo{journal}{science} \textbf{\bibinfo{volume}{320}},
  \bibinfo{pages}{1734} (\bibinfo{year}{2008}).

\bibitem[{\citenamefont{Krauser et~al.}(2012)\citenamefont{Krauser, Heinze,
  Fl{\"a}schner, G{\"o}tze, J{\"u}rgensen, L{\"u}hmann, Becker, and
  Sengstock}}]{krauser2012coherent}
\bibinfo{author}{\bibfnamefont{J.~S.} \bibnamefont{Krauser}},
  \bibinfo{author}{\bibfnamefont{J.}~\bibnamefont{Heinze}},
  \bibinfo{author}{\bibfnamefont{N.}~\bibnamefont{Fl{\"a}schner}},
  \bibinfo{author}{\bibfnamefont{S.}~\bibnamefont{G{\"o}tze}},
  \bibinfo{author}{\bibfnamefont{O.}~\bibnamefont{J{\"u}rgensen}},
  \bibinfo{author}{\bibfnamefont{D.-S.} \bibnamefont{L{\"u}hmann}},
  \bibinfo{author}{\bibfnamefont{C.}~\bibnamefont{Becker}}, \bibnamefont{and}
  \bibinfo{author}{\bibfnamefont{K.}~\bibnamefont{Sengstock}},
  \bibinfo{journal}{Nature Physics} \textbf{\bibinfo{volume}{8}},
  \bibinfo{pages}{813} (\bibinfo{year}{2012}).

\bibitem[{\citenamefont{Scazza et~al.}(2014)\citenamefont{Scazza, Hofrichter,
  H{\"o}fer, De~Groot, Bloch, and F{\"o}lling}}]{scazza2014observation}
\bibinfo{author}{\bibfnamefont{F.}~\bibnamefont{Scazza}},
  \bibinfo{author}{\bibfnamefont{C.}~\bibnamefont{Hofrichter}},
  \bibinfo{author}{\bibfnamefont{M.}~\bibnamefont{H{\"o}fer}},
  \bibinfo{author}{\bibfnamefont{P.}~\bibnamefont{De~Groot}},
  \bibinfo{author}{\bibfnamefont{I.}~\bibnamefont{Bloch}}, \bibnamefont{and}
  \bibinfo{author}{\bibfnamefont{S.}~\bibnamefont{F{\"o}lling}},
  \bibinfo{journal}{Nature Physics} \textbf{\bibinfo{volume}{10}},
  \bibinfo{pages}{779} (\bibinfo{year}{2014}).

\bibitem[{\citenamefont{Krauser et~al.}(2014)\citenamefont{Krauser, Ebling,
  Fl{\"a}schner, Heinze, Sengstock, Lewenstein, Eckardt, and
  Becker}}]{krauser2014giant}
\bibinfo{author}{\bibfnamefont{J.~S.} \bibnamefont{Krauser}},
  \bibinfo{author}{\bibfnamefont{U.}~\bibnamefont{Ebling}},
  \bibinfo{author}{\bibfnamefont{N.}~\bibnamefont{Fl{\"a}schner}},
  \bibinfo{author}{\bibfnamefont{J.}~\bibnamefont{Heinze}},
  \bibinfo{author}{\bibfnamefont{K.}~\bibnamefont{Sengstock}},
  \bibinfo{author}{\bibfnamefont{M.}~\bibnamefont{Lewenstein}},
  \bibinfo{author}{\bibfnamefont{A.}~\bibnamefont{Eckardt}}, \bibnamefont{and}
  \bibinfo{author}{\bibfnamefont{C.}~\bibnamefont{Becker}},
  \bibinfo{journal}{Science} \textbf{\bibinfo{volume}{343}},
  \bibinfo{pages}{157} (\bibinfo{year}{2014}).

\bibitem[{\citenamefont{Zhang et~al.}(2014{\natexlab{a}})\citenamefont{Zhang,
  Bishof, Bromley, Kraus, Safronova, Zoller, Rey, and
  Ye}}]{zhang2014spectroscopic}
\bibinfo{author}{\bibfnamefont{X.}~\bibnamefont{Zhang}},
  \bibinfo{author}{\bibfnamefont{M.}~\bibnamefont{Bishof}},
  \bibinfo{author}{\bibfnamefont{S.}~\bibnamefont{Bromley}},
  \bibinfo{author}{\bibfnamefont{C.}~\bibnamefont{Kraus}},
  \bibinfo{author}{\bibfnamefont{M.}~\bibnamefont{Safronova}},
  \bibinfo{author}{\bibfnamefont{P.}~\bibnamefont{Zoller}},
  \bibinfo{author}{\bibfnamefont{A.}~\bibnamefont{Rey}}, \bibnamefont{and}
  \bibinfo{author}{\bibfnamefont{J.}~\bibnamefont{Ye}},
  \bibinfo{journal}{science} \textbf{\bibinfo{volume}{345}},
  \bibinfo{pages}{1467} (\bibinfo{year}{2014}{\natexlab{a}}).

\bibitem[{\citenamefont{Mancini et~al.}(2015)\citenamefont{Mancini, Pagano,
  Cappellini, Livi, Rider, Catani, Sias, Zoller, Inguscio, Dalmonte
  et~al.}}]{mancini2015observation}
\bibinfo{author}{\bibfnamefont{M.}~\bibnamefont{Mancini}},
  \bibinfo{author}{\bibfnamefont{G.}~\bibnamefont{Pagano}},
  \bibinfo{author}{\bibfnamefont{G.}~\bibnamefont{Cappellini}},
  \bibinfo{author}{\bibfnamefont{L.}~\bibnamefont{Livi}},
  \bibinfo{author}{\bibfnamefont{M.}~\bibnamefont{Rider}},
  \bibinfo{author}{\bibfnamefont{J.}~\bibnamefont{Catani}},
  \bibinfo{author}{\bibfnamefont{C.}~\bibnamefont{Sias}},
  \bibinfo{author}{\bibfnamefont{P.}~\bibnamefont{Zoller}},
  \bibinfo{author}{\bibfnamefont{M.}~\bibnamefont{Inguscio}},
  \bibinfo{author}{\bibfnamefont{M.}~\bibnamefont{Dalmonte}},
  \bibnamefont{et~al.}, \bibinfo{journal}{Science}
  \textbf{\bibinfo{volume}{349}}, \bibinfo{pages}{1510} (\bibinfo{year}{2015}).

\bibitem[{\citenamefont{H{\"o}fer et~al.}(2015)\citenamefont{H{\"o}fer,
  Riegger, Scazza, Hofrichter, Fernandes, Parish, Levinsen, Bloch, and
  F{\"o}lling}}]{hofer2015observation}
\bibinfo{author}{\bibfnamefont{M.}~\bibnamefont{H{\"o}fer}},
  \bibinfo{author}{\bibfnamefont{L.}~\bibnamefont{Riegger}},
  \bibinfo{author}{\bibfnamefont{F.}~\bibnamefont{Scazza}},
  \bibinfo{author}{\bibfnamefont{C.}~\bibnamefont{Hofrichter}},
  \bibinfo{author}{\bibfnamefont{D.}~\bibnamefont{Fernandes}},
  \bibinfo{author}{\bibfnamefont{M.}~\bibnamefont{Parish}},
  \bibinfo{author}{\bibfnamefont{J.}~\bibnamefont{Levinsen}},
  \bibinfo{author}{\bibfnamefont{I.}~\bibnamefont{Bloch}}, \bibnamefont{and}
  \bibinfo{author}{\bibfnamefont{S.}~\bibnamefont{F{\"o}lling}},
  \bibinfo{journal}{Physical review letters} \textbf{\bibinfo{volume}{115}},
  \bibinfo{pages}{265302} (\bibinfo{year}{2015}).

\bibitem[{\citenamefont{Taie et~al.}(2012)\citenamefont{Taie, Yamazaki, Sugawa,
  and Takahashi}}]{taie20126}
\bibinfo{author}{\bibfnamefont{S.}~\bibnamefont{Taie}},
  \bibinfo{author}{\bibfnamefont{R.}~\bibnamefont{Yamazaki}},
  \bibinfo{author}{\bibfnamefont{S.}~\bibnamefont{Sugawa}}, \bibnamefont{and}
  \bibinfo{author}{\bibfnamefont{Y.}~\bibnamefont{Takahashi}},
  \bibinfo{journal}{Nature Physics} \textbf{\bibinfo{volume}{8}},
  \bibinfo{pages}{825} (\bibinfo{year}{2012}).

\bibitem[{\citenamefont{Zhang et~al.}(2014{\natexlab{b}})\citenamefont{Zhang,
  Bishof, Bromley, Kraus, Safronova, Zoller, Rey, and Ye}}]{Zhang2014}
\bibinfo{author}{\bibfnamefont{X.}~\bibnamefont{Zhang}},
  \bibinfo{author}{\bibfnamefont{M.}~\bibnamefont{Bishof}},
  \bibinfo{author}{\bibfnamefont{S.~L.} \bibnamefont{Bromley}},
  \bibinfo{author}{\bibfnamefont{C.~V.} \bibnamefont{Kraus}},
  \bibinfo{author}{\bibfnamefont{M.~S.} \bibnamefont{Safronova}},
  \bibinfo{author}{\bibfnamefont{P.}~\bibnamefont{Zoller}},
  \bibinfo{author}{\bibfnamefont{A.~M.} \bibnamefont{Rey}}, \bibnamefont{and}
  \bibinfo{author}{\bibfnamefont{J.}~\bibnamefont{Ye}},
  \bibinfo{journal}{Science (New York, N.Y.)} \textbf{\bibinfo{volume}{345}},
  \bibinfo{pages}{1467} (\bibinfo{year}{2014}{\natexlab{b}}), ISSN
  \bibinfo{issn}{1095-9203},
  \urlprefix\url{http://www.sciencemag.org/content/345/6203/1467.short}.

\bibitem[{\citenamefont{Gorshkov et~al.}(2010)\citenamefont{Gorshkov, Hermele,
  Gurarie, Xu, Julienne, Ye, Zoller, Demler, Lukin, and Rey}}]{gorshkov2010two}
\bibinfo{author}{\bibfnamefont{A.}~\bibnamefont{Gorshkov}},
  \bibinfo{author}{\bibfnamefont{M.}~\bibnamefont{Hermele}},
  \bibinfo{author}{\bibfnamefont{V.}~\bibnamefont{Gurarie}},
  \bibinfo{author}{\bibfnamefont{C.}~\bibnamefont{Xu}},
  \bibinfo{author}{\bibfnamefont{P.}~\bibnamefont{Julienne}},
  \bibinfo{author}{\bibfnamefont{J.}~\bibnamefont{Ye}},
  \bibinfo{author}{\bibfnamefont{P.}~\bibnamefont{Zoller}},
  \bibinfo{author}{\bibfnamefont{E.}~\bibnamefont{Demler}},
  \bibinfo{author}{\bibfnamefont{M.}~\bibnamefont{Lukin}}, \bibnamefont{and}
  \bibinfo{author}{\bibfnamefont{A.}~\bibnamefont{Rey}},
  \bibinfo{journal}{Nature Physics} \textbf{\bibinfo{volume}{6}},
  \bibinfo{pages}{289} (\bibinfo{year}{2010}).

\bibitem[{\citenamefont{Daley et~al.}(2008)\citenamefont{Daley, Boyd, Ye, and
  Zoller}}]{Daley2008}
\bibinfo{author}{\bibfnamefont{A.~J.} \bibnamefont{Daley}},
  \bibinfo{author}{\bibfnamefont{M.~M.} \bibnamefont{Boyd}},
  \bibinfo{author}{\bibfnamefont{J.}~\bibnamefont{Ye}}, \bibnamefont{and}
  \bibinfo{author}{\bibfnamefont{P.}~\bibnamefont{Zoller}},
  \bibinfo{journal}{Physical Review Letters} \textbf{\bibinfo{volume}{101}},
  \bibinfo{pages}{170504} (\bibinfo{year}{2008}), ISSN
  \bibinfo{issn}{0031-9007},
  \urlprefix\url{http://link.aps.org/doi/10.1103/PhysRevLett.101.170504}.

\bibitem[{\citenamefont{Taie et~al.}(2010)\citenamefont{Taie, Takasu, Sugawa,
  Yamazaki, Tsujimoto, Murakami, and Takahashi}}]{taie2010realization}
\bibinfo{author}{\bibfnamefont{S.}~\bibnamefont{Taie}},
  \bibinfo{author}{\bibfnamefont{Y.}~\bibnamefont{Takasu}},
  \bibinfo{author}{\bibfnamefont{S.}~\bibnamefont{Sugawa}},
  \bibinfo{author}{\bibfnamefont{R.}~\bibnamefont{Yamazaki}},
  \bibinfo{author}{\bibfnamefont{T.}~\bibnamefont{Tsujimoto}},
  \bibinfo{author}{\bibfnamefont{R.}~\bibnamefont{Murakami}}, \bibnamefont{and}
  \bibinfo{author}{\bibfnamefont{Y.}~\bibnamefont{Takahashi}},
  \bibinfo{journal}{Physical review letters} \textbf{\bibinfo{volume}{105}},
  \bibinfo{pages}{190401} (\bibinfo{year}{2010}).

\bibitem[{\citenamefont{Stellmer et~al.}(2011)\citenamefont{Stellmer, Grimm,
  and Schreck}}]{stellmer2011detection}
\bibinfo{author}{\bibfnamefont{S.}~\bibnamefont{Stellmer}},
  \bibinfo{author}{\bibfnamefont{R.}~\bibnamefont{Grimm}}, \bibnamefont{and}
  \bibinfo{author}{\bibfnamefont{F.}~\bibnamefont{Schreck}},
  \bibinfo{journal}{Physical Review A} \textbf{\bibinfo{volume}{84}},
  \bibinfo{pages}{043611} (\bibinfo{year}{2011}).

\bibitem[{\citenamefont{Pagano et~al.}(2014)\citenamefont{Pagano, Mancini,
  Cappellini, Lombardi, Sch{\"a}fer, Hu, Liu, Catani, Sias, Inguscio
  et~al.}}]{pagano2014one}
\bibinfo{author}{\bibfnamefont{G.}~\bibnamefont{Pagano}},
  \bibinfo{author}{\bibfnamefont{M.}~\bibnamefont{Mancini}},
  \bibinfo{author}{\bibfnamefont{G.}~\bibnamefont{Cappellini}},
  \bibinfo{author}{\bibfnamefont{P.}~\bibnamefont{Lombardi}},
  \bibinfo{author}{\bibfnamefont{F.}~\bibnamefont{Sch{\"a}fer}},
  \bibinfo{author}{\bibfnamefont{H.}~\bibnamefont{Hu}},
  \bibinfo{author}{\bibfnamefont{X.-J.} \bibnamefont{Liu}},
  \bibinfo{author}{\bibfnamefont{J.}~\bibnamefont{Catani}},
  \bibinfo{author}{\bibfnamefont{C.}~\bibnamefont{Sias}},
  \bibinfo{author}{\bibfnamefont{M.}~\bibnamefont{Inguscio}},
  \bibnamefont{et~al.}, \bibinfo{journal}{Nature Physics}
  \textbf{\bibinfo{volume}{10}}, \bibinfo{pages}{198} (\bibinfo{year}{2014}).

\bibitem[{\citenamefont{Busch et~al.}(1998)\citenamefont{Busch, Englert,
  Rza{\.z}ewski, and Wilkens}}]{busch_two_1998}
\bibinfo{author}{\bibfnamefont{T.}~\bibnamefont{Busch}},
  \bibinfo{author}{\bibfnamefont{B.-G.} \bibnamefont{Englert}},
  \bibinfo{author}{\bibfnamefont{K.}~\bibnamefont{Rza{\.z}ewski}},
  \bibnamefont{and} \bibinfo{author}{\bibfnamefont{M.}~\bibnamefont{Wilkens}},
  \bibinfo{journal}{Foundations of Physics} \textbf{\bibinfo{volume}{28}},
  \bibinfo{pages}{549} (\bibinfo{year}{1998}).

\bibitem[{\citenamefont{Zhang et~al.}(2016)\citenamefont{Zhang, Zhang, Cheng,
  Chen, Zhang, and Zhai}}]{zhang2016kondo}
\bibinfo{author}{\bibfnamefont{R.}~\bibnamefont{Zhang}},
  \bibinfo{author}{\bibfnamefont{D.}~\bibnamefont{Zhang}},
  \bibinfo{author}{\bibfnamefont{Y.}~\bibnamefont{Cheng}},
  \bibinfo{author}{\bibfnamefont{W.}~\bibnamefont{Chen}},
  \bibinfo{author}{\bibfnamefont{P.}~\bibnamefont{Zhang}}, \bibnamefont{and}
  \bibinfo{author}{\bibfnamefont{H.}~\bibnamefont{Zhai}},
  \bibinfo{journal}{Physical Review A} \textbf{\bibinfo{volume}{93}},
  \bibinfo{pages}{043601} (\bibinfo{year}{2016}).

\bibitem[{\citenamefont{Nakagawa and
  Kawakami}(2015)}]{nakagawa2015LaserInducedKondo}
\bibinfo{author}{\bibfnamefont{M.}~\bibnamefont{Nakagawa}} \bibnamefont{and}
  \bibinfo{author}{\bibfnamefont{N.}~\bibnamefont{Kawakami}},
  \bibinfo{journal}{Physical review letters} \textbf{\bibinfo{volume}{115}},
  \bibinfo{pages}{165303} (\bibinfo{year}{2015}).

\bibitem[{\citenamefont{Pekker et~al.}(2011)\citenamefont{Pekker, Babadi,
  Sensarma, Zinner, Pollet, Zwierlein, and Demler}}]{Pekker2011Competition}
\bibinfo{author}{\bibfnamefont{D.}~\bibnamefont{Pekker}},
  \bibinfo{author}{\bibfnamefont{M.}~\bibnamefont{Babadi}},
  \bibinfo{author}{\bibfnamefont{R.}~\bibnamefont{Sensarma}},
  \bibinfo{author}{\bibfnamefont{N.}~\bibnamefont{Zinner}},
  \bibinfo{author}{\bibfnamefont{L.}~\bibnamefont{Pollet}},
  \bibinfo{author}{\bibfnamefont{M.~W.} \bibnamefont{Zwierlein}},
  \bibnamefont{and} \bibinfo{author}{\bibfnamefont{E.}~\bibnamefont{Demler}},
  \bibinfo{journal}{Phys. Rev. Lett.} \textbf{\bibinfo{volume}{106}},
  \bibinfo{pages}{050402} (\bibinfo{year}{2011}),
  \urlprefix\url{https://link.aps.org/doi/10.1103/PhysRevLett.106.050402}.

\bibitem[{\citenamefont{Zhang et~al.}(2015)\citenamefont{Zhang, Cheng, Zhai,
  and Zhang}}]{Zhang2015}
\bibinfo{author}{\bibfnamefont{R.}~\bibnamefont{Zhang}},
  \bibinfo{author}{\bibfnamefont{Y.}~\bibnamefont{Cheng}},
  \bibinfo{author}{\bibfnamefont{H.}~\bibnamefont{Zhai}}, \bibnamefont{and}
  \bibinfo{author}{\bibfnamefont{P.}~\bibnamefont{Zhang}},
  \bibinfo{journal}{Physical Review Letters} \textbf{\bibinfo{volume}{115}},
  \bibinfo{pages}{135301} (\bibinfo{year}{2015}), ISSN
  \bibinfo{issn}{0031-9007},
  \urlprefix\url{http://link.aps.org/doi/10.1103/PhysRevLett.115.135301}.

\bibitem[{\citenamefont{Schrieffer and Wolff}(1966)}]{Schrieffer1966}
\bibinfo{author}{\bibfnamefont{J.~R.} \bibnamefont{Schrieffer}}
  \bibnamefont{and} \bibinfo{author}{\bibfnamefont{P.~A.} \bibnamefont{Wolff}},
  \bibinfo{journal}{Physical Review} \textbf{\bibinfo{volume}{149}},
  \bibinfo{pages}{491} (\bibinfo{year}{1966}), ISSN \bibinfo{issn}{0031-899X},
  \urlprefix\url{http://link.aps.org/doi/10.1103/PhysRev.149.491}.

\bibitem[{\citenamefont{Struck et~al.}(2011)\citenamefont{Struck,
  {\"O}lschl{\"a}ger, Le~Targat, Soltan-Panahi, Eckardt, Lewenstein,
  Windpassinger, and Sengstock}}]{struck2011quantum}
\bibinfo{author}{\bibfnamefont{J.}~\bibnamefont{Struck}},
  \bibinfo{author}{\bibfnamefont{C.}~\bibnamefont{{\"O}lschl{\"a}ger}},
  \bibinfo{author}{\bibfnamefont{R.}~\bibnamefont{Le~Targat}},
  \bibinfo{author}{\bibfnamefont{P.}~\bibnamefont{Soltan-Panahi}},
  \bibinfo{author}{\bibfnamefont{A.}~\bibnamefont{Eckardt}},
  \bibinfo{author}{\bibfnamefont{M.}~\bibnamefont{Lewenstein}},
  \bibinfo{author}{\bibfnamefont{P.}~\bibnamefont{Windpassinger}},
  \bibnamefont{and}
  \bibinfo{author}{\bibfnamefont{K.}~\bibnamefont{Sengstock}},
  \bibinfo{journal}{Science} \textbf{\bibinfo{volume}{333}},
  \bibinfo{pages}{996} (\bibinfo{year}{2011}).

\bibitem[{\citenamefont{Aidelsburger et~al.}(2011)\citenamefont{Aidelsburger,
  Atala, Nascimb{\`e}ne, Trotzky, Chen, and
  Bloch}}]{aidelsburger2011experimental}
\bibinfo{author}{\bibfnamefont{M.}~\bibnamefont{Aidelsburger}},
  \bibinfo{author}{\bibfnamefont{M.}~\bibnamefont{Atala}},
  \bibinfo{author}{\bibfnamefont{S.}~\bibnamefont{Nascimb{\`e}ne}},
  \bibinfo{author}{\bibfnamefont{S.}~\bibnamefont{Trotzky}},
  \bibinfo{author}{\bibfnamefont{Y.-A.} \bibnamefont{Chen}}, \bibnamefont{and}
  \bibinfo{author}{\bibfnamefont{I.}~\bibnamefont{Bloch}},
  \bibinfo{journal}{Physical review letters} \textbf{\bibinfo{volume}{107}},
  \bibinfo{pages}{255301} (\bibinfo{year}{2011}).

\bibitem[{\citenamefont{Struck et~al.}(2012)\citenamefont{Struck,
  {\"O}lschl{\"a}ger, Weinberg, Hauke, Simonet, Eckardt, Lewenstein, Sengstock,
  and Windpassinger}}]{struck2012tunable}
\bibinfo{author}{\bibfnamefont{J.}~\bibnamefont{Struck}},
  \bibinfo{author}{\bibfnamefont{C.}~\bibnamefont{{\"O}lschl{\"a}ger}},
  \bibinfo{author}{\bibfnamefont{M.}~\bibnamefont{Weinberg}},
  \bibinfo{author}{\bibfnamefont{P.}~\bibnamefont{Hauke}},
  \bibinfo{author}{\bibfnamefont{J.}~\bibnamefont{Simonet}},
  \bibinfo{author}{\bibfnamefont{A.}~\bibnamefont{Eckardt}},
  \bibinfo{author}{\bibfnamefont{M.}~\bibnamefont{Lewenstein}},
  \bibinfo{author}{\bibfnamefont{K.}~\bibnamefont{Sengstock}},
  \bibnamefont{and}
  \bibinfo{author}{\bibfnamefont{P.}~\bibnamefont{Windpassinger}},
  \bibinfo{journal}{Physical review letters} \textbf{\bibinfo{volume}{108}},
  \bibinfo{pages}{225304} (\bibinfo{year}{2012}).

\bibitem[{\citenamefont{Hauke et~al.}(2012)\citenamefont{Hauke, Tieleman, Celi,
  {\"O}lschl{\"a}ger, Simonet, Struck, Weinberg, Windpassinger, Sengstock,
  Lewenstein et~al.}}]{hauke2012non}
\bibinfo{author}{\bibfnamefont{P.}~\bibnamefont{Hauke}},
  \bibinfo{author}{\bibfnamefont{O.}~\bibnamefont{Tieleman}},
  \bibinfo{author}{\bibfnamefont{A.}~\bibnamefont{Celi}},
  \bibinfo{author}{\bibfnamefont{C.}~\bibnamefont{{\"O}lschl{\"a}ger}},
  \bibinfo{author}{\bibfnamefont{J.}~\bibnamefont{Simonet}},
  \bibinfo{author}{\bibfnamefont{J.}~\bibnamefont{Struck}},
  \bibinfo{author}{\bibfnamefont{M.}~\bibnamefont{Weinberg}},
  \bibinfo{author}{\bibfnamefont{P.}~\bibnamefont{Windpassinger}},
  \bibinfo{author}{\bibfnamefont{K.}~\bibnamefont{Sengstock}},
  \bibinfo{author}{\bibfnamefont{M.}~\bibnamefont{Lewenstein}},
  \bibnamefont{et~al.}, \bibinfo{journal}{Physical review letters}
  \textbf{\bibinfo{volume}{109}}, \bibinfo{pages}{145301}
  (\bibinfo{year}{2012}).

\bibitem[{\citenamefont{Aidelsburger et~al.}(2013)\citenamefont{Aidelsburger,
  Atala, Lohse, Barreiro, Paredes, and Bloch}}]{aidelsburger2013realization}
\bibinfo{author}{\bibfnamefont{M.}~\bibnamefont{Aidelsburger}},
  \bibinfo{author}{\bibfnamefont{M.}~\bibnamefont{Atala}},
  \bibinfo{author}{\bibfnamefont{M.}~\bibnamefont{Lohse}},
  \bibinfo{author}{\bibfnamefont{J.~T.} \bibnamefont{Barreiro}},
  \bibinfo{author}{\bibfnamefont{B.}~\bibnamefont{Paredes}}, \bibnamefont{and}
  \bibinfo{author}{\bibfnamefont{I.}~\bibnamefont{Bloch}},
  \bibinfo{journal}{Physical review letters} \textbf{\bibinfo{volume}{111}},
  \bibinfo{pages}{185301} (\bibinfo{year}{2013}).

\bibitem[{\citenamefont{Celi et~al.}(2014)\citenamefont{Celi, Massignan,
  Ruseckas, Goldman, Spielman, Juzeli{\=u}nas, and
  Lewenstein}}]{celi2014synthetic}
\bibinfo{author}{\bibfnamefont{A.}~\bibnamefont{Celi}},
  \bibinfo{author}{\bibfnamefont{P.}~\bibnamefont{Massignan}},
  \bibinfo{author}{\bibfnamefont{J.}~\bibnamefont{Ruseckas}},
  \bibinfo{author}{\bibfnamefont{N.}~\bibnamefont{Goldman}},
  \bibinfo{author}{\bibfnamefont{I.~B.} \bibnamefont{Spielman}},
  \bibinfo{author}{\bibfnamefont{G.}~\bibnamefont{Juzeli{\=u}nas}},
  \bibnamefont{and}
  \bibinfo{author}{\bibfnamefont{M.}~\bibnamefont{Lewenstein}},
  \bibinfo{journal}{Physical review letters} \textbf{\bibinfo{volume}{112}},
  \bibinfo{pages}{043001} (\bibinfo{year}{2014}).

\bibitem[{\citenamefont{Aidelsburger et~al.}(2015)\citenamefont{Aidelsburger,
  Lohse, Schweizer, Atala, Barreiro, Nascimbene, Cooper, Bloch, and
  Goldman}}]{aidelsburger2014measuring}
\bibinfo{author}{\bibfnamefont{M.}~\bibnamefont{Aidelsburger}},
  \bibinfo{author}{\bibfnamefont{M.}~\bibnamefont{Lohse}},
  \bibinfo{author}{\bibfnamefont{C.}~\bibnamefont{Schweizer}},
  \bibinfo{author}{\bibfnamefont{M.}~\bibnamefont{Atala}},
  \bibinfo{author}{\bibfnamefont{J.~T.} \bibnamefont{Barreiro}},
  \bibinfo{author}{\bibfnamefont{S.}~\bibnamefont{Nascimbene}},
  \bibinfo{author}{\bibfnamefont{N.}~\bibnamefont{Cooper}},
  \bibinfo{author}{\bibfnamefont{I.}~\bibnamefont{Bloch}}, \bibnamefont{and}
  \bibinfo{author}{\bibfnamefont{N.}~\bibnamefont{Goldman}},
  \bibinfo{journal}{Nature Physics} \textbf{\bibinfo{volume}{11}},
  \bibinfo{pages}{162} (\bibinfo{year}{2015}).

\bibitem[{\citenamefont{Jotzu et~al.}(2014)\citenamefont{Jotzu, Messer,
  Desbuquois, Lebrat, Uehlinger, Greif, and Esslinger}}]{jotzu2014experimental}
\bibinfo{author}{\bibfnamefont{G.}~\bibnamefont{Jotzu}},
  \bibinfo{author}{\bibfnamefont{M.}~\bibnamefont{Messer}},
  \bibinfo{author}{\bibfnamefont{R.}~\bibnamefont{Desbuquois}},
  \bibinfo{author}{\bibfnamefont{M.}~\bibnamefont{Lebrat}},
  \bibinfo{author}{\bibfnamefont{T.}~\bibnamefont{Uehlinger}},
  \bibinfo{author}{\bibfnamefont{D.}~\bibnamefont{Greif}}, \bibnamefont{and}
  \bibinfo{author}{\bibfnamefont{T.}~\bibnamefont{Esslinger}},
  \bibinfo{journal}{Nature} \textbf{\bibinfo{volume}{515}},
  \bibinfo{pages}{237} (\bibinfo{year}{2014}).

\bibitem[{\citenamefont{Jim\'enez-Garc\'{\i}a
  et~al.}(2015)\citenamefont{Jim\'enez-Garc\'{\i}a, LeBlanc, Williams, Beeler,
  Qu, Gong, Zhang, and Spielman}}]{Garcia2015SpinOrbit}
\bibinfo{author}{\bibfnamefont{K.}~\bibnamefont{Jim\'enez-Garc\'{\i}a}},
  \bibinfo{author}{\bibfnamefont{L.~J.} \bibnamefont{LeBlanc}},
  \bibinfo{author}{\bibfnamefont{R.~A.} \bibnamefont{Williams}},
  \bibinfo{author}{\bibfnamefont{M.~C.} \bibnamefont{Beeler}},
  \bibinfo{author}{\bibfnamefont{C.}~\bibnamefont{Qu}},
  \bibinfo{author}{\bibfnamefont{M.}~\bibnamefont{Gong}},
  \bibinfo{author}{\bibfnamefont{C.}~\bibnamefont{Zhang}}, \bibnamefont{and}
  \bibinfo{author}{\bibfnamefont{I.~B.} \bibnamefont{Spielman}},
  \bibinfo{journal}{Phys. Rev. Lett.} \textbf{\bibinfo{volume}{114}},
  \bibinfo{pages}{125301} (\bibinfo{year}{2015}),
  \urlprefix\url{https://link.aps.org/doi/10.1103/PhysRevLett.114.125301}.

\bibitem[{\citenamefont{Eckardt et~al.}(2005)\citenamefont{Eckardt, Weiss, and
  Holthaus}}]{eckardt2005superfluid}
\bibinfo{author}{\bibfnamefont{A.}~\bibnamefont{Eckardt}},
  \bibinfo{author}{\bibfnamefont{C.}~\bibnamefont{Weiss}}, \bibnamefont{and}
  \bibinfo{author}{\bibfnamefont{M.}~\bibnamefont{Holthaus}},
  \bibinfo{journal}{Physical review letters} \textbf{\bibinfo{volume}{95}},
  \bibinfo{pages}{260404} (\bibinfo{year}{2005}).

\bibitem[{\citenamefont{Kaufman et~al.}(2009)\citenamefont{Kaufman, Anderson,
  Hanna, Tiesinga, Julienne, and Hall}}]{kaufman2009radio}
\bibinfo{author}{\bibfnamefont{A.}~\bibnamefont{Kaufman}},
  \bibinfo{author}{\bibfnamefont{R.}~\bibnamefont{Anderson}},
  \bibinfo{author}{\bibfnamefont{T.~M.} \bibnamefont{Hanna}},
  \bibinfo{author}{\bibfnamefont{E.}~\bibnamefont{Tiesinga}},
  \bibinfo{author}{\bibfnamefont{P.}~\bibnamefont{Julienne}}, \bibnamefont{and}
  \bibinfo{author}{\bibfnamefont{D.}~\bibnamefont{Hall}},
  \bibinfo{journal}{Physical Review A} \textbf{\bibinfo{volume}{80}},
  \bibinfo{pages}{050701} (\bibinfo{year}{2009}).

\bibitem[{\citenamefont{Tscherbul et~al.}(2010)\citenamefont{Tscherbul,
  Calarco, Lesanovsky, Krems, Dalgarno, and Schmiedmayer}}]{tscherbul2010rf}
\bibinfo{author}{\bibfnamefont{T.~V.} \bibnamefont{Tscherbul}},
  \bibinfo{author}{\bibfnamefont{T.}~\bibnamefont{Calarco}},
  \bibinfo{author}{\bibfnamefont{I.}~\bibnamefont{Lesanovsky}},
  \bibinfo{author}{\bibfnamefont{R.~V.} \bibnamefont{Krems}},
  \bibinfo{author}{\bibfnamefont{A.}~\bibnamefont{Dalgarno}}, \bibnamefont{and}
  \bibinfo{author}{\bibfnamefont{J.}~\bibnamefont{Schmiedmayer}},
  \bibinfo{journal}{Physical Review A} \textbf{\bibinfo{volume}{81}},
  \bibinfo{pages}{050701} (\bibinfo{year}{2010}).

\bibitem[{\citenamefont{Shirley}(1965)}]{Shirley1965Floquet}
\bibinfo{author}{\bibfnamefont{J.~H.} \bibnamefont{Shirley}},
  \bibinfo{journal}{Phys. Rev.} \textbf{\bibinfo{volume}{138}},
  \bibinfo{pages}{B979} (\bibinfo{year}{1965}),
  \urlprefix\url{https://link.aps.org/doi/10.1103/PhysRev.138.B979}.

\bibitem[{\citenamefont{Sambe}(1973)}]{Sambe1973Floquet}
\bibinfo{author}{\bibfnamefont{H.}~\bibnamefont{Sambe}},
  \bibinfo{journal}{Phys. Rev. A} \textbf{\bibinfo{volume}{7}},
  \bibinfo{pages}{2203} (\bibinfo{year}{1973}),
  \urlprefix\url{https://link.aps.org/doi/10.1103/PhysRevA.7.2203}.

\bibitem[{\citenamefont{Rahav et~al.}(2003)\citenamefont{Rahav, Gilary, and
  Fishman}}]{Rahav2003Floquet}
\bibinfo{author}{\bibfnamefont{S.}~\bibnamefont{Rahav}},
  \bibinfo{author}{\bibfnamefont{I.}~\bibnamefont{Gilary}}, \bibnamefont{and}
  \bibinfo{author}{\bibfnamefont{S.}~\bibnamefont{Fishman}},
  \bibinfo{journal}{Phys. Rev. A} \textbf{\bibinfo{volume}{68}},
  \bibinfo{pages}{013820} (\bibinfo{year}{2003}),
  \urlprefix\url{https://link.aps.org/doi/10.1103/PhysRevA.68.013820}.

\bibitem[{\citenamefont{G{\"o}rg et~al.}(2017)\citenamefont{G{\"o}rg, Messer,
  Sandholzer, Jotzu, Desbuquois, and Esslinger}}]{Esslinger2017Floquet}
\bibinfo{author}{\bibfnamefont{F.}~\bibnamefont{G{\"o}rg}},
  \bibinfo{author}{\bibfnamefont{M.}~\bibnamefont{Messer}},
  \bibinfo{author}{\bibfnamefont{K.}~\bibnamefont{Sandholzer}},
  \bibinfo{author}{\bibfnamefont{G.}~\bibnamefont{Jotzu}},
  \bibinfo{author}{\bibfnamefont{R.}~\bibnamefont{Desbuquois}},
  \bibnamefont{and}
  \bibinfo{author}{\bibfnamefont{T.}~\bibnamefont{Esslinger}},
  \bibinfo{journal}{arXiv preprint arXiv:1708.06751}  (\bibinfo{year}{2017}).

\bibitem[{\citenamefont{T{\'o}th et~al.}(2008)\citenamefont{T{\'o}th, Moca,
  Legeza, and Zar{\'a}nd}}]{toth2008BudapestNRG}
\bibinfo{author}{\bibfnamefont{A.}~\bibnamefont{T{\'o}th}},
  \bibinfo{author}{\bibfnamefont{C.}~\bibnamefont{Moca}},
  \bibinfo{author}{\bibfnamefont{{\"O}.}~\bibnamefont{Legeza}},
  \bibnamefont{and}
  \bibinfo{author}{\bibfnamefont{G.}~\bibnamefont{Zar{\'a}nd}},
  \bibinfo{journal}{Physical Review B} \textbf{\bibinfo{volume}{78}},
  \bibinfo{pages}{245109} (\bibinfo{year}{2008}).

\bibitem[{\citenamefont{Knap et~al.}(2013)\citenamefont{Knap, Kantian,
  Giamarchi, Bloch, Lukin, and Demler}}]{Knap2013CorrelationFunctionsRamsey}
\bibinfo{author}{\bibfnamefont{M.}~\bibnamefont{Knap}},
  \bibinfo{author}{\bibfnamefont{A.}~\bibnamefont{Kantian}},
  \bibinfo{author}{\bibfnamefont{T.}~\bibnamefont{Giamarchi}},
  \bibinfo{author}{\bibfnamefont{I.}~\bibnamefont{Bloch}},
  \bibinfo{author}{\bibfnamefont{M.~D.} \bibnamefont{Lukin}}, \bibnamefont{and}
  \bibinfo{author}{\bibfnamefont{E.}~\bibnamefont{Demler}},
  \bibinfo{journal}{Phys. Rev. Lett.} \textbf{\bibinfo{volume}{111}},
  \bibinfo{pages}{147205} (\bibinfo{year}{2013}),
  \urlprefix\url{https://link.aps.org/doi/10.1103/PhysRevLett.111.147205}.

\bibitem[{\citenamefont{Mora et~al.}(2015)\citenamefont{Mora, Moca, Von~Delft,
  and Zar{\'a}nd}}]{mora2015fermi}
\bibinfo{author}{\bibfnamefont{C.}~\bibnamefont{Mora}},
  \bibinfo{author}{\bibfnamefont{C.~P.} \bibnamefont{Moca}},
  \bibinfo{author}{\bibfnamefont{J.}~\bibnamefont{Von~Delft}},
  \bibnamefont{and}
  \bibinfo{author}{\bibfnamefont{G.}~\bibnamefont{Zar{\'a}nd}},
  \bibinfo{journal}{Physical Review B} \textbf{\bibinfo{volume}{92}},
  \bibinfo{pages}{075120} (\bibinfo{year}{2015}).

\bibitem[{\citenamefont{Mehta et~al.}(2005)\citenamefont{Mehta, Andrei,
  Coleman, Borda, and Zarand}}]{Zarand.2005SFL}
\bibinfo{author}{\bibfnamefont{P.}~\bibnamefont{Mehta}},
  \bibinfo{author}{\bibfnamefont{N.}~\bibnamefont{Andrei}},
  \bibinfo{author}{\bibfnamefont{P.}~\bibnamefont{Coleman}},
  \bibinfo{author}{\bibfnamefont{L.}~\bibnamefont{Borda}}, \bibnamefont{and}
  \bibinfo{author}{\bibfnamefont{G.}~\bibnamefont{Zarand}},
  \bibinfo{journal}{Phys. Rev. B} \textbf{\bibinfo{volume}{72}},
  \bibinfo{pages}{014430} (\bibinfo{year}{2005}),
  \urlprefix\url{https://link.aps.org/doi/10.1103/PhysRevB.72.014430}.

\bibitem[{\citenamefont{Eisert et~al.}(2014)\citenamefont{Eisert, Friesdorf,
  and Gogolin}}]{Eisert2014QuantumManyBodyOutOfEquilibrium}
\bibinfo{author}{\bibfnamefont{J.}~\bibnamefont{Eisert}},
  \bibinfo{author}{\bibfnamefont{M.}~\bibnamefont{Friesdorf}},
  \bibnamefont{and} \bibinfo{author}{\bibfnamefont{C.}~\bibnamefont{Gogolin}},
  \bibinfo{journal}{arXiv preprint arXiv:1408.5148}  (\bibinfo{year}{2014}).

\bibitem[{\citenamefont{Miranda et~al.}(2015)\citenamefont{Miranda, Inoue,
  Okuyama, Nakamoto, and Kozuma}}]{miranda2015site}
\bibinfo{author}{\bibfnamefont{M.}~\bibnamefont{Miranda}},
  \bibinfo{author}{\bibfnamefont{R.}~\bibnamefont{Inoue}},
  \bibinfo{author}{\bibfnamefont{Y.}~\bibnamefont{Okuyama}},
  \bibinfo{author}{\bibfnamefont{A.}~\bibnamefont{Nakamoto}}, \bibnamefont{and}
  \bibinfo{author}{\bibfnamefont{M.}~\bibnamefont{Kozuma}},
  \bibinfo{journal}{Physical Review A} \textbf{\bibinfo{volume}{91}},
  \bibinfo{pages}{063414} (\bibinfo{year}{2015}).

\bibitem[{\citenamefont{Yamamoto et~al.}(2016)\citenamefont{Yamamoto,
  Kobayashi, Kuno, Kato, and Takahashi}}]{yamamoto2016ytterbium}
\bibinfo{author}{\bibfnamefont{R.}~\bibnamefont{Yamamoto}},
  \bibinfo{author}{\bibfnamefont{J.}~\bibnamefont{Kobayashi}},
  \bibinfo{author}{\bibfnamefont{T.}~\bibnamefont{Kuno}},
  \bibinfo{author}{\bibfnamefont{K.}~\bibnamefont{Kato}}, \bibnamefont{and}
  \bibinfo{author}{\bibfnamefont{Y.}~\bibnamefont{Takahashi}},
  \bibinfo{journal}{New Journal of Physics} \textbf{\bibinfo{volume}{18}},
  \bibinfo{pages}{023016} (\bibinfo{year}{2016}).

\bibitem[{\citenamefont{Korringa}(1950)}]{Korringa1950}
\bibinfo{author}{\bibfnamefont{J.}~\bibnamefont{Korringa}},
  \bibinfo{journal}{Physica} \textbf{\bibinfo{volume}{16}},
  \bibinfo{pages}{601} (\bibinfo{year}{1950}), ISSN \bibinfo{issn}{00318914},
  \urlprefix\url{http://www.sciencedirect.com/science/article/pii/0031891450901054}.

\bibitem[{\citenamefont{Gupta}(2012)}]{gupta2012KorringaExpt}
\bibinfo{author}{\bibfnamefont{L.~C.} \bibnamefont{Gupta}},
  \emph{\bibinfo{title}{Theoretical and experimental aspects of valence
  fluctuations and heavy fermions}} (\bibinfo{publisher}{Springer Science \&
  Business Media}, \bibinfo{year}{2012}).

\bibitem[{\citenamefont{Zwerger}(1983{\natexlab{b}})}]{Zwerger1983}
\bibinfo{author}{\bibfnamefont{W.}~\bibnamefont{Zwerger}},
  \bibinfo{journal}{Zeitschrift f\"{u}r Physik B Condensed Matter}
  \textbf{\bibinfo{volume}{53}}, \bibinfo{pages}{53}
  (\bibinfo{year}{1983}{\natexlab{b}}), ISSN \bibinfo{issn}{0722-3277},
  \urlprefix\url{http://link.springer.com/10.1007/BF01578247}.

\bibitem[{\citenamefont{Andrei}(1980{\natexlab{b}})}]{andrei1981Kondo}
\bibinfo{author}{\bibfnamefont{N.}~\bibnamefont{Andrei}},
  \bibinfo{journal}{Phys. Rev. Lett.} \textbf{\bibinfo{volume}{45}},
  \bibinfo{pages}{379} (\bibinfo{year}{1980}{\natexlab{b}}),
  \urlprefix\url{https://link.aps.org/doi/10.1103/PhysRevLett.45.379}.

\bibitem[{\citenamefont{Wiegmann}(1981)}]{wiegmann1981exact}
\bibinfo{author}{\bibfnamefont{P.}~\bibnamefont{Wiegmann}},
  \bibinfo{journal}{Journal of Physics C: Solid State Physics}
  \textbf{\bibinfo{volume}{14}}, \bibinfo{pages}{1463} (\bibinfo{year}{1981}).

\bibitem[{\citenamefont{Wigman}(1982)}]{wigman1982exact}
\bibinfo{author}{\bibfnamefont{P.~B.} \bibnamefont{Wigman}},
  \bibinfo{journal}{Physics-Uspekhi} \textbf{\bibinfo{volume}{25}},
  \bibinfo{pages}{183} (\bibinfo{year}{1982}).

\bibitem[{\citenamefont{Andrei}(1995)}]{andrei1995integrable}
\bibinfo{author}{\bibfnamefont{N.}~\bibnamefont{Andrei}}, in
  \emph{\bibinfo{booktitle}{Low-Dimensional Quantum Field Theories for
  Condensed Matter Physicists}} (\bibinfo{publisher}{World Scientific},
  \bibinfo{year}{1995}), pp. \bibinfo{pages}{457--551}.

\bibitem[{\citenamefont{Tsvelick and Wiegmann}(1984)}]{tsvelick1984solution}
\bibinfo{author}{\bibfnamefont{A.}~\bibnamefont{Tsvelick}} \bibnamefont{and}
  \bibinfo{author}{\bibfnamefont{P.}~\bibnamefont{Wiegmann}},
  \bibinfo{journal}{Zeitschrift f{\"u}r Physik B Condensed Matter}
  \textbf{\bibinfo{volume}{54}}, \bibinfo{pages}{201} (\bibinfo{year}{1984}).

\bibitem[{\citenamefont{Affleck and Ludwig}(1993)}]{affleck1993exact}
\bibinfo{author}{\bibfnamefont{I.}~\bibnamefont{Affleck}} \bibnamefont{and}
  \bibinfo{author}{\bibfnamefont{A.~W.} \bibnamefont{Ludwig}},
  \bibinfo{journal}{Physical Review B} \textbf{\bibinfo{volume}{48}},
  \bibinfo{pages}{7297} (\bibinfo{year}{1993}).

\bibitem[{\citenamefont{Fendley et~al.}(1996)\citenamefont{Fendley, Lesage, and
  Saleur}}]{fendley1996unified}
\bibinfo{author}{\bibfnamefont{P.}~\bibnamefont{Fendley}},
  \bibinfo{author}{\bibfnamefont{F.}~\bibnamefont{Lesage}}, \bibnamefont{and}
  \bibinfo{author}{\bibfnamefont{H.}~\bibnamefont{Saleur}},
  \bibinfo{journal}{Journal of statistical physics}
  \textbf{\bibinfo{volume}{85}}, \bibinfo{pages}{211} (\bibinfo{year}{1996}).

\bibitem[{\citenamefont{LeClair and Ludwig}(1999)}]{leclair1999minimal}
\bibinfo{author}{\bibfnamefont{A.}~\bibnamefont{LeClair}} \bibnamefont{and}
  \bibinfo{author}{\bibfnamefont{A.~W.} \bibnamefont{Ludwig}},
  \bibinfo{journal}{Nuclear Physics B} \textbf{\bibinfo{volume}{549}},
  \bibinfo{pages}{546} (\bibinfo{year}{1999}).

\bibitem[{\citenamefont{Schmidt et~al.}(2008)\citenamefont{Schmidt, Werner,
  M\"uhlbacher, and Komnik}}]{Schmidt2008MonteCarlo}
\bibinfo{author}{\bibfnamefont{T.~L.} \bibnamefont{Schmidt}},
  \bibinfo{author}{\bibfnamefont{P.}~\bibnamefont{Werner}},
  \bibinfo{author}{\bibfnamefont{L.}~\bibnamefont{M\"uhlbacher}},
  \bibnamefont{and} \bibinfo{author}{\bibfnamefont{A.}~\bibnamefont{Komnik}},
  \bibinfo{journal}{Phys. Rev. B} \textbf{\bibinfo{volume}{78}},
  \bibinfo{pages}{235110} (\bibinfo{year}{2008}),
  \urlprefix\url{https://link.aps.org/doi/10.1103/PhysRevB.78.235110}.

\bibitem[{\citenamefont{White and Feiguin}(2004)}]{White2004DMRG}
\bibinfo{author}{\bibfnamefont{S.~R.} \bibnamefont{White}} \bibnamefont{and}
  \bibinfo{author}{\bibfnamefont{A.~E.} \bibnamefont{Feiguin}},
  \bibinfo{journal}{Phys. Rev. Lett.} \textbf{\bibinfo{volume}{93}},
  \bibinfo{pages}{076401} (\bibinfo{year}{2004}),
  \urlprefix\url{https://link.aps.org/doi/10.1103/PhysRevLett.93.076401}.

\bibitem[{\citenamefont{Schmitteckert}(2004)}]{Schmitteckert2004DMRG}
\bibinfo{author}{\bibfnamefont{P.}~\bibnamefont{Schmitteckert}},
  \bibinfo{journal}{Phys. Rev. B} \textbf{\bibinfo{volume}{70}},
  \bibinfo{pages}{121302} (\bibinfo{year}{2004}),
  \urlprefix\url{https://link.aps.org/doi/10.1103/PhysRevB.70.121302}.

\bibitem[{\citenamefont{Anders and Schiller}(2005)}]{Anders2005TDNRG}
\bibinfo{author}{\bibfnamefont{F.~B.} \bibnamefont{Anders}} \bibnamefont{and}
  \bibinfo{author}{\bibfnamefont{A.}~\bibnamefont{Schiller}},
  \bibinfo{journal}{Phys. Rev. Lett.} \textbf{\bibinfo{volume}{95}},
  \bibinfo{pages}{196801} (\bibinfo{year}{2005}),
  \urlprefix\url{https://link.aps.org/doi/10.1103/PhysRevLett.95.196801}.

\bibitem[{\citenamefont{Anders and Schiller}(2006)}]{anders2006spin}
\bibinfo{author}{\bibfnamefont{F.~B.} \bibnamefont{Anders}} \bibnamefont{and}
  \bibinfo{author}{\bibfnamefont{A.}~\bibnamefont{Schiller}},
  \bibinfo{journal}{Physical Review B} \textbf{\bibinfo{volume}{74}},
  \bibinfo{pages}{245113} (\bibinfo{year}{2006}).

\bibitem[{\citenamefont{Lobaskin and
  Kehrein}(2005{\natexlab{a}})}]{lobaskin2005CrossoverFromNonequlibriumToEquilibrium_FlowEquation}
\bibinfo{author}{\bibfnamefont{D.}~\bibnamefont{Lobaskin}} \bibnamefont{and}
  \bibinfo{author}{\bibfnamefont{S.}~\bibnamefont{Kehrein}},
  \bibinfo{journal}{Physical Review B} \textbf{\bibinfo{volume}{71}},
  \bibinfo{pages}{193303} (\bibinfo{year}{2005}{\natexlab{a}}).

\bibitem[{\citenamefont{Hackl et~al.}(2009)\citenamefont{Hackl, Roosen,
  Kehrein, and Hofstetter}}]{Hackl2009FM_Kondo_FlowEquation}
\bibinfo{author}{\bibfnamefont{A.}~\bibnamefont{Hackl}},
  \bibinfo{author}{\bibfnamefont{D.}~\bibnamefont{Roosen}},
  \bibinfo{author}{\bibfnamefont{S.}~\bibnamefont{Kehrein}}, \bibnamefont{and}
  \bibinfo{author}{\bibfnamefont{W.}~\bibnamefont{Hofstetter}},
  \bibinfo{journal}{Phys. Rev. Lett.} \textbf{\bibinfo{volume}{102}},
  \bibinfo{pages}{196601} (\bibinfo{year}{2009}),
  \urlprefix\url{https://link.aps.org/doi/10.1103/PhysRevLett.102.196601}.

\bibitem[{\citenamefont{Nuss et~al.}(2015)\citenamefont{Nuss, Ganahl, Arrigoni,
  von~der Linden, and Evertz}}]{nuss2015nonequilibriumKondoTEBD}
\bibinfo{author}{\bibfnamefont{M.}~\bibnamefont{Nuss}},
  \bibinfo{author}{\bibfnamefont{M.}~\bibnamefont{Ganahl}},
  \bibinfo{author}{\bibfnamefont{E.}~\bibnamefont{Arrigoni}},
  \bibinfo{author}{\bibfnamefont{W.}~\bibnamefont{von~der Linden}},
  \bibnamefont{and} \bibinfo{author}{\bibfnamefont{H.~G.}
  \bibnamefont{Evertz}}, \bibinfo{journal}{Physical Review B}
  \textbf{\bibinfo{volume}{91}}, \bibinfo{pages}{085127}
  (\bibinfo{year}{2015}).

\bibitem[{\citenamefont{D{\'o}ra et~al.}(2017)\citenamefont{D{\'o}ra, Werner,
  and Moca}}]{dora2017information}
\bibinfo{author}{\bibfnamefont{B.}~\bibnamefont{D{\'o}ra}},
  \bibinfo{author}{\bibfnamefont{M.~A.} \bibnamefont{Werner}},
  \bibnamefont{and} \bibinfo{author}{\bibfnamefont{C.~P.} \bibnamefont{Moca}},
  \bibinfo{journal}{Physical Review B} \textbf{\bibinfo{volume}{96}},
  \bibinfo{pages}{155116} (\bibinfo{year}{2017}).

\bibitem[{\citenamefont{Guinea et~al.}(1985)\citenamefont{Guinea, Hakim, and
  Muramatsu}}]{guinea1985bosonization}
\bibinfo{author}{\bibfnamefont{F.}~\bibnamefont{Guinea}},
  \bibinfo{author}{\bibfnamefont{V.}~\bibnamefont{Hakim}}, \bibnamefont{and}
  \bibinfo{author}{\bibfnamefont{A.}~\bibnamefont{Muramatsu}},
  \bibinfo{journal}{Physical Review B} \textbf{\bibinfo{volume}{32}},
  \bibinfo{pages}{4410} (\bibinfo{year}{1985}).

\bibitem[{\citenamefont{Affleck and
  Ludwig}(1991)}]{Affleck1991KondoConformalFieldTheory}
\bibinfo{author}{\bibfnamefont{I.}~\bibnamefont{Affleck}} \bibnamefont{and}
  \bibinfo{author}{\bibfnamefont{A.~W.} \bibnamefont{Ludwig}},
  \bibinfo{journal}{Nuclear Physics B} \textbf{\bibinfo{volume}{352}},
  \bibinfo{pages}{849} (\bibinfo{year}{1991}).

\bibitem[{\citenamefont{Affleck et~al.}(1992)\citenamefont{Affleck, Ludwig,
  Pang, and Cox}}]{Affleck1992KondoConformalFieldTheory}
\bibinfo{author}{\bibfnamefont{I.}~\bibnamefont{Affleck}},
  \bibinfo{author}{\bibfnamefont{A.~W.~W.} \bibnamefont{Ludwig}},
  \bibinfo{author}{\bibfnamefont{H.-B.} \bibnamefont{Pang}}, \bibnamefont{and}
  \bibinfo{author}{\bibfnamefont{D.~L.} \bibnamefont{Cox}},
  \bibinfo{journal}{Phys. Rev. B} \textbf{\bibinfo{volume}{45}},
  \bibinfo{pages}{7918} (\bibinfo{year}{1992}),
  \urlprefix\url{https://link.aps.org/doi/10.1103/PhysRevB.45.7918}.

\bibitem[{\citenamefont{Lesage et~al.}(1996)\citenamefont{Lesage, Saleur, and
  Skorik}}]{Lesagne1996TimeCorrelationsExact}
\bibinfo{author}{\bibfnamefont{F.}~\bibnamefont{Lesage}},
  \bibinfo{author}{\bibfnamefont{H.}~\bibnamefont{Saleur}}, \bibnamefont{and}
  \bibinfo{author}{\bibfnamefont{S.}~\bibnamefont{Skorik}},
  \bibinfo{journal}{Phys. Rev. Lett.} \textbf{\bibinfo{volume}{76}},
  \bibinfo{pages}{3388} (\bibinfo{year}{1996}),
  \urlprefix\url{https://link.aps.org/doi/10.1103/PhysRevLett.76.3388}.

\bibitem[{\citenamefont{Lesage and Saleur}(1998)}]{Lesage1998BoundaryExact}
\bibinfo{author}{\bibfnamefont{F.}~\bibnamefont{Lesage}} \bibnamefont{and}
  \bibinfo{author}{\bibfnamefont{H.}~\bibnamefont{Saleur}},
  \bibinfo{journal}{Physical review letters} \textbf{\bibinfo{volume}{80}},
  \bibinfo{pages}{4370} (\bibinfo{year}{1998}).

\bibitem[{\citenamefont{Medvedyeva et~al.}(2013)\citenamefont{Medvedyeva,
  Hoffmann, and Kehrein}}]{medvedyeva2013spatiotemporal}
\bibinfo{author}{\bibfnamefont{M.}~\bibnamefont{Medvedyeva}},
  \bibinfo{author}{\bibfnamefont{A.}~\bibnamefont{Hoffmann}}, \bibnamefont{and}
  \bibinfo{author}{\bibfnamefont{S.}~\bibnamefont{Kehrein}},
  \bibinfo{journal}{Physical Review B} \textbf{\bibinfo{volume}{88}},
  \bibinfo{pages}{094306} (\bibinfo{year}{2013}).

\bibitem[{\citenamefont{Fowler and Zawadowski}(1971)}]{Fowler1971ScalingKondo}
\bibinfo{author}{\bibfnamefont{M.}~\bibnamefont{Fowler}} \bibnamefont{and}
  \bibinfo{author}{\bibfnamefont{A.}~\bibnamefont{Zawadowski}},
  \bibinfo{journal}{Solid State Communications} \textbf{\bibinfo{volume}{9}},
  \bibinfo{pages}{471} (\bibinfo{year}{1971}).

\bibitem[{\citenamefont{Nordlander et~al.}(1999)\citenamefont{Nordlander,
  Pustilnik, Meir, Wingreen, and
  Langreth}}]{nordlander1999HowLongDoesItTakeForTheKondoEffectToDevelop}
\bibinfo{author}{\bibfnamefont{P.}~\bibnamefont{Nordlander}},
  \bibinfo{author}{\bibfnamefont{M.}~\bibnamefont{Pustilnik}},
  \bibinfo{author}{\bibfnamefont{Y.}~\bibnamefont{Meir}},
  \bibinfo{author}{\bibfnamefont{N.~S.} \bibnamefont{Wingreen}},
  \bibnamefont{and} \bibinfo{author}{\bibfnamefont{D.~C.}
  \bibnamefont{Langreth}}, \bibinfo{journal}{Physical review letters}
  \textbf{\bibinfo{volume}{83}}, \bibinfo{pages}{808} (\bibinfo{year}{1999}).

\bibitem[{\citenamefont{Keil and
  Schoeller}(2001)}]{Keil2001PerturbativeRGSpinBoson}
\bibinfo{author}{\bibfnamefont{M.}~\bibnamefont{Keil}} \bibnamefont{and}
  \bibinfo{author}{\bibfnamefont{H.}~\bibnamefont{Schoeller}},
  \bibinfo{journal}{Phys. Rev. B} \textbf{\bibinfo{volume}{63}},
  \bibinfo{pages}{180302} (\bibinfo{year}{2001}),
  \urlprefix\url{https://link.aps.org/doi/10.1103/PhysRevB.63.180302}.

\bibitem[{\citenamefont{Lobaskin and
  Kehrein}(2005{\natexlab{b}})}]{Lobaskin2005PerturbativeRG_FM_Kondo}
\bibinfo{author}{\bibfnamefont{D.}~\bibnamefont{Lobaskin}} \bibnamefont{and}
  \bibinfo{author}{\bibfnamefont{S.}~\bibnamefont{Kehrein}},
  \bibinfo{journal}{Phys. Rev. B} \textbf{\bibinfo{volume}{71}},
  \bibinfo{pages}{193303} (\bibinfo{year}{2005}{\natexlab{b}}),
  \urlprefix\url{https://link.aps.org/doi/10.1103/PhysRevB.71.193303}.

\bibitem[{\citenamefont{Hackl and
  Kehrein}(2008)}]{Hackl2008PerturbativeRGSpinBoson}
\bibinfo{author}{\bibfnamefont{A.}~\bibnamefont{Hackl}} \bibnamefont{and}
  \bibinfo{author}{\bibfnamefont{S.}~\bibnamefont{Kehrein}},
  \bibinfo{journal}{Phys. Rev. B} \textbf{\bibinfo{volume}{78}},
  \bibinfo{pages}{092303} (\bibinfo{year}{2008}),
  \urlprefix\url{https://link.aps.org/doi/10.1103/PhysRevB.78.092303}.

\bibitem[{\citenamefont{Pletyukhov et~al.}(2010)\citenamefont{Pletyukhov,
  Schuricht, and
  Schoeller}}]{pletyukhov2010RelaxationPerturbativeRG_WeakCoupling}
\bibinfo{author}{\bibfnamefont{M.}~\bibnamefont{Pletyukhov}},
  \bibinfo{author}{\bibfnamefont{D.}~\bibnamefont{Schuricht}},
  \bibnamefont{and}
  \bibinfo{author}{\bibfnamefont{H.}~\bibnamefont{Schoeller}},
  \bibinfo{journal}{Physical review letters} \textbf{\bibinfo{volume}{104}},
  \bibinfo{pages}{106801} (\bibinfo{year}{2010}).

\bibitem[{\citenamefont{Shi et~al.}(2017)\citenamefont{Shi, Demler, and
  Cirac}}]{TaoPaper}
\bibinfo{author}{\bibfnamefont{T.}~\bibnamefont{Shi}},
  \bibinfo{author}{\bibfnamefont{E.}~\bibnamefont{Demler}}, \bibnamefont{and}
  \bibinfo{author}{\bibfnamefont{J.~I.} \bibnamefont{Cirac}},
  \bibinfo{journal}{arXiv preprint arXiv:1707.05902}  (\bibinfo{year}{2017}).

\bibitem[{\citenamefont{Ashida et~al.}()\citenamefont{Ashida, Shi, Ba{\~n}uls,
  Cirac, and Demler}}]{yuto2017ToBePublished}
\bibinfo{author}{\bibfnamefont{Y.}~\bibnamefont{Ashida}},
  \bibinfo{author}{\bibfnamefont{T.}~\bibnamefont{Shi}},
  \bibinfo{author}{\bibfnamefont{M.-C.} \bibnamefont{Ba{\~n}uls}},
  \bibinfo{author}{\bibfnamefont{J.~I.} \bibnamefont{Cirac}}, \bibnamefont{and}
  \bibinfo{author}{\bibfnamefont{E.}~\bibnamefont{Demler}}, \bibinfo{note}{in
  preparation}.

\bibitem[{\citenamefont{Knap et~al.}(2012)\citenamefont{Knap, Shashi, Nishida,
  Imambekov, Abanin, and Demler}}]{Knap2012}
\bibinfo{author}{\bibfnamefont{M.}~\bibnamefont{Knap}},
  \bibinfo{author}{\bibfnamefont{A.}~\bibnamefont{Shashi}},
  \bibinfo{author}{\bibfnamefont{Y.}~\bibnamefont{Nishida}},
  \bibinfo{author}{\bibfnamefont{A.}~\bibnamefont{Imambekov}},
  \bibinfo{author}{\bibfnamefont{D.~A.} \bibnamefont{Abanin}},
  \bibnamefont{and} \bibinfo{author}{\bibfnamefont{E.}~\bibnamefont{Demler}},
  \bibinfo{journal}{Physical Review X} \textbf{\bibinfo{volume}{2}},
  \bibinfo{pages}{041020} (\bibinfo{year}{2012}), ISSN
  \bibinfo{issn}{2160-3308},
  \urlprefix\url{http://link.aps.org/doi/10.1103/PhysRevX.2.041020}.

\bibitem[{\citenamefont{Recati et~al.}(2005)\citenamefont{Recati, Fedichev,
  Zwerger, Von~Delft, and Zoller}}]{recati2005atomic}
\bibinfo{author}{\bibfnamefont{A.}~\bibnamefont{Recati}},
  \bibinfo{author}{\bibfnamefont{P.}~\bibnamefont{Fedichev}},
  \bibinfo{author}{\bibfnamefont{W.}~\bibnamefont{Zwerger}},
  \bibinfo{author}{\bibfnamefont{J.}~\bibnamefont{Von~Delft}},
  \bibnamefont{and} \bibinfo{author}{\bibfnamefont{P.}~\bibnamefont{Zoller}},
  \bibinfo{journal}{Physical review letters} \textbf{\bibinfo{volume}{94}},
  \bibinfo{pages}{040404} (\bibinfo{year}{2005}).

\bibitem[{\citenamefont{Bauer et~al.}(2013)\citenamefont{Bauer, Salomon, and
  Demler}}]{Bauer2013}
\bibinfo{author}{\bibfnamefont{J.}~\bibnamefont{Bauer}},
  \bibinfo{author}{\bibfnamefont{C.}~\bibnamefont{Salomon}}, \bibnamefont{and}
  \bibinfo{author}{\bibfnamefont{E.}~\bibnamefont{Demler}},
  \bibinfo{journal}{Physical Review Letters} \textbf{\bibinfo{volume}{111}},
  \bibinfo{pages}{215304} (\bibinfo{year}{2013}), ISSN
  \bibinfo{issn}{0031-9007},
  \urlprefix\url{http://link.aps.org/doi/10.1103/PhysRevLett.111.215304}.

\bibitem[{\citenamefont{Lamacraft}(2008)}]{Lamacraft2008KondoIn1D}
\bibinfo{author}{\bibfnamefont{A.}~\bibnamefont{Lamacraft}},
  \bibinfo{journal}{Physical review letters} \textbf{\bibinfo{volume}{101}},
  \bibinfo{pages}{225301} (\bibinfo{year}{2008}).

\bibitem[{\citenamefont{Withoff and
  Fradkin}(1990{\natexlab{a}})}]{withoff1990phase}
\bibinfo{author}{\bibfnamefont{D.}~\bibnamefont{Withoff}} \bibnamefont{and}
  \bibinfo{author}{\bibfnamefont{E.}~\bibnamefont{Fradkin}},
  \bibinfo{journal}{Physical review letters} \textbf{\bibinfo{volume}{64}},
  \bibinfo{pages}{1835} (\bibinfo{year}{1990}{\natexlab{a}}).

\bibitem[{\citenamefont{Withoff and
  Fradkin}(1990{\natexlab{b}})}]{WithoffFradkin1990Kondo}
\bibinfo{author}{\bibfnamefont{D.}~\bibnamefont{Withoff}} \bibnamefont{and}
  \bibinfo{author}{\bibfnamefont{E.}~\bibnamefont{Fradkin}},
  \bibinfo{journal}{Physical review letters} \textbf{\bibinfo{volume}{64}},
  \bibinfo{pages}{1835} (\bibinfo{year}{1990}{\natexlab{b}}).

\bibitem[{\citenamefont{Borkowski and Hirschfeld}(1992)}]{borkowski1992kondo}
\bibinfo{author}{\bibfnamefont{L.~S.} \bibnamefont{Borkowski}}
  \bibnamefont{and}
  \bibinfo{author}{\bibfnamefont{P.}~\bibnamefont{Hirschfeld}},
  \bibinfo{journal}{Physical Review B} \textbf{\bibinfo{volume}{46}},
  \bibinfo{pages}{9274} (\bibinfo{year}{1992}).

\bibitem[{\citenamefont{Chen and Jayaprakash}(1995)}]{chen1995kondo}
\bibinfo{author}{\bibfnamefont{K.}~\bibnamefont{Chen}} \bibnamefont{and}
  \bibinfo{author}{\bibfnamefont{C.}~\bibnamefont{Jayaprakash}},
  \bibinfo{journal}{Journal of Physics: Condensed Matter}
  \textbf{\bibinfo{volume}{7}}, \bibinfo{pages}{L491} (\bibinfo{year}{1995}).

\bibitem[{\citenamefont{Ingersent}(1996)}]{ingersent1996gaplessFermi}
\bibinfo{author}{\bibfnamefont{K.}~\bibnamefont{Ingersent}},
  \bibinfo{journal}{Phys. Rev. B} \textbf{\bibinfo{volume}{54}},
  \bibinfo{pages}{11936} (\bibinfo{year}{1996}),
  \urlprefix\url{https://link.aps.org/doi/10.1103/PhysRevB.54.11936}.

\bibitem[{\citenamefont{Hentschel and
  Guinea}(2007)}]{hentschel2007orthogonality}
\bibinfo{author}{\bibfnamefont{M.}~\bibnamefont{Hentschel}} \bibnamefont{and}
  \bibinfo{author}{\bibfnamefont{F.}~\bibnamefont{Guinea}},
  \bibinfo{journal}{Physical Review B} \textbf{\bibinfo{volume}{76}},
  \bibinfo{pages}{115407} (\bibinfo{year}{2007}).

\bibitem[{\citenamefont{Sengupta and Baskaran}(2008)}]{sengupta2008tuning}
\bibinfo{author}{\bibfnamefont{K.}~\bibnamefont{Sengupta}} \bibnamefont{and}
  \bibinfo{author}{\bibfnamefont{G.}~\bibnamefont{Baskaran}},
  \bibinfo{journal}{Physical Review B} \textbf{\bibinfo{volume}{77}},
  \bibinfo{pages}{045417} (\bibinfo{year}{2008}).

\bibitem[{\citenamefont{Jones and
  Varma}(1987)}]{JonesVarma1987TwoKondoImpurities}
\bibinfo{author}{\bibfnamefont{B.~A.} \bibnamefont{Jones}} \bibnamefont{and}
  \bibinfo{author}{\bibfnamefont{C.~M.} \bibnamefont{Varma}},
  \bibinfo{journal}{Phys. Rev. Lett.} \textbf{\bibinfo{volume}{58}},
  \bibinfo{pages}{843} (\bibinfo{year}{1987}),
  \urlprefix\url{https://link.aps.org/doi/10.1103/PhysRevLett.58.843}.

\bibitem[{\citenamefont{Jones et~al.}(1988)\citenamefont{Jones, Varma, and
  Wilkins}}]{JonesVarma1988TwoKondoImpurities}
\bibinfo{author}{\bibfnamefont{B.~A.} \bibnamefont{Jones}},
  \bibinfo{author}{\bibfnamefont{C.~M.} \bibnamefont{Varma}}, \bibnamefont{and}
  \bibinfo{author}{\bibfnamefont{J.~W.} \bibnamefont{Wilkins}},
  \bibinfo{journal}{Phys. Rev. Lett.} \textbf{\bibinfo{volume}{61}},
  \bibinfo{pages}{125} (\bibinfo{year}{1988}),
  \urlprefix\url{https://link.aps.org/doi/10.1103/PhysRevLett.61.125}.

\bibitem[{\citenamefont{Affleck and
  Ludwig}(1992)}]{Affleck1992TwoKondoImpurities}
\bibinfo{author}{\bibfnamefont{I.}~\bibnamefont{Affleck}} \bibnamefont{and}
  \bibinfo{author}{\bibfnamefont{A.~W.~W.} \bibnamefont{Ludwig}},
  \bibinfo{journal}{Phys. Rev. Lett.} \textbf{\bibinfo{volume}{68}},
  \bibinfo{pages}{1046} (\bibinfo{year}{1992}),
  \urlprefix\url{https://link.aps.org/doi/10.1103/PhysRevLett.68.1046}.

\bibitem[{\citenamefont{Affleck et~al.}(1995)\citenamefont{Affleck, Ludwig, and
  Jones}}]{Affleck1995TwoKondoImpurities}
\bibinfo{author}{\bibfnamefont{I.}~\bibnamefont{Affleck}},
  \bibinfo{author}{\bibfnamefont{A.~W.} \bibnamefont{Ludwig}},
  \bibnamefont{and} \bibinfo{author}{\bibfnamefont{B.~A.} \bibnamefont{Jones}},
  \bibinfo{journal}{Physical Review B} \textbf{\bibinfo{volume}{52}},
  \bibinfo{pages}{9528} (\bibinfo{year}{1995}).

\bibitem[{\citenamefont{Zar{\'a}nd et~al.}(2006)\citenamefont{Zar{\'a}nd,
  Chung, Simon, and Vojta}}]{zarand2006quantum}
\bibinfo{author}{\bibfnamefont{G.}~\bibnamefont{Zar{\'a}nd}},
  \bibinfo{author}{\bibfnamefont{C.-H.} \bibnamefont{Chung}},
  \bibinfo{author}{\bibfnamefont{P.}~\bibnamefont{Simon}}, \bibnamefont{and}
  \bibinfo{author}{\bibfnamefont{M.}~\bibnamefont{Vojta}},
  \bibinfo{journal}{Physical review letters} \textbf{\bibinfo{volume}{97}},
  \bibinfo{pages}{166802} (\bibinfo{year}{2006}).

\bibitem[{\citenamefont{Billy et~al.}(2008)\citenamefont{Billy, Josse, Zuo,
  Bernard, Hambrecht, Lugan, Cl{\'e}ment, Sanchez-Palencia, Bouyer, and
  Aspect}}]{billy2008direct}
\bibinfo{author}{\bibfnamefont{J.}~\bibnamefont{Billy}},
  \bibinfo{author}{\bibfnamefont{V.}~\bibnamefont{Josse}},
  \bibinfo{author}{\bibfnamefont{Z.}~\bibnamefont{Zuo}},
  \bibinfo{author}{\bibfnamefont{A.}~\bibnamefont{Bernard}},
  \bibinfo{author}{\bibfnamefont{B.}~\bibnamefont{Hambrecht}},
  \bibinfo{author}{\bibfnamefont{P.}~\bibnamefont{Lugan}},
  \bibinfo{author}{\bibfnamefont{D.}~\bibnamefont{Cl{\'e}ment}},
  \bibinfo{author}{\bibfnamefont{L.}~\bibnamefont{Sanchez-Palencia}},
  \bibinfo{author}{\bibfnamefont{P.}~\bibnamefont{Bouyer}}, \bibnamefont{and}
  \bibinfo{author}{\bibfnamefont{A.}~\bibnamefont{Aspect}},
  \bibinfo{journal}{Nature} \textbf{\bibinfo{volume}{453}},
  \bibinfo{pages}{891} (\bibinfo{year}{2008}).

\bibitem[{\citenamefont{Roati et~al.}(2008)\citenamefont{Roati, D’Errico,
  Fallani, Fattori, Fort, Zaccanti, Modugno, Modugno, and
  Inguscio}}]{roati2008anderson}
\bibinfo{author}{\bibfnamefont{G.}~\bibnamefont{Roati}},
  \bibinfo{author}{\bibfnamefont{C.}~\bibnamefont{D’Errico}},
  \bibinfo{author}{\bibfnamefont{L.}~\bibnamefont{Fallani}},
  \bibinfo{author}{\bibfnamefont{M.}~\bibnamefont{Fattori}},
  \bibinfo{author}{\bibfnamefont{C.}~\bibnamefont{Fort}},
  \bibinfo{author}{\bibfnamefont{M.}~\bibnamefont{Zaccanti}},
  \bibinfo{author}{\bibfnamefont{G.}~\bibnamefont{Modugno}},
  \bibinfo{author}{\bibfnamefont{M.}~\bibnamefont{Modugno}}, \bibnamefont{and}
  \bibinfo{author}{\bibfnamefont{M.}~\bibnamefont{Inguscio}},
  \bibinfo{journal}{Nature} \textbf{\bibinfo{volume}{453}},
  \bibinfo{pages}{895} (\bibinfo{year}{2008}).

\bibitem[{\citenamefont{Jendrzejewski et~al.}(2012)\citenamefont{Jendrzejewski,
  Bernard, Mueller, Cheinet, Josse, Piraud, Pezz{\'e}, Sanchez-Palencia,
  Aspect, and Bouyer}}]{jendrzejewski2012three}
\bibinfo{author}{\bibfnamefont{F.}~\bibnamefont{Jendrzejewski}},
  \bibinfo{author}{\bibfnamefont{A.}~\bibnamefont{Bernard}},
  \bibinfo{author}{\bibfnamefont{K.}~\bibnamefont{Mueller}},
  \bibinfo{author}{\bibfnamefont{P.}~\bibnamefont{Cheinet}},
  \bibinfo{author}{\bibfnamefont{V.}~\bibnamefont{Josse}},
  \bibinfo{author}{\bibfnamefont{M.}~\bibnamefont{Piraud}},
  \bibinfo{author}{\bibfnamefont{L.}~\bibnamefont{Pezz{\'e}}},
  \bibinfo{author}{\bibfnamefont{L.}~\bibnamefont{Sanchez-Palencia}},
  \bibinfo{author}{\bibfnamefont{A.}~\bibnamefont{Aspect}}, \bibnamefont{and}
  \bibinfo{author}{\bibfnamefont{P.}~\bibnamefont{Bouyer}},
  \bibinfo{journal}{Nature Physics} \textbf{\bibinfo{volume}{8}},
  \bibinfo{pages}{398} (\bibinfo{year}{2012}).

\bibitem[{\citenamefont{Dobrosavljevi\ifmmode~\acute{c}\else \'{c}\fi{}
  et~al.}(1992)\citenamefont{Dobrosavljevi\ifmmode~\acute{c}\else \'{c}\fi{},
  Kirkpatrick, and Kotliar}}]{Dobrosavljeviifmmode1992DisorderedKondo}
\bibinfo{author}{\bibfnamefont{V.}~\bibnamefont{Dobrosavljevi\ifmmode~\acute{c}\else
  \'{c}\fi{}}}, \bibinfo{author}{\bibfnamefont{T.~R.}
  \bibnamefont{Kirkpatrick}}, \bibnamefont{and}
  \bibinfo{author}{\bibfnamefont{B.~G.} \bibnamefont{Kotliar}},
  \bibinfo{journal}{Phys. Rev. Lett.} \textbf{\bibinfo{volume}{69}},
  \bibinfo{pages}{1113} (\bibinfo{year}{1992}),
  \urlprefix\url{https://link.aps.org/doi/10.1103/PhysRevLett.69.1113}.

\bibitem[{\citenamefont{Jones}(1998)}]{JonesBookSymmetries}
\bibinfo{author}{\bibfnamefont{H.~F.} \bibnamefont{Jones}},
  \emph{\bibinfo{title}{Groups, representations and physics}}
  (\bibinfo{publisher}{CRC Press}, \bibinfo{year}{1998}).

\bibitem[{\citenamefont{Vladar and
  Zawadowski}(1983)}]{VladarZawadowski1983theory1}
\bibinfo{author}{\bibfnamefont{K.}~\bibnamefont{Vladar}} \bibnamefont{and}
  \bibinfo{author}{\bibfnamefont{A.}~\bibnamefont{Zawadowski}},
  \bibinfo{journal}{Physical Review B} \textbf{\bibinfo{volume}{28}},
  \bibinfo{pages}{1564} (\bibinfo{year}{1983}).

\bibitem[{\citenamefont{Vlad{\'a}r and
  Zawadowski}(1983)}]{VladarZawadowski1983theory2}
\bibinfo{author}{\bibfnamefont{K.}~\bibnamefont{Vlad{\'a}r}} \bibnamefont{and}
  \bibinfo{author}{\bibfnamefont{A.}~\bibnamefont{Zawadowski}},
  \bibinfo{journal}{Physical Review B} \textbf{\bibinfo{volume}{28}},
  \bibinfo{pages}{1582} (\bibinfo{year}{1983}).

\bibitem[{\citenamefont{Zar\'and and Vlad\'ar}(1996)}]{VladarZarand1996Low}
\bibinfo{author}{\bibfnamefont{G.}~\bibnamefont{Zar\'and}} \bibnamefont{and}
  \bibinfo{author}{\bibfnamefont{K.}~\bibnamefont{Vlad\'ar}},
  \bibinfo{journal}{Phys. Rev. Lett.} \textbf{\bibinfo{volume}{76}},
  \bibinfo{pages}{2133} (\bibinfo{year}{1996}),
  \urlprefix\url{https://link.aps.org/doi/10.1103/PhysRevLett.76.2133}.

\end{thebibliography}

\end{document}